%% file: dtWL.tex
\documentclass[12pt]{article}

\pdfoutput=1
\usepackage[colorlinks=true,linkcolor=darkblue,citecolor=darkblue,urlcolor=darkblue]{hyperref}
\usepackage[T1]{fontenc}
\usepackage[ansinew]{inputenc}
\usepackage[english]{babel}
\usepackage{color}
\usepackage{epsfig, palatino}
\usepackage{pstricks,pst-node,pst-tree}
\usepackage{epic}
\usepackage{mathrsfs}
\usepackage{ae} 
\usepackage{amsmath}
\usepackage{amssymb}
\usepackage{graphicx}
\usepackage{ulem}
\usepackage{color}
\definecolor{darkblue}{cmyk}{0.9,0.9,0,0}
\usepackage{cite}
\usepackage{wasysym}
\usepackage{varioref}
\usepackage{makeidx}
\usepackage{simplewick}
\usepackage{array}
\usepackage{mathrsfs}
\usepackage{feynmf} 
\usepackage{tikz}
\usepackage{dsfont}
\usepackage{slashed}
\usepackage{subfig}
\usepackage{float}

\usetikzlibrary{decorations.pathmorphing}

\newcommand{\comment}[1]{}

\newcommand{\beq}{\begin{equation}}
\newcommand{\eeq}{\end{equation}}
\newcommand{\beqq}{\begin{equation*}}
\newcommand{\eeqq}{\end{equation*}}
\newcommand\beqa{\begin{eqnarray}}
\newcommand\eeqa{\end{eqnarray}}
\newcommand\beqaa{\begin{eqnarray*}}
\newcommand\eeqaa{\end{eqnarray*}}
\newcommand\bea{\begin{array}}
\newcommand\eea{\end{array}}

\newcommand{\nn}{\nonumber}

\newcommand{\neqa}{\nonumber\end{eqnarray}} 
\newcommand{\la}[1]{\label{#1}}

\renewcommand{\d}{\partial}

\newcommand{\<}{{\langle}}
\renewcommand{\>}{{\rangle}}

\newcommand{\cA}{{\cal A}}
\newcommand{\cB}{{\cal B}}

\newcommand{\cL}{{\cal L}}

\newcommand{\re}{\relax{\rm I\kern-.18em R}}

\renewcommand{\sp}{p\hspace{-.40em}/}

\newcommand{\Blue}[1]{{\color{blue}#1\color{blue}}}
\newcommand{\Red}[1]{{\color{red}#1\color{black}}}

\newcommand{\planarbox}{\raisebox{-4.5ex}{$\mathord{\includegraphics[height=10ex]{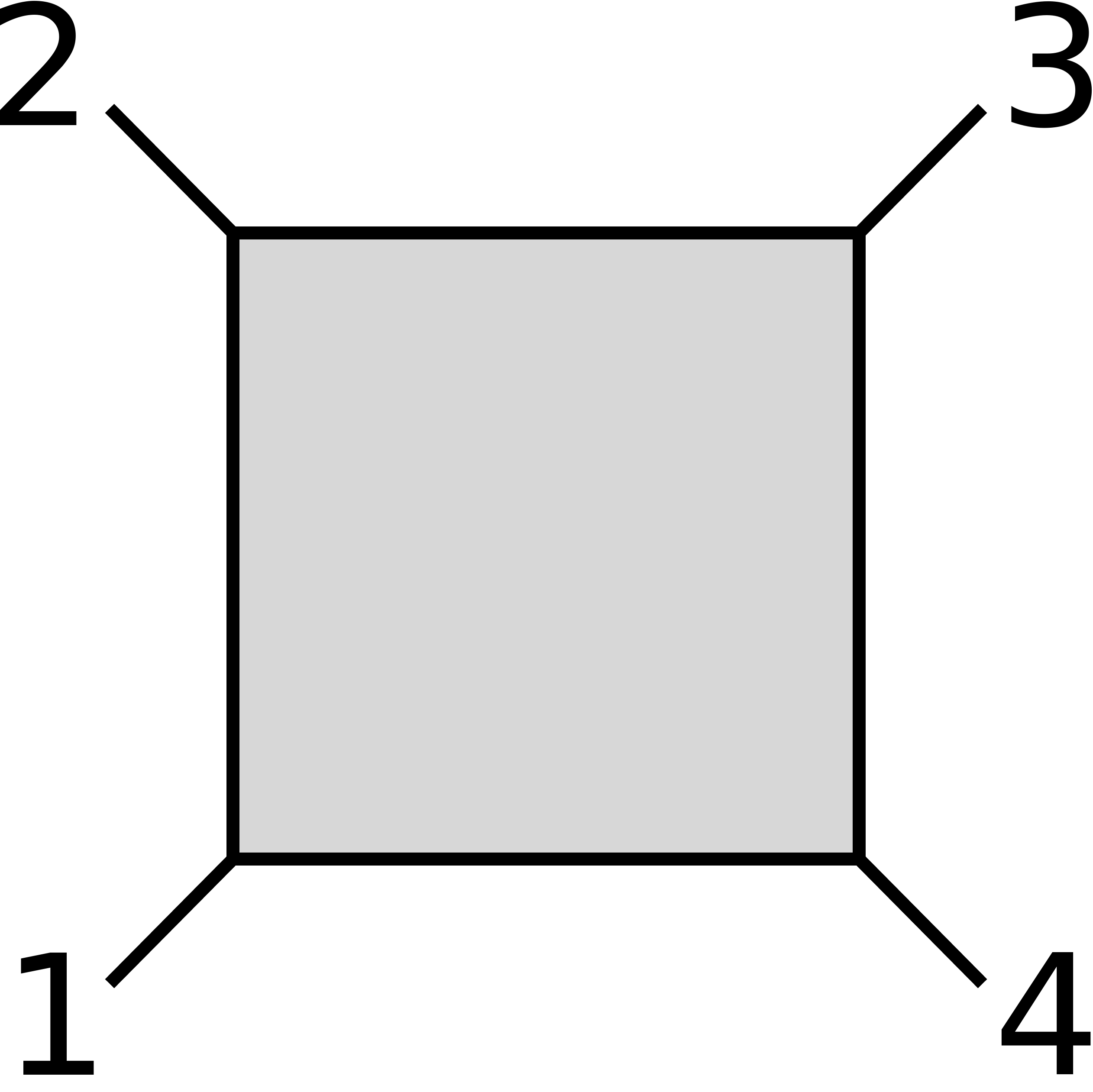}}$}}

\newcommand{\boxa}{\raisebox{-4.5ex}{$\mathord{\includegraphics[height=10ex]{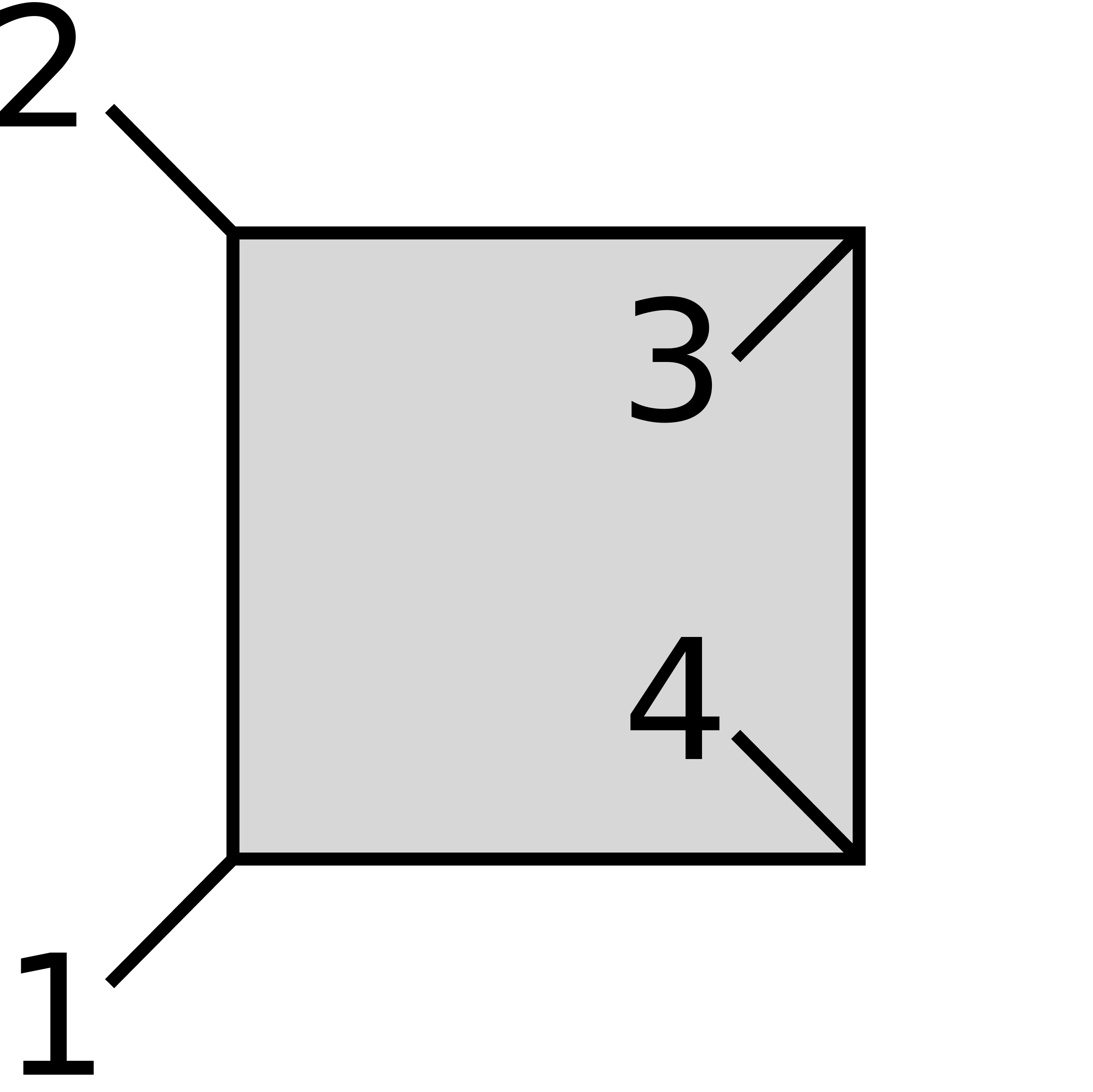}}$}}
\newcommand{\boxb}{\raisebox{-4.5ex}{$\mathord{\includegraphics[height=10ex]{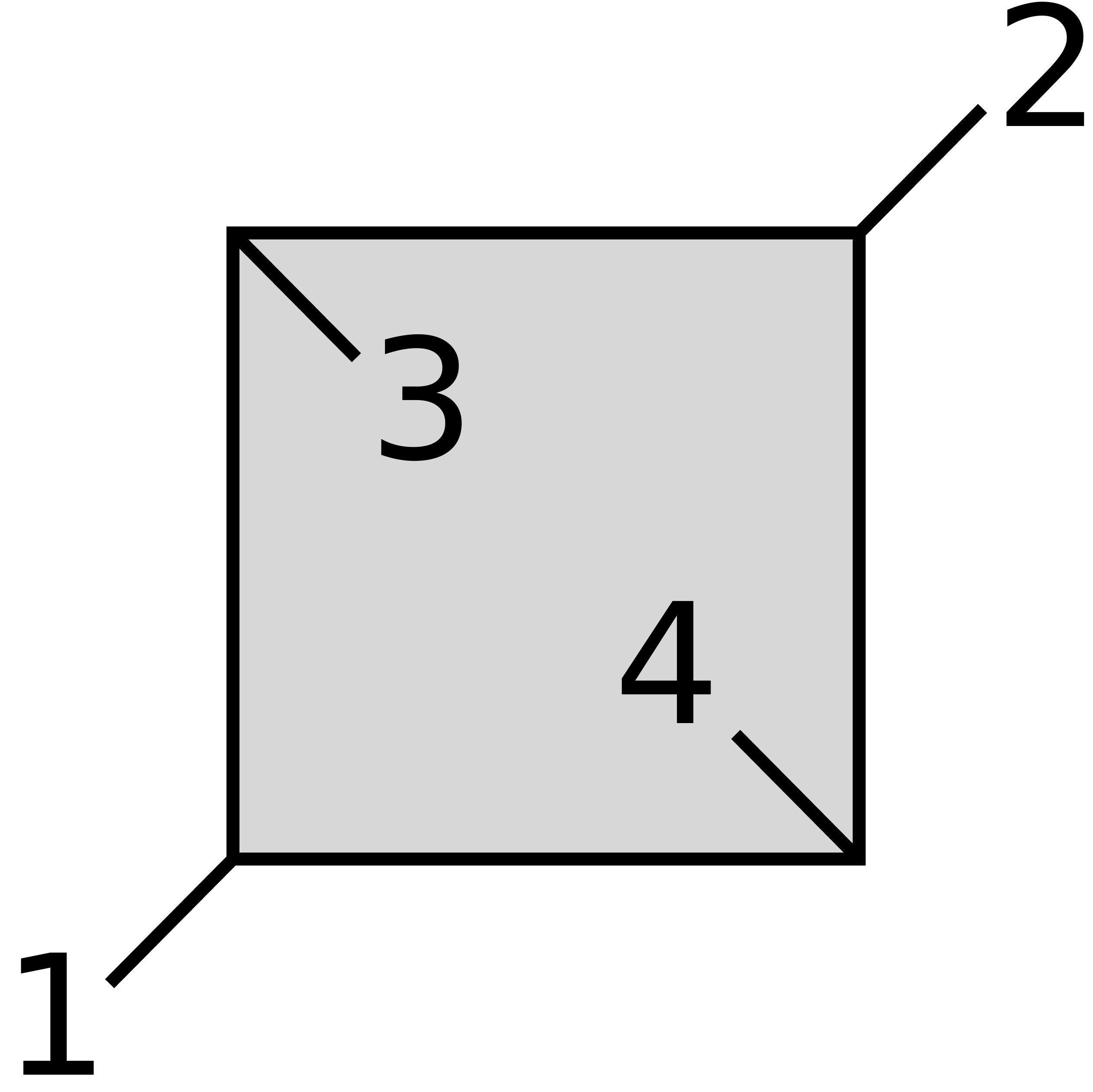}}$}}
\newcommand{\boxc}{\raisebox{-4.5ex}{$\mathord{\includegraphics[height=10ex]{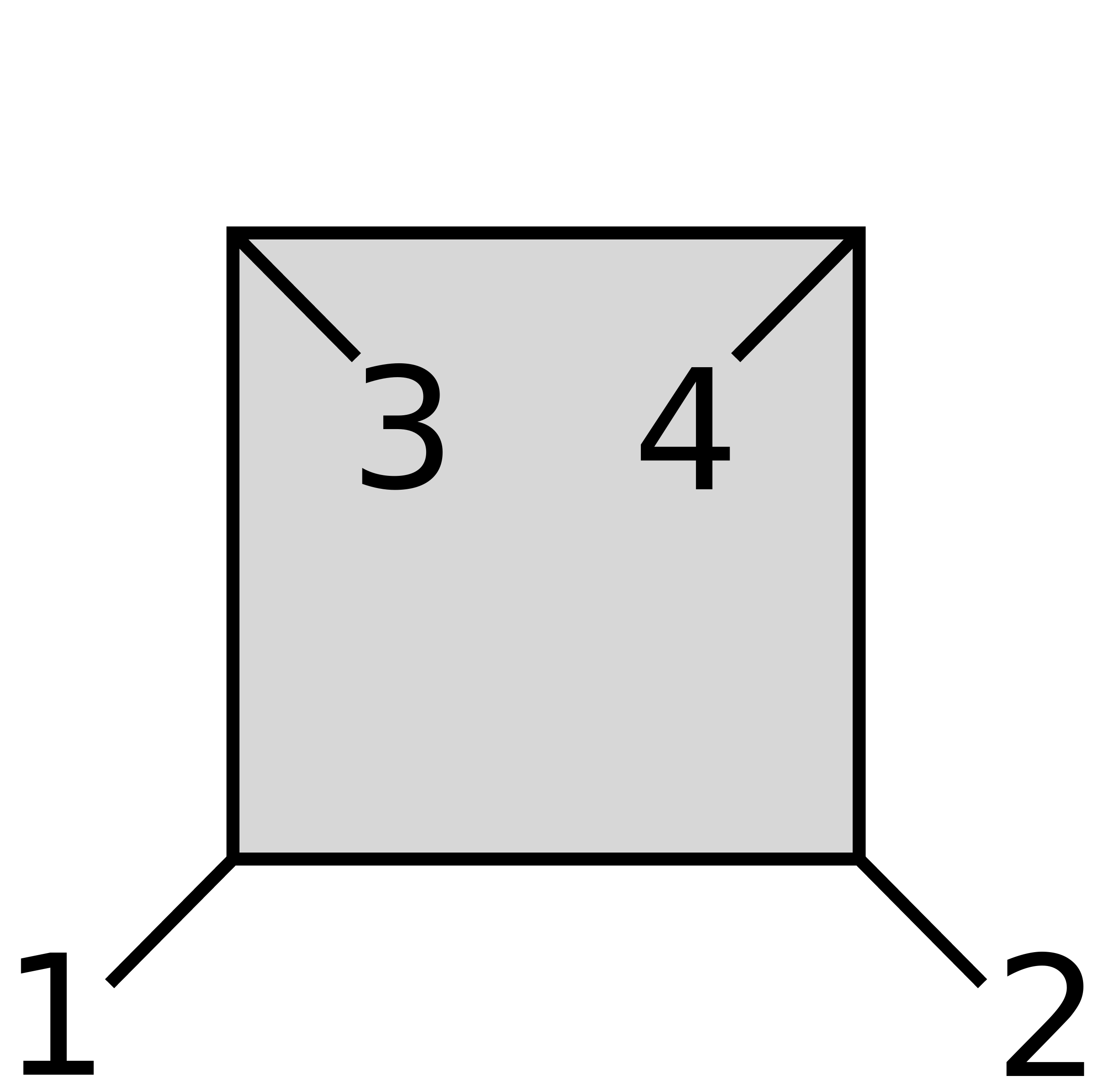}}$}}
\newcommand{\cboxa}{\raisebox{-4.5ex}{$\mathord{\includegraphics[height=10ex]{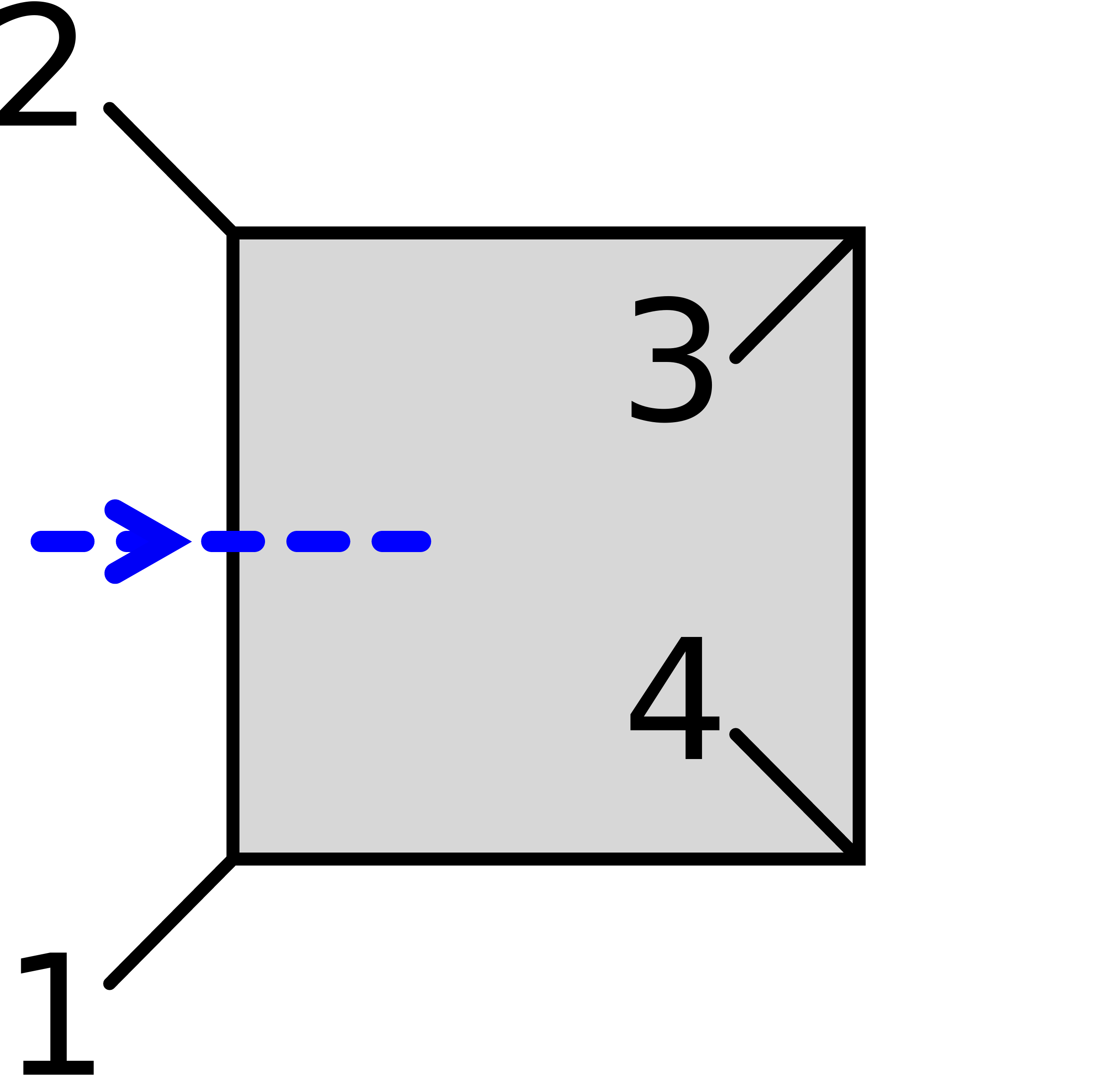}}$}}
\newcommand{\cboxb}{\raisebox{-4.5ex}{$\mathord{\includegraphics[height=10ex]{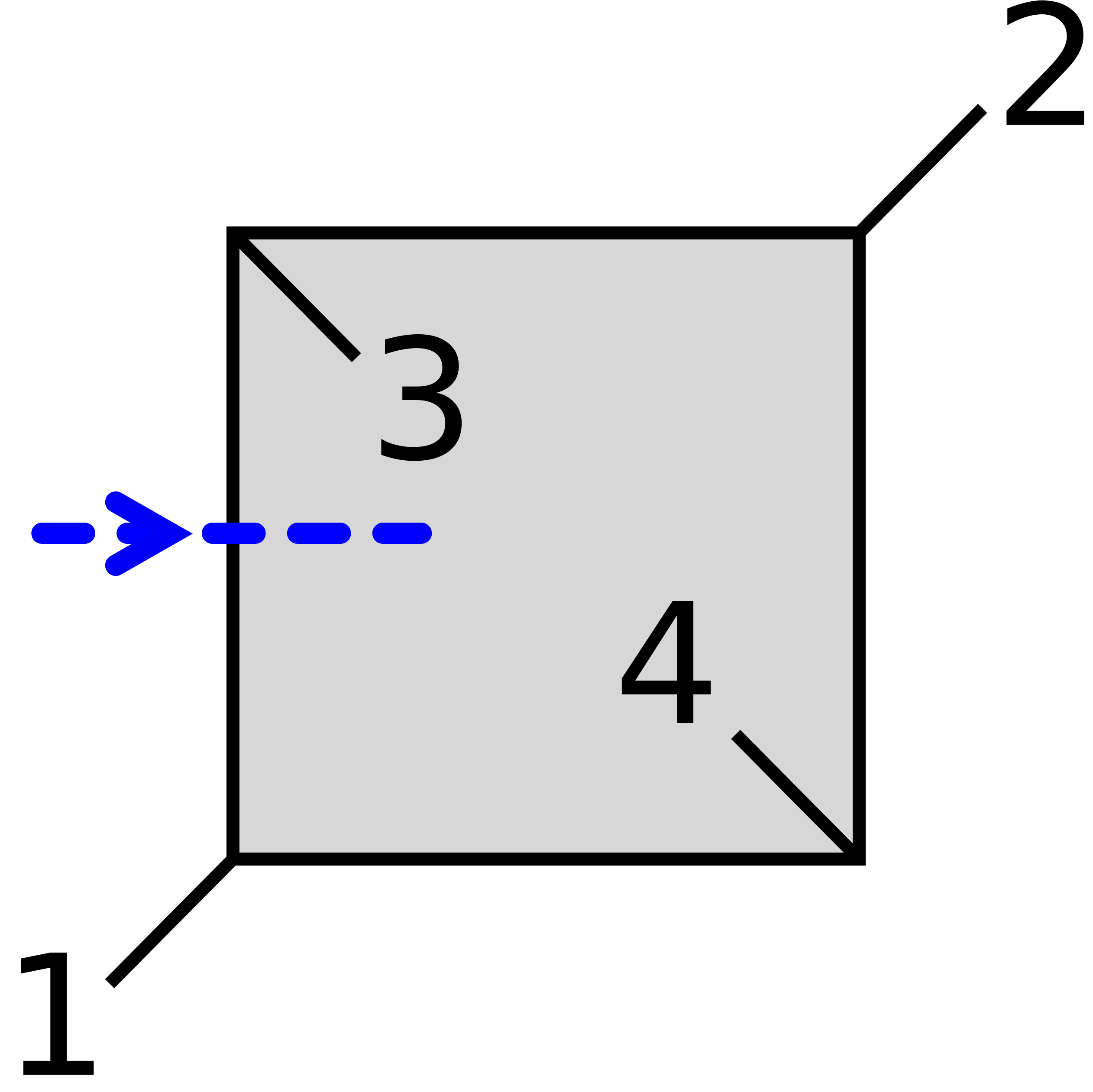}}$}}
\newcommand{\cboxc}{\raisebox{-4.5ex}{$\mathord{\includegraphics[height=10ex]{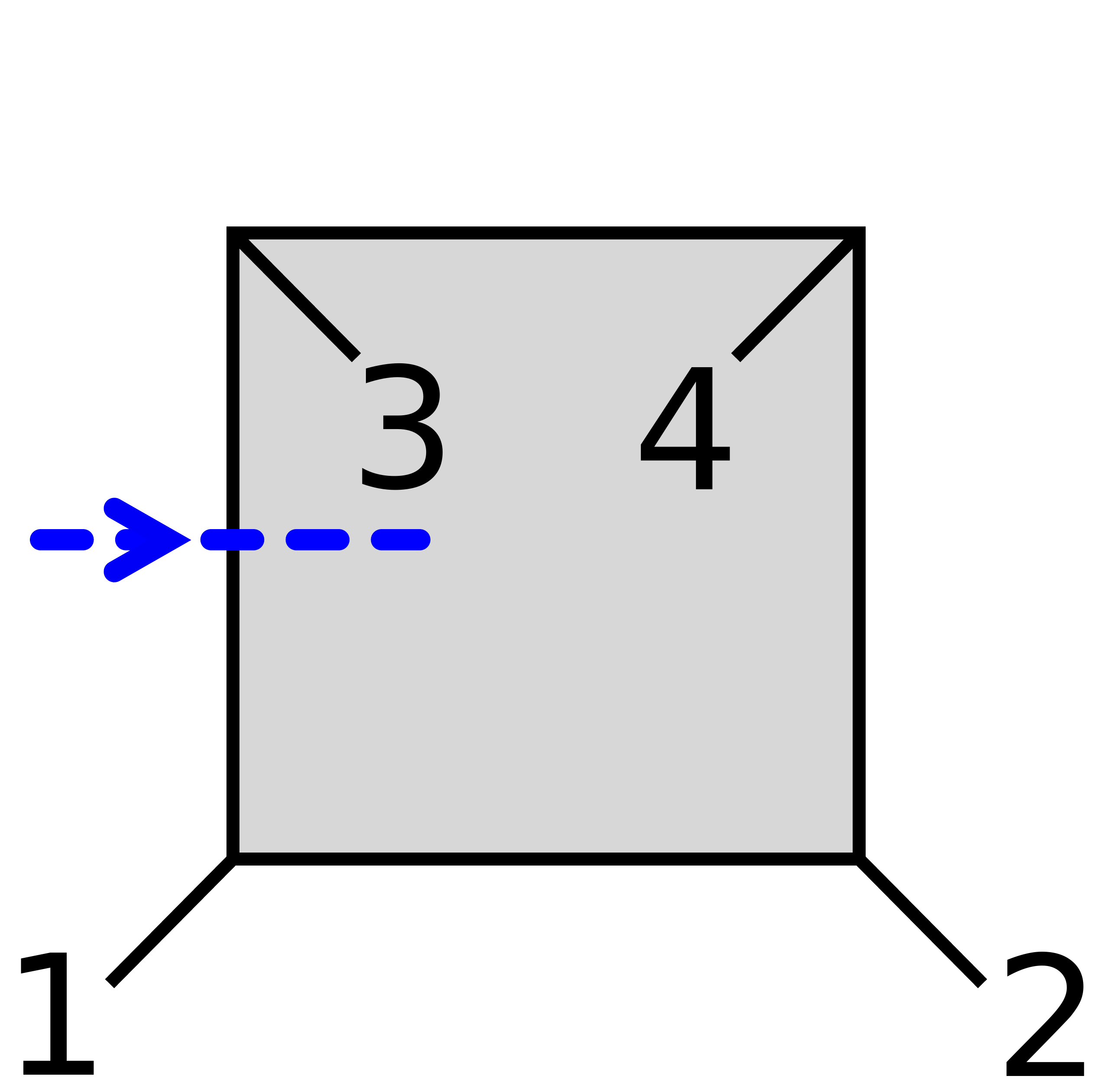}}$}}
\newcommand{\cboxaa}{\raisebox{-4.5ex}{$\mathord{\includegraphics[height=10ex]{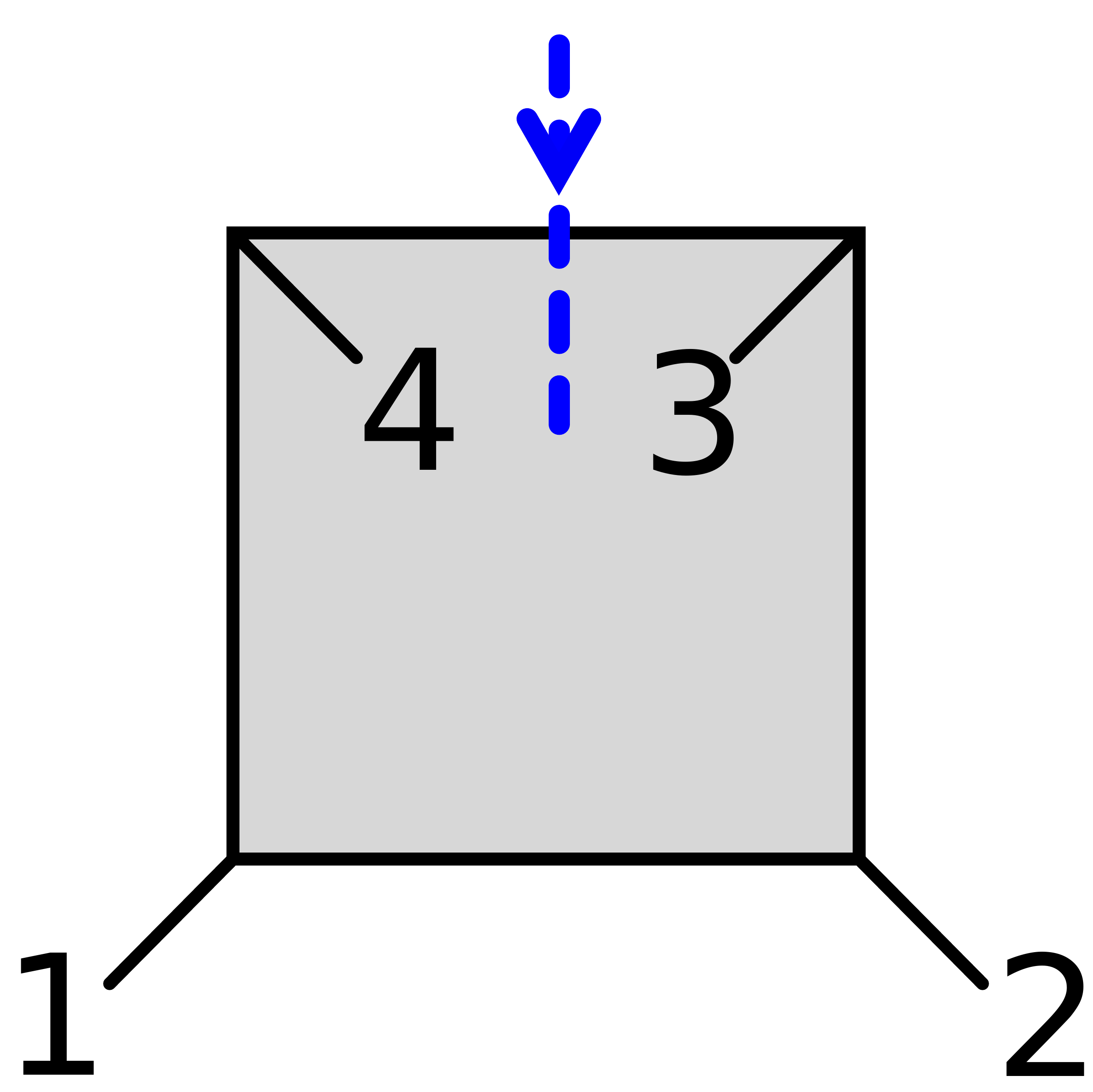}}$}}
\newcommand{\cboxbb}{\raisebox{-4.5ex}{$\mathord{\includegraphics[height=10ex]{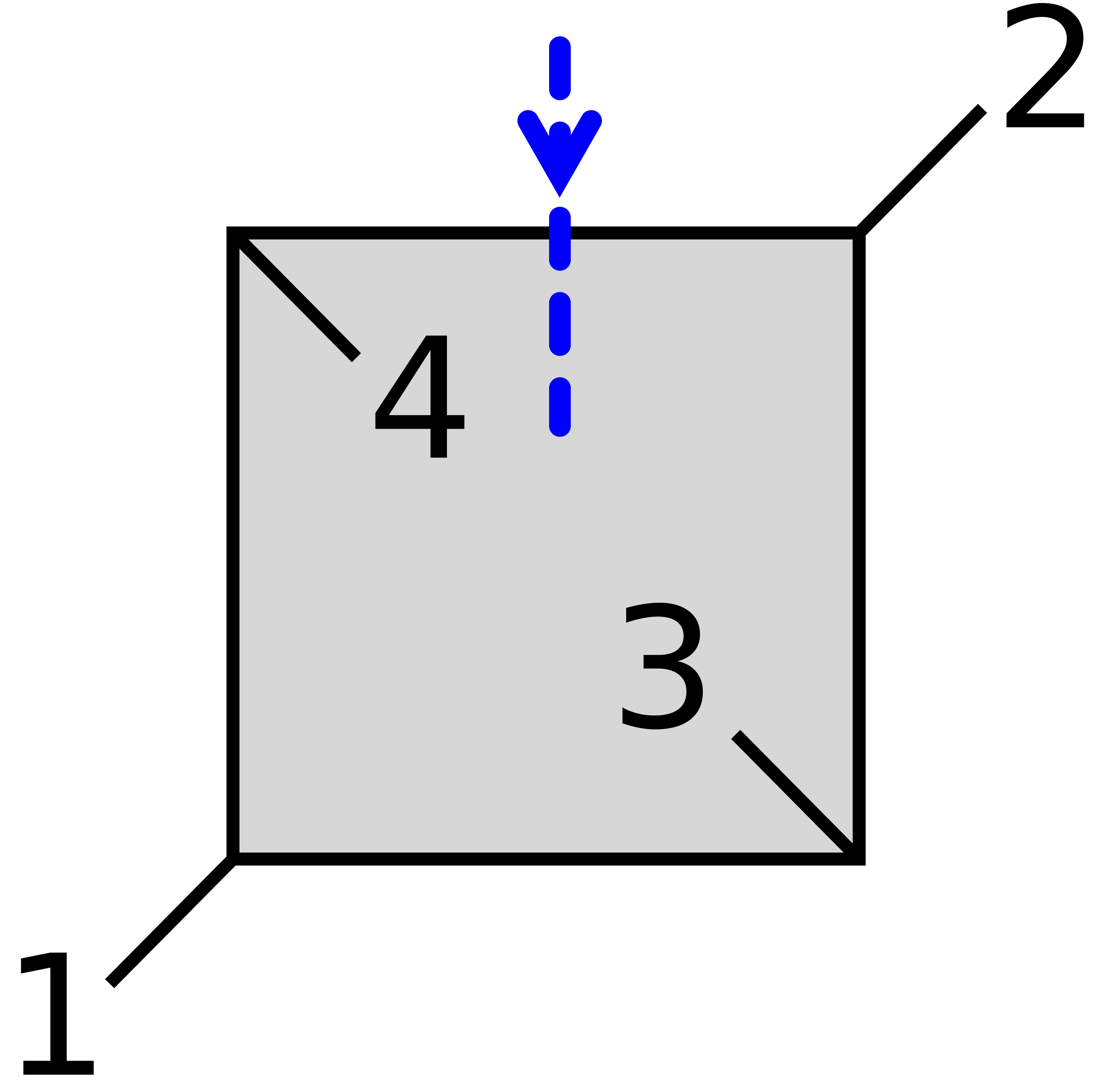}}$}}
\newcommand{\cboxcc}{\raisebox{-4.5ex}{$\mathord{\includegraphics[height=10ex]{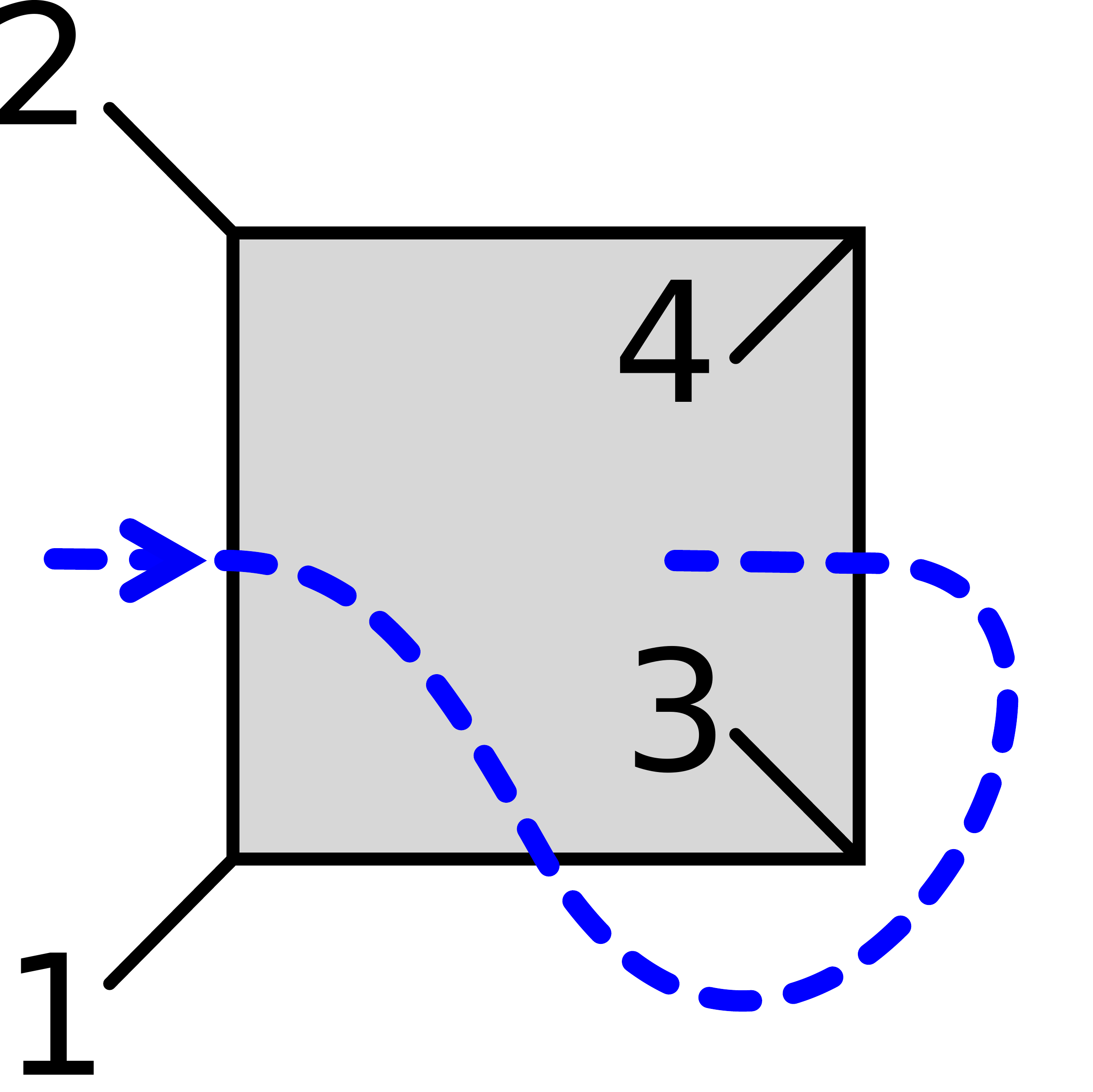}}$}}

\newcommand{\cc}{\raisebox{-5.3ex}{$\def\svgwidth{.2\textwidth}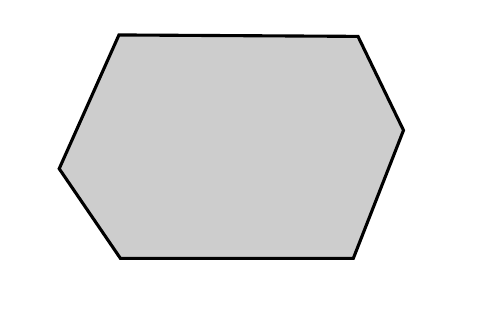$}}
\newcommand{\ce}{\raisebox{-5.3ex}{$\def\svgwidth{.2\textwidth}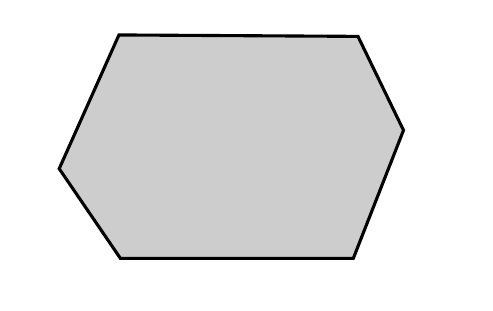$}}
\newcommand{\ee}{\raisebox{-5.3ex}{$\def\svgwidth{.2\textwidth}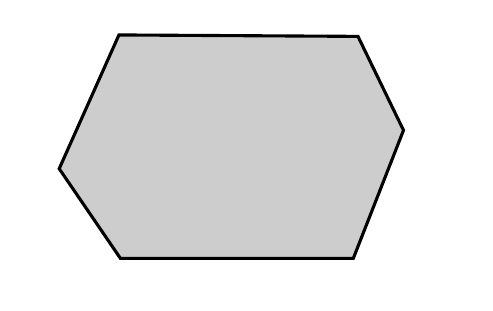$}}

\def\XXint#1#2#3{{\setbox0=\hbox{$#1{#2#3}{\int}$}
\vcenter{\hbox{$#2#3$}}\kern-.5\wd0}}

\def\[{\left[}
\def\]{\right]}

\def\({\left(}
\def\){\right)}
\def\[{\left[}
\def\]{\right]}

\def\<{\langle}
\def\>{\rangle}

\def\i2{\frac{i}{2}}

\def\spi{\relax{\rm \pi\kern-0.5em /}}
\def\sA{\relax{\rm A\kern-0.5em /}}
\def\sp{\relax{\rm p\kern-0.5em /}}
\def\sd{\relax{\rm \d\kern-0.5em /}}
\def\sk{\relax{\rm k\kern-0.5em /}}
\def\sn{\relax{\rm n\kern-0.5em /}}
\def\sl{\relax{\rm l\kern-0.5em /}}
\def\sP{\relax{\rm P\kern-0.7em /}}
\def\sBethe{\relax{\rm \Bethe\kern-0.5em /}}

\def\One{1\hskip-.16cm1}

\def\cM{{\cal M}}
\def\cO{{\cal O}}
\def\cK{{\cal K}}

\def\cP{{\cal P}}

\def\cW{{\cal W}}
\def\cZ{{\cal Z}}

\def\cQ{{\cal Q}}
\def\2F1{\,_2{\rm F}_1}

\def \Ascr {\cA^{\,^{\kern-0.2em (\infty)}}}

\makeindex

\numberwithin{figure}{section}

\begin{document}

\thispagestyle{empty}

\renewcommand{\thefootnote}{\fnsymbol{footnote}}
\setcounter{page}{1}
\setcounter{footnote}{0}
\setcounter{figure}{0}
\begin{center}
$$$$
{\Large\textbf{\mathversion{bold}
Scattering Amplitudes -- Wilson Loops Duality\\ for the First Non-planar Correction
}\par}

\vspace{1.0cm}

\textrm{Roy Ben-Israel$^{\tiny a}$, Alexander G. Tumanov$^{\tiny a}$ and Amit Sever$^{\tiny a,b}$}
\\ \vspace{1.2cm}
{\textit{
$^\text{\tiny a}$School of Physics and Astronomy, Tel Aviv University, Ramat Aviv 69978, Israel\\
$^\text{\tiny b}$CERN, Theoretical Physics Department, 1211 Geneva 23, Switzerland
}  
\vspace{4mm}
}

\par\vspace{1.5cm}

\textbf{Abstract}\vspace{2mm}
\end{center}
We study the first non-planar correction to gluon scattering amplitudes in ${\cal N}=4$ SYM theory. The correction takes the form of a double trace partial amplitude and is suppressed by one power of $1/N$ with respect to the leading single trace contribution. We extend the duality between planar scattering amplitudes and null polygonal Wilson loops to the double trace amplitude. The new duality relates the amplitude to the correlation function of two infinite null polygonal Wilson lines that are subject to a quantum periodicity constraint. We test the duality perturbatively at one-loop order and demonstrate it for the dual string in AdS. The duality allows us to extend the notion of the loop integrand beyond the planar limit and to determine it using recursion relations. It also allows one to apply the integrability-based pentagon operator product expansion approach to the first non-planar order. 
\noindent

\setcounter{page}{1}
\renewcommand{\thefootnote}{\arabic{footnote}}
\setcounter{footnote}{0}

 \def\nref#1{{(\ref{#1})}}

\newpage

\tableofcontents

\newpage

\section{Introduction}

The gluon S-matrix of an interacting gauge theory in four spacetime dimensions is an extremely rich and interesting object. 
A major simplification emerges in the large $N$ 't Hooft limit, in which the amplitude is reorganized in terms of 2-dimensional topologies of the 't Hooft diagrams. The leading order of this expansion is known as the planar amplitude, for which the diagrams have the topology of a disk. In the past ten years there has been major progress in the computation of planar scattering amplitudes in general and in ${\cal N}=4$ SYM theory in particular, see \cite{Elvang:2013cua} for a recent review. 

This progress has been driven, to a large extent, by a surprising duality between planar scattering amplitudes and polygonal Wilson loops in ${\cal N}=4$ SYM theory \cite{Alday:2007hr,Berkovits:2008ic,AmplitudeWilson,CaronHuot:2010ek}. This duality is directly tied to the Yangian symmetry of the amplitude \cite{Beisert:2008iq,CaronHuot:2011kk} and has been useful in two major ways. First, the duality allows one to relate the momentum loop integrands of different Feynman diagrams in a physically meaningful way. The corresponding loop integrand can be recursively constructed at any loop order \cite{ArkaniHamed:2010kv}. The second way is the application of non-perturbative integrability methods to calculating amplitudes. It turns out that integrability is most useful for computing the planar amplitudes using their dual representation in terms of polygonal Wilson loops \cite{Alday:2010ku,Basso:2013vsa}. In this approach the Wilson loop expectation value takes the form of a sum over two-dimensional excitations that propagate on top of the Gubser-Klebanov-Polyakov (GKP) flux tube \cite{Gubser:2002tv,POPEprogram}. As the dynamics of GKP excitations are well understood at finite coupling \cite{Basso:2010in}, the duality between the amplitudes and the Wilson loops opens the door for the finite coupling computation of the amplitude.

Evaluation of the full amplitude also requires being able to calculate the higher order corrections in the $1/N$ expansion, which, in turn, requires new ideas and techniques. One such idea is presented in this paper. It allows one to apply the techniques outlined above to the non-planar corrections. 

Let us outline the main idea. 
One way of obtaining higher genus Riemann surfaces is to start with a disk and glue different segments of its boundary together. Going in the opposite direction, one may start with a Riemann surface and cut it open along all its cycles back into a disk. This is exactly the approach we will take, but instead of working in position space, we will be working in dual momentum space.\footnote{See \cite{Bargheer:2017nne} for a different approach of cutting and gluing that works directly in position space (see also \cite{Eden:2017}).} The crucial point that allows us to do it is the fact that the spacetime momentum that flows around any cycle of the 't Hooft diagram can be defined in a physically meaningful way. Fixing these cycle loop momenta has the effect of cutting the Riemann surface open. The integration over these momenta is related to the integration over the moduli space of a dual two-dimensional surface. The resulting sum of the disk diagrams has a dual description in terms of a Wilson lines correlator with certain periodicity constraints that correspond to gluing at the disk boundary. Once the amplitude has been mapped onto a disk at the expense of stripping away the cycles' loop integrations, all the planar techniques can be applied to it. This will be done in detail for the leading non-planar correction, for which the 't Hooft diagrams have a cylindrical topology. The generalization to higher orders in the $1/N$ expansion will be briefly described in the discussion section.

The paper is organized as follows. In section  \ref{dualitysec} the cylindrical duality is stated and the conventions are set. Section 3 is dedicated to demonstrating how the duality comes about in a simple toy model -- the double scaling limit of the $\gamma$-deformed ${\cal N}=4$ SYM theory. Next, in section \ref{Tdualitysec} a T-duality transformation is performed on a string in AdS with the topology of a cylinder and the holographic dual is found. In section  \ref{oneloopsec} an explicit check of the duality in  ${\cal N}=4$ SYM is carried out at leading order in perturbation theory. In section \ref{BCFWsec} the BCFW recursion relation for the cylindrically cut amplitude is derived at the Born level. In section  \ref{loopintegrandsec} the planar loop integrand is generalized for the case of the cylindrically cut amplitude and in section \ref{BCFWloopsec} a recursion relation is derived for it. In section \ref{symmetriessec} the role played by superconformal symmetry is clarified. Finally, in section \ref{discussionsec} a discussion of future applications and extensions of the duality is presented.

\section{The cylindrical duality}\la{dualitysec}

Here, we put forward a precise conjecture for a duality between double trace amplitude and a correlation function between two infinite null polygonal Wilson lines subject to a quantum periodicity constraint. Let us first outline the reason why these two observables are related to each other. The string dual of the double trace amplitude has the topology of a cylinder. If one considers the univesral cover of the cylinder, which is a strip, then the double trace amplitude appears as a single trace one with infinitely many external particles subject to a periodicity constraint. The standard duality map between single trace amplitudes and null polygonal Wilson loops is then applied to this object. 
\begin{figure}[t]
\centering
\def\svgwidth{12cm}
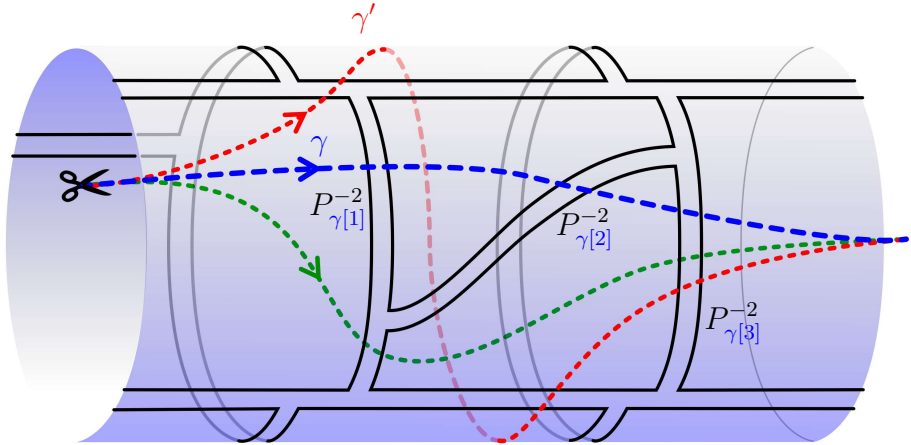
\caption{\small A Feynman diagram that contributes to the double trace partial amplitude $A_{3,2}$. The dashed lines indicate its various cylinder cuts, which correspond to the diagram with fixed momenta flowing around the cylinder, $l$. The red dashed line wraps around the cylinder once more than the blue one, which implies that the cylinder cut momenta, $l$, is only defined modulo a shift by the momentum flowing through the cylinder, $l\simeq l+q$.}\label{cutampfig}
\end{figure}

There are a few important subtleties to consider. The first one has to do with the relative ordering of the traces. We consider the double trace partial amplitude with $n$ ordered particles in one trace and $m$ in the other, denoted $A_{n,m}$, \cite{Bern:1990ux}. In the double line notation, the leading color Feynman diagrams that contribute to this partial amplitude have the topology of a cylinder. The two sets of ordered external momenta from the color traces, $k_1,k_2,\dots,k_n$ and $k_{n+1},k_{n+2},\dots,k_{n+m}$, are inserted on the two boundaries of the cylinder and their orderings are correlated through the cylinder. The coefficient of ${\rm Tr}(T_1T_2\dots T_n)\,{\rm Tr}(T_{n+1}T_{n+2}\dots T_{n+m})$ in the color decomposition of the full amplitude is given by the sum of two different partial amplitudes that correspond to the two different relative orderings. The duality can be demonstrated independently for either relative orderings and therefore only one will be considered throughout this paper, see figure \ref{cutampfig}.

For example, the coefficient of ${\rm Tr}(T_1T_2T_3T_4)\,{\rm Tr}(T_5T_6T_7)$ in the color decomposition of the full amplitude $A_{4,3}$ is the sum of two partial amplitudes. In the first, $A_{4,3}(1, 2, 3, 4; 5, 6, 7)$, the color ordering $1\to 2\to 3\to 4\to1$ in the first trace is correlated through the cylinder with the color ordering $5\to6\to7\to5$ in the second trace. In the second partial amplitude, $A_{4,3}(1, 2, 3, 4; 7, 6, 5)$, it is correlated with the reversed color ordering $7\to6\to5\to7$. If one of the traces has only two particles in it, there is no way to define the relative ordering and, therefore, there is only one partial amplitude associated with this color structure.

The second subtlety is that we do not establish the duality for the full double trace partial amplitude, but for its cylinder cut, which is defined as follows. At $L$ loops, any Feynman diagram that contributes to the amplitude has $L$ internal momenta, $\{l_1,\dots,l_L\}$. We consider a curve on the diagram, $\gamma$, that crosses a certain number of internal propagators, $\{1/P^2_{\gamma(j)}\}$, see the blue dashed line in figure \ref{cutampfig}. It starts between $k_1$ and $k_n$ in one trace and ends between $k_{n+1}$ and $k_{n+m}$ in the other.\footnote{The case of $n=m=2$ requires some extra clarification and will be discussed below.} The unintegrated Feynman diagram $G(l_1,\dots,l_L)$ is then multiplied by the $\delta$-function,
\beq\la{deltainsert}
\int \(\prod_{i=1}^Ld^4l_i\)\,G(l_1,\dots,l_L)=\int d^4l\[\int \(\prod_{i=1}^Ld^4l_i\)\,G(l_1,\dots,l_L)\times\delta^4(l-\sum_jP_{\gamma(j)})\] \ .
\eeq
where the sign of $P_{\gamma(j)}$ is defined such that $P_{\gamma(j)}$ is the momentum that crosses the cut in the direction that coincides with the external particle ordering $(1,2,\dots,n)$, see figure \ref{cutampfig}. Here, $l$ is interpreted as the momentum flow around the cylinder. The cylindrically cut amplitude, $\cA^\gamma(l)$, is obtained by stripping off the integral over $l$ and a factor of $\lambda/N$, and summing over all Feynman diagrams. When integrated over $l$, it gives the full amplitude, 
\beq\la{cutamplitude}
A_{n,m}={\lambda\over N}\int d^4l\, \cA_{n,m}^\gamma(l) \ .
\eeq
The superscript $\gamma$ of $\cA$ indicates that it depends on the choice of the cylinder cut for any diagram. Because of this dependence $\cA_{n,m}^\gamma(l)$ is not yet a physical object.

Due to momentum conservation at the interaction vertices any continuous deformation of the cut of any diagram does not change the $\delta$-function in (\ref{deltainsert}) and hence does not affect $\cA_{n,m}^\gamma(l)$. For example, the green and blue curves in figure \ref{cutampfig} result in the same $\cA_{n,m}^\gamma(l)$. However, the red curve $\gamma'$ in figure \ref{cutampfig} winds around the cylinder and thus cannot be continuously deformed into $\gamma$, resulting in a shift of $l$ by the total momentum flow through the cylinder, $q$,
\beq\la{qdefinition}
\cA_{n,m}^{\gamma'}(l)=\cA_{n,m}^{\gamma}(l+q)\ ,\qquad q=\sum_{i=1}^nk_i=-\sum_{j=1}^mk_{n+j}\ .
\eeq
Similarly, a curve that winds around the cylinder $a$ times more than $\gamma$ does would correspond to a shift in the momentum flow around the cylinder, $l\to l+a\,q$. In other words, the momentum flow around the cylinder $l$ is only well-defined modulo a shift by the total momentum flow through the cylinder,
\beq\la{modq}
l\simeq l+q\ .
\eeq
This ambiguity in the definition of $l$ is washed off by the integration in (\ref{cutamplitude}).

To construct the cylindrically cut amplitude ${\mathbb A}_{n,m}(l)$ which does not depends on $\gamma$ one has to sum over all possible shifts of $l$ by an integer number of $q$'s. This is equivalent to summing over all winding numbers of the curves $\gamma$ around the cylinder, 
\beq\la{cutsum}
{\mathbb A}_{n,m}(l)=\sum_a\cA_{n,m}\left(l+a\,q\right)\ ,\qquad A_{n,m}={\lambda\over N}\int\limits_{l\,\simeq\,l+q}\!\!\! d^4 l\,{\mathbb A}_{n,m}(l)\ ,
\eeq
where the superscript $\gamma$ of $\cA_{n,m}(l)$ was dropped because the sum is independent of the curve. This sum is convergent because the summand decays at least as $1/a^4$ at large $a$. One may view this sum as part of the $l$ integration in (\ref{cutamplitude}). The sum converges because the amplitude is UV finite. Unlike $\cA_{n,m}(l)$, the cylindrically cut amplitude ${\mathbb A}_{n,m}(l)$ is unambiguously defined and hence is a physical quantity. Moreover, it consists only of planar diagrams.

{If one of the traces has only two particles in it, say $n=2$ and $m>2$, then there is no relative ordering of the two traces and, therefore, there is only one partial amplitude instead of two. Still, the definition of ${\mathbb A}$ remains the same and is unambiguous. In the special case where $n=m=2$ there is no ordering at all and, therefore, no orientation on the cylinder. In this case every Feynman diagram is associated with the average over two cylinder cuts. These cylinder cuts are defined as above and are related to each other by turning the cylinder inside-out. One way to derive this prescription is to start with an amplitude with more than two particles in one of the traces and take a soft limit. An example of the $n=m=2$ amplitude will be studied explicitly in section \ref{oneloopsec}. Appendix \ref{nm2app} further expands on this point.}

The cylindrically cut amplitude ${\mathbb A}_{n,m}(l)$ is sensitive to integration by parts in momentum space. That is, the cylinder cut of a given diagram is not the same before and after momentum integration by parts as these can shift the momentum flow around the cylinder by a total derivative. According to the definition of the cylinder cut, one has to perform the cutting (\ref{deltainsert}) and the summation (\ref{cutsum}) prior to any integration by parts at the level of the Feynman diagram.\footnote{At two loops and higher, this technical detail rules out the use of reduction techniques to scalar integrals before cutting.}

To specify the duality with the Wilson lines object, it is convenient to further strip off the momentum and supersymmetry $\delta$-functions and a Parke-Taylor-like factor as 
\beq\la{Rdefinition}
{\mathbb A}_{n,m}(l)={g^{n+m-2}_\text{YM} \,\delta^4\(\sum k_i\)\delta^8\(\sum \lambda_i\tilde\eta_i\)
\over(\<1\,2\>\dots\<n\,1\>)\,(\<n+1\,m+2\>\dots\<n+m\,n+1\>)}\times{\mathbb M}_{n,m}(l)\ .
\eeq 
The conjectured duality then reads
\beq\la{cutduality}
{\mathbb M}_{n,m}(l)=\int\!d^8\theta\,\widehat\cW_{n,m}(l,\theta)\equiv{\mathbb W}_{n,m}(l)\ ,
\eeq
where $\widehat\cW_{n,m}(l,\theta)$ and the eight Grassmann variables $\theta_\alpha^A$ are defined below. By plugging (\ref{cutduality}) into (\ref{cutamplitude}) one can go back to the full amplitude, for which the duality takes the form
\beq\la{duality1}
A_{n,m}={g^{n+m-2}_\text{YM}\,\delta^4\(\sum k_i\)\delta^8\(\sum \lambda_i\tilde\eta_i\)
\over(\<1\,2\>\dots\<n\,1\>)\,(\<n+1\,m+2\>\dots\<n+m\,n+1\>)}
\times{\lambda\over N}\times\!\!\!\int\limits_{l\,\simeq\,l+q}\!\!\! d^4 l\int\!d^8\theta\,\widehat\cW_{n,m}(l,\theta)\ .
\eeq
We will now define the cylinder Wilson lines correlator $\widehat\cW_{n,m}(l,\theta)$ entering the duality (\ref{cutduality}).

\begin{figure}[t]
\centering
\def\svgwidth{5cm}
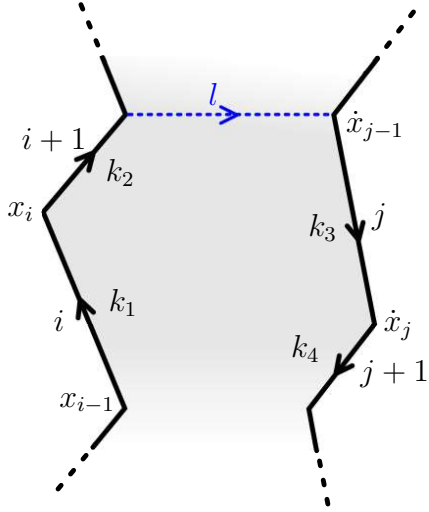
\caption{\small The periodic Wilson lines configuration that is dual to the cylindrically cut double trace amplitude. Each line consists of the ordered gluon momenta in one of the two traces. Because the total momentum in each trace is non-zero, the dual line is not closed. Instead, it is repeated periodically and forms the boundary of the universal cover of the cylinder. The separation between the two Wilson lines is equal to the momentum that flows around the cylinder, $l$.}\label{WL^2}
\end{figure}

The Wilson lines consist of two infinite sets of null edges $(\dots,k_1,k_2,\dots,k_n,k_1,\dots)$ and $(\dots,k_{n+1},k_{n+2},\dots,k_{n+m},k_{n+1},\dots)$. Cusps between $k_{i\, \text{mod}\,n}$ and $k_{i+1\, \text{mod}\,n}$ are denoted by $x_i$ and cusps between $k_{n+(j\,\text{mod}\,m)}$ and $k_{n+(j+1\,\text{mod}\, m)}$ by $\dot x_j$,
\beq\la{xdef}
x_{i}-x_{i-1}=k_{i\, \text{mod}\,n}\ ,\qquad \dot x_j-\dot x_{j-1}=k_{n+(j\,\text{mod}\,m)}\ .
\eeq
A convenient notation to use is 
\beq\la{blockshift}
x_i^{[a]}=x_{i+an}=x_i+a\,q\ ,\quad x_i^{[\pm]}=x_{i\pm n}\ ,
\qquad\dot x_j^{[a]}=\dot x_{j-am}=\dot x_j+a\,q\ ,\quad\dot x_j^{[\pm]}=\dot x_{j\mp m}\ .
\eeq
In (\ref{duality1}) the momentum flow around the double trace cylinder $l$ is integrated over. This momentum is dual to the separation between the two lines,
\beq\la{distance}
l=(\dot x_m-x_n)\text{ mod }q=(\dot x_0-x_{n+m})\text{ mod }q=\dots\ ,
\eeq
and is only defined modulo $q$, as in (\ref{qdefinition}). 

It is convenient to represent these periodic coordinates using dual momentum twistors. Under the periodicity constraint, the first two components of the twistors $Z_i^{[a]}$ are independent of the shift index $a$,
\beq\la{totwistors}
Z_i^{[a]}=Z_{i+an}=\(\!\!\begin{array}{cc}\lambda_i\\ x_i^{[a]}\lambda_i\end{array}\!\!\)\ ,\qquad \dot Z_j^{[a]}=\dot Z_{j+am}=\(\!\!\begin{array}{cc}\dot\lambda_j\\ \dot x_j^{[a]}\dot\lambda_j\end{array}\!\!\)\ ,
\eeq
that is, the helicity weight of $Z_i$ ($\dot Z_j$) is equated with that of its periodic image $Z_{i+n}$ ($\dot Z_{j+m}$).

Note that the geometry of the Wilson lines is invariant under a shift of $l$ by $q$. Hence, it makes perfect sense to relate it with ${\mathbb A}(l)$, as defined by the sum in (\ref{cutsum}). 

We denote the supersymmetric counterparts of the bosonic coordinates of the cusps, $x_i$ and $\dot x_j$, by $\theta_i$ and $\dot\theta_j$ respectively. These supercoordinates are related to the particles' helicities on each of the infinite Wilson lines in the standard way, $\theta_{i}-\theta_{i-1}=\lambda_i\tilde\eta_i$ and $\dot\theta_{j}-\dot\theta_{j-1}=\dot\lambda_j\dot{\tilde\eta}_j$. It follows from supersymmetry that the total R-charge entering one side of the cylinder is equal to the one exiting from the other,
\beq\la{Qperiod}
\theta_{i+n}-\theta_i=\dot\theta_{j+m}-\dot\theta_j=Q\ ,\qquad\text{where}\qquad Q=\sum_{i=1}^n\lambda_i\tilde\eta_i=-\sum_{j=1}^m\dot\lambda_i\dot{\tilde\eta}_j\ .
\eeq
A notation similar to (\ref{blockshift}) will be used for the supercoordinates,
\beq\la{thetaperiodicity}
\theta_i^{[a]}=\theta_{i+an}=\theta_i+a\,Q\ ,\qquad \dot\theta_j^{[a]}=\dot\theta_{j-an}=\dot\theta_j+a\,Q\ . 
\eeq

Similarly to the cut momentum $l$ in (\ref{distance}), the total supersymmetry charge 
that flows around the cylinder (or equivalently, the relative separation of the two Wilson lines in superspace) is defined as 
\beq\la{superdistance}
\theta=\dot\theta_m-\theta_n\ .
\eeq
Prior to integration, this separation is only defined modulo a shift,
\beq
(l,\theta)\simeq(l+q,\theta+Q)\ .
\eeq
We did not find it useful to  strip off the $\theta$ integration as we did for the bosonic variable $l$ in (\ref{cutsum}). Instead, (\ref{cutduality}) contains the Grassmann integration and there is no need for summing over its $Q$-shifts.

The superperiodicity of the Wilson lines geometry, (\ref{blockshift}) and (\ref{thetaperiodicity}), can be made manifest using supertwistors, 
\beq\la{supertwistors}
\cZ_i^{[a]}=\(\!\!\begin{array}{cc}\lambda_i\\ x_i^{[a]}\lambda_i\\\theta_i^{[a]}\lambda_i\end{array}\!\!\)\ ,\qquad \dot{\cZ}_j^{[a]}=\(\!\!\begin{array}{cc}\dot\lambda_j\\ \dot x_j^{[a]}\dot\lambda_j\\ \dot\theta_j^{[a]}\dot\lambda_j\end{array}\!\!\)\ .
\eeq

For this periodic external data the $\delta^8(\cQ)$ for the infinite loop reduces to the one of a single periodic block of the Wilson lines. Similarly, the infinite cover of the Parke-Taylor-like factor in (\ref{Rdefinition}) reproduces the Parke-Taylor factor of the corresponding infinite single trace Wilson loop. 

The final subtlety is that the periodicity constraint has to be imposed not only on the external data but also at the quantum level -- on all the planar diagrams. It is important to note that this constraint is only defined for the leading color diagrams. One particular way the periodically constrained Wilson lines correlator can be defined is to start with ${\cal N}=4$ SYM theory compactified on a circle of radius $q/(2\pi)$ and consider two closed null polygonal Wilson loops that wrap around the circle with edges $(k_1,\dots,k_n)$ and $(k_{n+1},\dots,k_{n+m})$. Then all diagrams that contribute to the expectation value of the correlator between these Wilson loops in the leading order of the $1/N$ expansion are considered. Every individual propagator in these diagrams is periodic and there is no correlation between the color contractions and a spacetime shift around the circle. Next, these two are correlated by replacing the periodic propagators of the compactified theory with those of the non-compact flat spacetime. After this replacement, the diagrams in position space no longer close. Instead, as one follows the propagators around the cylinder, one finds a mismatch by one period, $q$. Hence, the new diagrams with non-compact propagators cannot live on the cylinder because they do not respect the cylinder periodicity. Instead, they live on its universal cover. It is important to point out that every propagator in each of these diagrams has infinitely many images but still contributes only once. The result is, by construction, gauge invariant -- every gauge transformation $g(x)$ has infinitely many periodic images, $g^{[a]}(x)\equiv g(x-a\,q)$, under which the images of the propagators transform. Finally, the interaction points are integrated on the full non-compact spacetime. This last step requires regularization of the cusp operators.

An alternative way the periodically constrained Wilson lines correlator can be defined is to start with all the planar Feynman diagrams that contribute to the correlation function between the two infinite Wilson lines in position space, then consider only the diagrams that, prior to integration, are individually invariant under a simultaneous relabeling of all external cusps, $x_i^{[a]}\to x_i^{[a+1]}$, $\dot x_j^{[a]}\to\dot x_j^{[a+1]}$ and $\theta_i^{[a]}\to \theta_i^{[a+1]}$, $\dot \theta_j^{[a]}\to\dot\theta_j^{[a+1]}$. 
Any propagator in such a diagram, $G(y_1-y_2)$, has infinitely many images, $G(y_1^{[a]}-y_2^{[a]})=G(y_1-y_2)$, but should be counted only once. Finally, the interaction points are integrated over. Several examples are given in the next section.

The main differences between the duality (\ref{duality1}) and the standard duality between single-trace amplitudes and null polygonal Wilson loops are the presence of the integration over $\theta$ and the need to impose a periodicity constraint on both the external data and the internal planar diagrams. Because of this periodicity constraint the Wilson lines correlator will be referred to as a {\it cylinder} correlator. The subscript {\it cylinder} is added to the definition of the expectation value to indicate the imposition of the periodicity constraint,
\beqa\la{superlines}
\widehat\cW_{n,m}(l,\theta)&=&\left<{\rm Tr}\(\dots\cP e^{\int_0^1dt\,{\cal E}_1(t)}{\cal V}_{12} e^{\int_0^1dt\,{\cal E}_2(t)}{\cal V}_{23}\dots e^{\int_0^1dt\,{\cal E}_n(t)}{\cal V}_{n1^{[+]}}\dots\)\right.\\
&&\left.\times{\rm Tr}\(\dots\cP e^{\int_0^1dt\,{\cal E}_{\dot 1}(t)}{\cal V}_{\dot 1\dot 2} e^{\int_0^1dt\,{\cal E}_{\dot 2}(t)}{\cal V}_{\dot 2\dot 3}\dots e^{\int_0^1dt\,{\cal E}_{\dot m}(t)}{\cal V}_{\dot m\dot 1^{[-]}}\dots\)\right>_\text{cylinder}\ ,\nn
\eeqa
where
\beq
{\cal E}_i(t)=-k_i\cdot A+\dots\qquad\text{and}\qquad {\cal V}_{i\,i+1}=1+\dots\ ,
\eeq
are the edge and vertex operators defined in \cite{CaronHuot:2010ek}. It is convenient to normalize the Wlison lines without $1/N$ factor in front of the trace. 

The periodicity constraint may look unnatural from the point of view of Feynman perturbation theory. However, it has very useful implications. First, it enables the computation of the double trace amplitude at finite coupling using the integrability-based POPE approach \cite{Basso:2013vsa}. In this framework, it simply becomes the very natural periodicity of a string that has a cylindrical topology, see section \ref{discussionsec}. Second, it allows one to generalize the notion of the loop integrand to the double trace amplitude and determine it using recursion relations, see sections \ref{loopintegrandsec} and \ref{BCFWloopsec}.

In order to illustrate how perturbation theory works under this constraint, a toy model will be considered in the next section. We will perform explicit perturbative computations of a double trace amplitude and its dual cylinder Wilson lines correlator in the the double scaling limit of the $\gamma$-deformed ${\cal N}=4$ SYM theory \cite{Gurdogan:2015csr}.

\section{A toy model example}\la{Kazsec}

Before studying the duality in detail for ${\cal N}=4$ SYM theory, it is helpful to first consider a simple limit of it. In this section, we will consider the duality in the double scaling limit of the $\gamma$-deformation of the theory, henceforth referred to as the {\it fishnet model} \cite{Gurdogan:2015csr}. After taking this limit, of all the ${\cal N}=4$ fields, one is left with a pair of complex scalars, $\phi_1$ and $\phi_2$. Their dynamics are dictated by the Lagrangian
\beq\la{Kazakovact}
\cL ={1\over g^2}{\rm Tr}\,(-\partial_\mu\phi_1^\dagger\partial^\mu\phi_1- \partial_\mu\phi_2^\dagger\partial^\mu\phi_2 +\phi_1^\dagger\phi_2^\dagger\phi_1\phi_2) \ .
\eeq
In the planar limit, $N$ is taken to infinity, while keeping the 't Hooft coupling $\lambda=g^2N$ fixed, see \cite{Gurdogan:2015csr} for details.\footnote{Our normalization is related to the one in \cite{Gurdogan:2015csr} by $\phi_{here}=g\sqrt{N/2}\,\phi_{there}$ and $\lambda_{here}=4\xi_{there}^2$.} 

\begin{figure}[t]
\centering
\def\svgwidth{.4\textwidth}
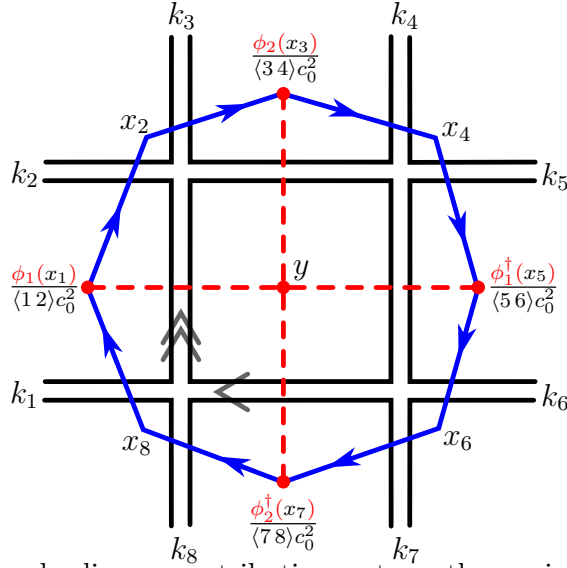
\caption{\small The leading contribution to the eight-point amplitude of $\{\phi_1(k_1),\phi_1(k_2),\phi_2(k_3),\phi_2(k_4),\phi_1^\dagger(k_5),\phi_1^\dagger(k_6),\phi_2^\dagger(k_7),\phi_2^\dagger(k_8)\}$ in the fishnet model. Here black lines correspond to the propagators in the double line notation and the grey arrows indicates the flow of the two conserved $U(1)$ charges. In the dual space one finds a null octagon Wilson loop, represented by the blue line. It consists of four scalars inserted at $\{x_1,x_3,x_5,x_7\}$. The dashed red lines represent the single Feynman diagram that contributes to the expectation value of the dual octagon. It has a single interaction vertex at $y=l+x_1$.}\label{KZPlanar}
\end{figure}

Being a deformation of ${\cal N}=4$ SYM theory that preserves the Yangian symmetry in the planar limit, this theory also exhibits dual conformal invariance. Hence, planar scattering amplitudes in this theory are expected to have a dual description in terms of Wilson loop like objects, as will now be demonstrated.\footnote{See \cite{Chicherin:2017cns} for a discussion of scattering amplitudes and their Yangian symmetry in this theory.} The Wilson loop dual of the planar ${\cal N}=4$ amplitude contains insertions that depend on the type of particles that are being scattered. In the double scaling limit of \cite{Gurdogan:2015csr} the gauge field decouples and one is left with only the scalar insertions along the polygonal loop. However, the corresponding object will still be referred to as a ``Wilson loop". This duality can be illustrated by considering the planar amplitude of eight ordered scalars, ${\rm Tr}(\phi_1\,\phi_1\,\phi_2^\dagger\,\phi_2^\dagger\,\phi_1^\dagger\,\phi_1^\dagger\,\phi_2\,\phi_2)$. There is only one planar diagram that contributes to this process, see figure \ref{KZPlanar}. 
It is given by
\beq\la{steightpoint}
A_8^\text{fishnet}=g^6\lambda\,\delta^4\left(\sum k_i\right)\int \frac{d^4l}{l^2(l + k_2 + k_3)^2(l + k_2 + k_3 + k_4 + k_5)^2(l - k_1 - k_8)^2}\ .
\eeq

This amplitude can be obtained from a N$^2$MHV amplitude in ${\cal N}=4$ SYM by taking the $\gamma$-deformation and the double scaling limit of \cite{Gurdogan:2015csr}. By taking the same limit of the dual polygonal Wilson loop one ends up with its representation in dual momentum space. It is a null octagon with its cusps at $x_{i=1,\dots,8}$, where $x_i-x_{i-1}=k_i$. The octagon has four scalar insertions, see figure \ref{KZPlanar},
\beq\la{WKaz}
{\mathbb W}_8^\text{fishnet}={1\over N}{\rm Tr}\,{\phi_2^\dagger(x_7)\over c_0^2\<7\,8\>}{\phi_1^\dagger(x_5)\over c_0^2\<5\,6\>}{\phi_2(x_3)\over c_0^2\<3\,4\>}{\phi_1(x_1)\over c_0^2\<1\,2\>}\ ,
\eeq
where we have chosen to dress each scalar insertion by the two-bracket $\<a\,b\>=\epsilon_{\alpha\dot\alpha}\lambda_a^\alpha\lambda_b^{\dot\alpha}$, with $k_a^{\alpha\dot\alpha}=\lambda_a^\alpha\tilde\lambda_a^{\dot\alpha}$ for $k_a^2=0$, and by $c_0^2=g\sqrt{N}/(2\pi)$. This choice is made to match the convention in ${\cal N}=4$ SYM. As in the amplitude case, there is only one planar Feynman diagram that contributes to the expectation value of this configuration. It has one interaction vertex at $y$, which is related to the internal momentum in (\ref{steightpoint}) as $y-x_1=l$. This leads to the dual representation of the amplitude, 
\beqa
A_8^\text{fishnet}&=&g^6(2\pi)^4\delta^4\left(\sum k_i\right)\<1\,2\>\<3\,4\>\<5\,6\>\<7\,8\>\times\<{\mathbb W}_8\>\ ,\la{Kazduality}\\
\<{\mathbb W}_8^\text{fishnet}\>&=&{\lambda\over\<1\,2\>\<3\,4\>\<5\,6\>\<7\,8\>}\int \frac{d^4y}{(2\pi)^4}{1\over(y-x_1)^2(y-x_3)^2(y-x_5)^2(y-x_7)^2}\ .\la{octagonWL}
\eeqa
The Wilson loop (\ref{octagonWL}) is invariant under dual conformal transformations that act as regular conformal transformations on the dual $x$ space.\footnote{This integral can be computed analytically \cite{Usyukina:1992jd}, $$A_8 =g^6\lambda\,\delta^4\left(\sum k_i\right){\pi^2\over x_{1\,5}^2x_{3\,7}^2}{2{\rm Li}_2(z)-2{\rm Li}_2(\bar z)+\log(z\bar z)\log{1-z\over1-\bar z}\over z-\bar z}\ ,$$ with two independent dual conformal cross ratios $z\bar z={x_{1\,3}^2x_{5\,7}^2\over x_{1\,5}^2x_{3\,7}^2}$ and $(1-z)(1-\bar z)={x_{1\,7}^2x_{3\,5}^2\over x_{1\,5}^2x_{3\,7}^2}$.}

\begin{figure}[t]
\centering
\def\svgwidth{.9\textwidth}
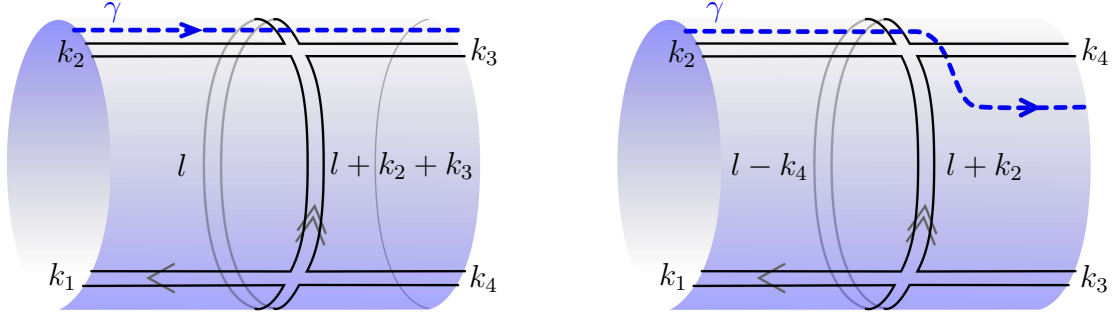
\caption{\small The two one-loop Feynman diagrams that contribute to the four-point amplitude of $\{\phi_1\left(k_1\right),\phi_1\left(k_2\right),\phi_1^\dagger(k_3),\phi_1^\dagger(k_4)\}$.}\label{KZ1L}
\end{figure}

The next step is to generalize this duality beyond the planar limit by considering $1/N$ corrections to the scattering amplitudes. These are not the same as non-planar corrections to the closed polygonal Wilson loop duals, such as (\ref{WKaz}). To demonstrate the duality in the most simple setting we consider the 4-point amplitude of the scalars $\{\phi_1(k_1),\phi_1(k_2),\phi_1^\dagger(k_3),\phi_1^\dagger(k_4)\}$. This amplitude does not receive any planar contributions. The first non-trivial contribution comes from the double trace color contraction, at order $1/N$ in the 't Hooft expansion,
\begin{align}
{\rm Tr}\left[T_1T_2\right]\,{\rm Tr}\left[T_3T_4\right]\,A_{2,2}\left(1,2;3,4\right)\ .\label{22DefKaz}
\end{align}
The partial amplitude $A_{2,2}$ receives its first perturbative contribution at one-loop order. There are two Feynman diagrams that have the topology of a cylinder, see figure \ref{KZ1L}.\footnote{Here the action given strictly by (\ref{Kazakovact}), with no double trace term that is needed for conformal invariance in all sectors \cite{Sieg:2016vap}.}

We try to follow the same steps as in the case of the planar eight-point amplitude and the octagon Wilson loop that was presented above. The Feynman diagrams are drawn on a cylinder with the $\phi_1$'s on the left boundary and the $\phi_1^\dagger$'s on the right one, see figure \ref{KZ1L}. For the special case considered here, where there are only two particles on each side, there is no ordering of the traces. Therefore, the traces do not induce an orientation on the cylinder. Instead, the orientation is determined by the the $\phi_2$ charge flow, which is chosen to go counterclockwise, as viewed from the left boundary, see figure \ref{KZ1L}. This allows one to distinguish between the interior and the exterior of the cylinder. The ability to do so is related to the fact that the double scaled theory is not CPT symmetric and its corresponding 't Hooft string is not orientable.\footnote{In ${\cal N}=4$ SYM theory the planar diagrams are orientable and the special case of $A_{2,2}$ requires averaging over the two possibilities, see section \ref{cutoneloop}.} The two cusps, $x_1$ and $x_2$, are associated with faces of the diagram and are defined such that $x_2-x_1=k_2$. As opposed to the planar case, these coordinates satisfy $x_1-x_2=k_1-q$ instead of $x_1-x_2=k_1$, where $q=k_1+k_2$ is the total momentum in the first trace. Since $q\ne0$, the dual Wilson line is not closed. Going around the cylinder takes $x_i$ to $x_i^{[+]}=x_i+q$. Consequently, $k_1=x_1-x_2^{[-]}=x_1^{[+]}-x_2$.

\begin{figure}[t]
\centering
\def\svgwidth{.8\textwidth}
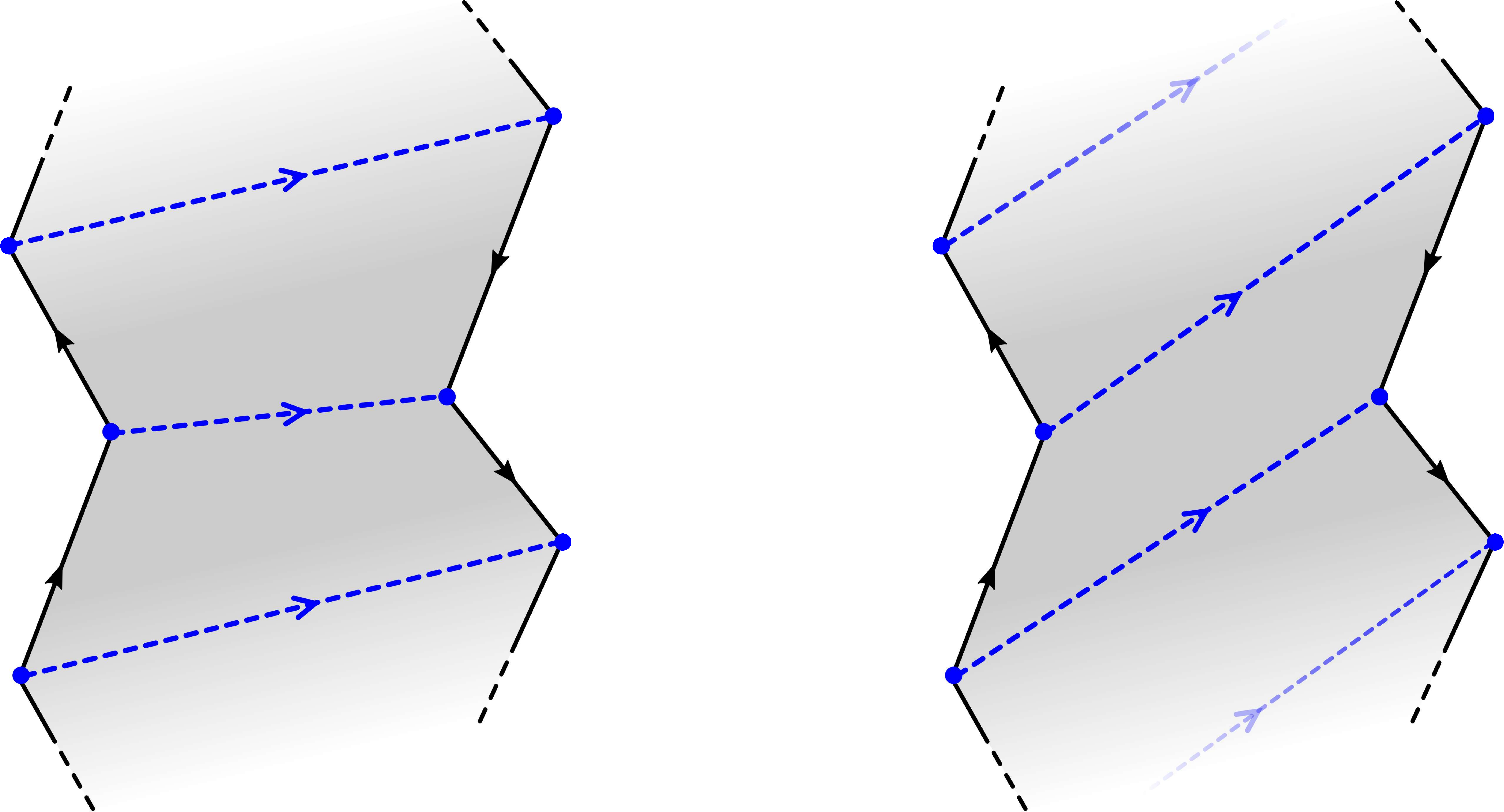
\caption{\small The two types of diagrams that contribute to the Wilson lines correlator (\ref{KazWL}). The dual space picture consists of two infinite null polygonal Wilson lines composed of two sets of scalars, $\{\phi_2(x_1^{[a]})/(\<1\,2\>c_0^2),\phi_2(x_2^{[a]})/(\<2\,1\>c_0^2)\}$ and $\{\phi^\dagger_2(\dot{x}_1^{[a]})/(\<3\,4\>c_0^2),\phi^\dagger_2(\dot{x}_2^{[a]})/(\<4\,3\>c_0^2)\}$, contracted along the edges. Each propagator between $x_i$ and $\dot x_j$ has infinitely many identical images running between $x_i^{[a]}$ and $\dot x_j^{[a]}$, but is counted only once. Apart from these two diagrams, there are two infinite sets of diagrams that are related to these by shifting one of the lines by an integer number of periods.}\label{DRCNP}
\end{figure}

One ends up with two infinite periodic null polygonal Wilson lines with cusps at\\ $\{\dots,x_1^{[-]},x_2^{[-]},x_1,x_2,x_1^{[+]},x_2^{[+]},\dots\}$ and $\{\dots,\dot x_1^{[-]},\dot x_2^{[-]},\dot x_1,\dot x_2,\dot x_1^{[+]},\dot x_2^{[+]},\dots\}$ that correspond to the two traces in (\ref{22DefKaz}), see figure \ref{DRCNP}. The separation between the two infinite lines is only defined modulo a shift by $q$, $l=\dot x_2- x_2\ \text{mod}\ q$, (\ref{distance}). It is equal to the loop momentum flowing around the cylinder.

The cylindrically cut amplitude $\mathcal{A}_{2,2}(l)$, defined in (\ref{cutamplitude}), is constructed by cutting the cylinder across the blue dashed line in figure \ref{KZ1L}. Despite the fact that the cut on the right looks different from the one on the left, both of them start at $x_2$ and end at $\dot x_2$. The ambiguity in the definition of the cut is eliminated by summing over the shifts of $l$ by $q$, according to equation (\ref{cutsum}). This results in the following expression for the cylindrically cut double trace amplitude, 
\begin{align}
&\mathcal{A}_{2,2\,\text{tree}}^{\text{fishnet}}\left(l\right) = g^2\,\delta^4\left(\sum k_i\right)\left(\frac{1}{l^2(l + k_2 + k_3)^2} + \frac{1}{(l + k_2)^2(l - k_4)^2}\right)\ ,\label{KAZAmp1LC}\\
&\mathbb{A}_{2,2}^\text{fishnet}\left(l\right) = \sum\limits_{a\,=\,-\infty}^\infty\mathcal{A}_{2,2}^\text{fishnet}\left(l + a\,q\right),\qquad A_{2,2}^\text{fishnet} ={\lambda\over N}\!\int\limits_{l\,\simeq\,l+q}\!d^4l\,\mathbb{A}_{2,2}^\text{fishnet}(l)\ .\nn
\end{align}

The dual cylindrical Wilson lines correlator $\mathbb{W}_{2,2}(l)$ has two scalar insertions on each of the lines that are repeated periodically, see figure \ref{DRCNP},
\beq\la{KazWL}
\mathbb{W}_{2,2}^\text{fishnet}(l)=\<1\,2\>^2\<3\,4\>^2\left<{\rm Tr}\(\dots{\phi_1(x_1)\over c_0^2\<1\,2\>}{\phi_1(x_2)\over c_0^2\<2\,1\>}\dots\)\,{\rm Tr}\(\dots{\phi_1^\dagger(\dot x_1)\over c_0^2\<3\,4\>}{\phi_1^\dagger(\dot x_2)\over c_0^2\<4\,3\>}\dots\)\right>_\text{cylinder}\ .
\eeq
Every propagator in any planar Feynman diagram that contributes to the correlator (\ref{KazWL}) has infinitely many images. The subscript {\it cylinder} of the expectation value indicates that such a propagator is counted only once, with the rest of them being the periodic images of the same propagator.

At tree level this expectation value is given by
\beq\la{WLsum}
{\mathbb W}_{2,2\,\text{tree}}^\text{fishnet}(l) = \sum\limits_{a\,=\,-\infty}^{\infty}\left(\frac{1}{(\dot{x}_1^{[a-1]} - x_1)^2(\dot{x}_2^{[a]} - x_2)^2} + \frac{1}{(\dot{x}_2^{[a]}  - x_1)^2(\dot{x}_1^{[a]} - x_2)^2}\right)\ ,
\eeq
where the two terms in the sum correspond to the two diagrams in figure \ref{DRCNP}, and the sum accounts for all diagrams that are related to these by a periodic shift of one of the two lines. One can see that at tree level,
\beq\la{dualityKaz}
{\mathbb A}_{2,2}^\text{fishnet}(l)=g^2\,\delta^4\left(\sum k_i\right)\,{\mathbb W}_{2,2}^\text{fishnet}(l)\ ,
\eeq
where the first (second) term in the sum in (\ref{WLsum}) corresponds to the first (second) term in the sum in (\ref{KAZAmp1LC}). This matching is specific to the fishnet model. In general, a single Feynman diagram cannot be isolated in a physically meaningful way and, hence, neither does its cylinder cut.

\begin{figure}[t]
\centering
\def\svgwidth{.9\textwidth}
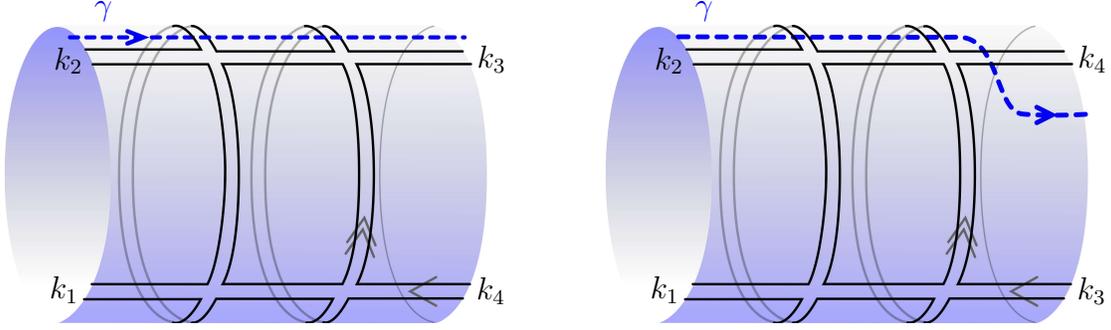
\caption{\small The two diagrams that contribute to the double trace amplitude (\ref{22DefKaz}) at three-loop order. The blue line represents the cylinder cut of these diagrams.}\label{KZ2L}
\end{figure}

The duality (\ref{dualityKaz}) between the cylindrically cut double trace amplitude and the cylinder Wilson lines correlator has been confirmed at leading order in perturbation theory. It is not hard to show that relation (\ref{dualityKaz}) holds to all orders in perturbation theory. At any non-vanishing loop order, there are only two diagrams that contribute to the four-point double trace amplitude. They are obtained from the two diagrams in figure \ref{KZ1L} by wrapping more $\phi_2$ loops around the cylinder. Each internal $\phi_2$ line that wraps around the cylinder increases the loop order by two. Hence, the next correction appears at three-loop order. This case will now be considered in detail to illustrate how the duality extends to higher loop orders. The corresponding two three-loop diagrams are given in figure \ref{KZ2L}. 
The cylindrical cut of this amplitude reduces the loop order by one. Specifically, it is a two-loop object that is obtained from the two diagrams in figure \ref{KZ2L} by cutting them open along the blue dashed lines. Note that the cut now goes through two propagators. Hence, only the sum of the momenta of these two propagators is fixed while the relative momentum is being integrated over.  

\begin{figure}[t]
\centering
\def\svgwidth{.7\textwidth}
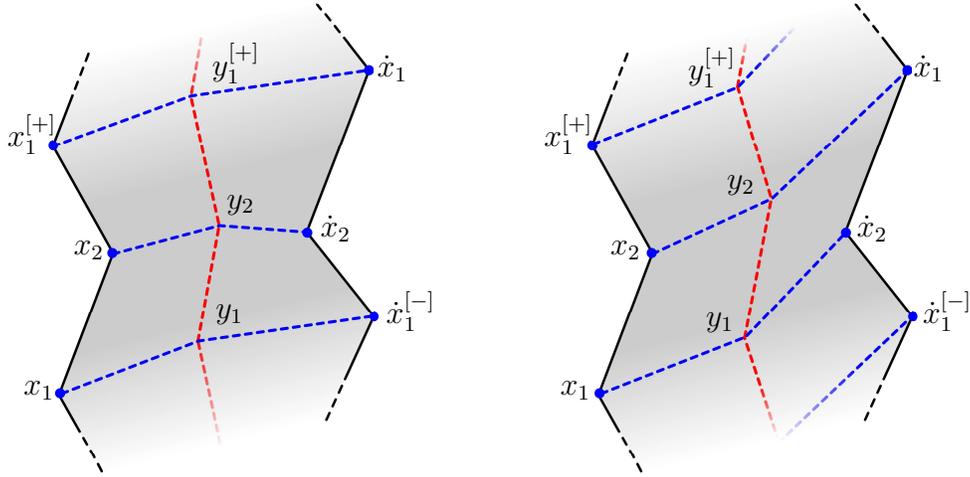
\caption{\small The two diagrams that contribute to $\mathbb{W}^{\text{fishnet}}_{2,2}(l)$ at two-loop order. Blue (red) dashed lines correspond to $\phi_1$ ($\phi_2$) propagators. These two diagrams come with an infinite set of diagrams that are related to them by a shift of one of the two Wilson lines by an integer number of periods, say $\dot x_j\to\dot x_j^{[a]}$.}\label{KZ2LD}
\end{figure}

On the Wilson lines correlator side of the duality there are two new bulk integration points, $y_1$ and $y_2$, along with their periodic images. They are connected by $\phi_2$ propagators that form an additional line that stretches vertically parallel to the Wilson lines (around the dual cylinder), see figure \ref{KZ2LD}. One finds that
\beqa
{\mathbb W}_{2,2\,\text{2-loop}}^\text{fishnet}(l)& =& \lambda^2\int{d^4y_1\over(2\pi)^4}\int{d^4y_2\over(2\pi)^4}\frac{1}{(x_1 - y_1)^2(x_2 - y_2)^2(y_2 - y_1)^2(y_1^{[+]} - y_2)^2}\times\nn\\
&&\times\sum\limits_{a=-\infty}^\infty\(\frac{1}{(y_1-\dot{x}^{[a-1]}_1 )^2(y_2-\dot{x}^{[a]}_2)^2} + \frac{1}{(y_1-\dot{x}^{[a]}_2)^2( y_2-\dot{x}_1^{[a]})^2}\)\ .\la{Kaz2loopAmp}
\eeqa

The integrations over $y_1$ and $y_2$ correspond to the loop integrations of the amplitude diagrams in figure \ref{KZ2L}. The resulting integrals precisely match the 2-loop correction to the cylindrically cut amplitude. There are only two diagrams at any given non-vanishing loop order. On the amplitude side more $\phi_2$ loops that wrap around the cylinder are added. On the Wilson lines correlator side one finds more vertical $\phi_2$ lines parallel to the Wilson lines. These expressions agree on the level of the integrand. The same also turns out to be true for ${\cal N}=4$ SYM theory. This will be discussed in detail in section \ref{loopintegrandsec}. 

Before performing a similar perturbative check for ${\cal N}=4$ SYM theory in section \ref{oneloopsec}, we discuss how T-duality on the holographic string side works when the string has the topology of a cylinder in the next section.

\section{T-duality of the cylindrical string amplitude}\la{Tdualitysec}

The duality between planar scattering amplitudes and closed polygonal Wilson loops was first observed at the strong coupling limit of ${\cal N}=4$ SYM \cite{Alday:2007hr}. There, it emerges by performing a T-duality on the string worldsheet in AdS spacetime. T-duality in general is a change of variables that relates strings propagating in different, T-dual, backgrounds. In the context of gluon scattering amplitudes the AdS background is non-compact and, therefore, this duality is restricted to the planar limit, in which the string has the topology of a disk. The planar duality maps the AdS background back to itself and the amplitude with disk topology to the Wilson loop. Here, we will generalize this correspondence to double trace amplitudes, for which the holographic dual string has the topology of a cylinder. Similar considerations were used for studying form factors at strong coupling in \cite{Maldacena:2010kp}. Our discussion will not be restricted to the strong coupling limit (where the string description becomes classical). However, only a simplified bosonic version of the worldsheet theory will be considered.

Gluon scattering amplitudes are holographically dual to an open string path integral in $AdS_5\times S^5$ \cite{Alday:2007hr}. The open string ends on an IR D3 brane close to the Poincaré horizon of $AdS_5$. While for the planar amplitude the string has the topology of a disk, for the double trace amplitude it has the topology of a cylinder. T-duality in a non-compact target space, however, is known to break down beyond the leading disk topology order. This puts the relationship between double trace scattering amplitudes and Wilson loops in question. Indeed, it will soon become evident that double trace scattering amplitudes and Wilson loops in ${\cal N}=4$ SYM theory are not dual to each other. Instead, as in the fishnet model discussed in the previous section, double trace amplitudes can be computed from the correlation function of two Wilson lines in ${\cal N}=4$ SYM theory only once a new periodicity constraint is imposed. In this section this constraint will be derived at the level of the worldsheet path integral for the bosonic string. We leave the generalization of this cylindrical duality to an exact fermionic T-duality for future work \cite{Berkovits:2008ic}. In the rest of the paper we will explain how the constraint can be imposed at the full quantum level of the gauge theory and how it leads to an exact duality between double trace amplitudes and Wilson loops.
\begin{figure}[t]
\centering
\def\svgwidth{8cm}
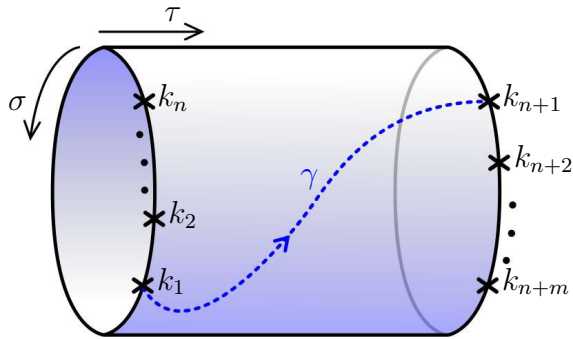
\caption{\small In AdS/CFT the double trace amplitude is mapped to a worldsheet amplitude in $AdS_5\times S^5$. The worldsheet has cylindrical topology and ends on an IR D3 brane close to the Poincaré horizon. It has $n$ ordered gluon vertex operators on one boundary, (\ref{leftvertex}), and $m$ on the other, (\ref{rightvertex}). Here, the cut $\gamma$ is a curve on the worldsheet that starts at $\sigma_1$ on the left boundary, crosses the cylinder and ends at $\dot\sigma_1$ on the right boundary.}\label{cylinder}
\end{figure}

We will be working in conformal gauge and parameterizing the Euclidean cylinder by a periodic coordinate $\sigma\in[0,2\pi]$ and a coordinate along the cylinder $\tau\in[0,L]$, where $iL/(2\pi)$ is the modular parameter of the cylinder, see figure \ref{cylinder}. The cylinder has two boundaries. A set of $n$ vertex operators is inserted on the boundary at $\tau = 0$, corresponding to a set of ordered gluon asymptotic states,
\beq\la{leftvertex}
V\left({k}_i;\,\sigma_i\right) \propto e^{i\,{k}_i\cdot\, {x}\left(0,\,\sigma_i\right)}\ , \qquad i = 1,\dots,n\ .
\eeq
Similarly, on the other boundary, at $\tau = L$, a set of $m$ ordered gluon vertex operators is inserted,
\beq\la{rightvertex}
V({k}_{n+j};\,\dot\sigma_{j}) \propto e^{i\,{k}_{n+j}\cdot\, {x}(L,\,\dot\sigma_{j})}\ , \qquad j= 1,\dots,m\ ,
\eeq
where the dot in $\dot\sigma_j$ indicates that it is located on the second boundary of the cylinder. The full double trace amplitude is given by a sum of the two relative orderings of the gluons vertex operators on the two boundaries of the cylinder. Only the ordering for which $\sigma_i<\sigma_{i+1}$ and $\dot\sigma_{j}<\dot\sigma_{j+1}$ on these two boundaries will be considered, see figure \ref{cylinder}. The other ordering is related to this one by a relabelling of the external gluons.

Contrary to the planar case, the total momentum going through each boundary is non-zero,
\beq\la{qmomenta}
{q}=\sum\limits_{i=1}^{n} {k}_i= - \sum\limits_{j=1}^{m} {k}_{n+j}\ .
\eeq
The Euclidean worldsheet action reads
\beq\la{wsaction}
S_1 = \frac{\sqrt{\lambda}}{4\pi}\,\int\limits_{\mathcal{D}} d\sigma d\tau\,\frac{\left(\partial_\alpha z\right)^2 + \left(\partial_\alpha{x}\right)^2}{z^2}
-i\sum\limits_{i=1}^{n} {k}_i\cdot x\(0,\sigma_i\)-i\sum\limits_{j=1}^{m} k_{n+j}\cdot x(L,\dot\sigma_{j})\ ,
\eeq
where $z$ is the radial AdS direction in Poincaré coordinates.

Using the same manipulations as in \cite{McGreevy:2007kt} one can rewrite the contribution to the action of all the vertex operators on one boundary as
\beq
i\sum\limits_{i=1}^{n} {k}_i\cdot x(0,\sigma_i)= i\,q\cdot x(0,\sigma_1)- i\sum\limits_{i=1}^{n} \int\limits_{\sigma_i}^{\sigma_{i+1}}d\sigma\,\d_\sigma x(0,\sigma)\cdot(\sum\limits_{p\leq i} k_p + c)\ ,
\eeq
where $c$ is an arbitrary constant four-vector and $\sigma_{n+1}=\sigma_1$. The only difference from the single boundary case is the new term $q\cdot x(0,\sigma_1)$. A similar contribution comes from the other boundary at $\tau=L$, given by $-q\cdot x(L,\dot\sigma_{1})$. One can rewrite the sum of these two new contributions as an integral of a total derivative in $\tau$,
\begin{align}
i\,q\cdot [x(0,\sigma_1)-x(L,\dot\sigma_{1})]= -\,i\,{q}\cdot\int\limits_0^{L}d\tau\,\partial_\tau{x}\left(\tau,\gamma\left(\tau\right)\right)\ ,\label{DefGamma}
\end{align}
where $\sigma=\gamma(\tau)$ is an arbitrary curve on the cylinder that stretches between the boundaries and obeys $\gamma(0) = \sigma_1$, $\gamma(L) =\dot\sigma_{1}$. 

The action is invariant under global translations of $x$, as is evident from the fact that $x$ now enters it only with derivatives. In order to construct the T-dual action we follow Buscher \cite{Buscher:1985kb} and gauge this translation symmetry. This is done by introducing a worldsheet gauge field ${A}_\alpha$ that transforms under a local translation $x\to x+a(\tau,\sigma)$ as ${A}_\alpha \rightarrow {A}_\alpha + \partial_\alpha{a}$. The extended action takes the following form,
\begin{align}
S_2 =\,\,& \frac{\sqrt{\lambda}}{4\pi}\,\int\limits_{\mathcal{D}} d\sigma d\tau\,\left(\frac{\left(\partial_\alpha z\right)^2 + \left(\partial_\alpha{x} - {A}_\alpha\right)^2}{z^2} - i\,{y}\cdot{F}\right)\nn\\
& + i\,{q}\cdot\int\limits_0^{L}d\tau\,[\partial_\tau{x}\left(\tau,\gamma(\tau)\right) - {A}_\tau\left(\tau,\gamma(\tau)\right)] - i\,l\cdot\oint d\sigma\,{A}_\sigma(\tau,\sigma)\nn\\
& + i\sum\limits_{i=1}^{n}\int\limits_{\sigma_i}^{\sigma_{i+1}}d\sigma\,\left(\partial_\sigma{x}\left(0,\sigma\right) - {A}_\sigma(0,\sigma)\right)\cdot(\sum\limits_{p\leq i}{k}_p + {c})\la{gaugedaction}\\
& + i\sum\limits_{j=1}^{m}\int\limits_{\dot\sigma_j}^{\dot\sigma_{j+1}}d\sigma\,\left(\partial_\sigma x(L,\sigma) - {A}_\sigma(L,\sigma)\right)\cdot(\sum\limits_{p\leq j}{k}_{n+p} - c)\ .\nn
\end{align}
Here, ${y}\left(\tau,\sigma\right)$ is a vector field Lagrange multiplier that sets ${F} = \partial_\tau{A}_\sigma - \partial_\sigma{A}_\tau= \partial_{[\tau}{A}_{\sigma]}$ to zero. The vector $l$ is another Lagrange multiplier, which ensures that the holonomy of ${A}_\alpha$ around the cylinder vanishes. Letting the curve $\gamma$ wrap around the cylinder one more time has the effect of shifting $l$ by $q$. Since ${F} = 0$, this holonomy is independent of $\tau$. Together these two constraints ensure that ${A}_\alpha$ is a flat connection on the cylinder and that the action in (\ref{gaugedaction}) is independent of the curve $\gamma(\tau)$.  

Since the connection is flat and $x$ is periodic, the gauge ${x} = 0$ can be chosen. In this gauge, the action becomes
\begin{align}
S =\,\,& \frac{\sqrt{\lambda}}{4\pi}\,\int\limits_{\mathcal{D}} d\sigma d\tau\,\left(\frac{\left(\partial_\alpha z\right)^2 +  {A}_\alpha^2}{z^2} - i\, {y}\cdot {F}\right) - i\, {q}\cdot\int\limits_0^{L}d\tau\,  {A}_\tau\left(\tau,\gamma\left(\tau\right)\right) - i\, l\cdot\oint d\sigma\, {A}_\sigma\left(\tau,\sigma\right)\nonumber\\
&- i\sum\limits_{i=1}^{n}\int\limits_{\sigma_i}^{\sigma_{i+1}}d\sigma\, A_\sigma\left(0,\sigma\right) \cdot(\sum\limits_{p\leq i}{k}_p + {c})
- i\sum\limits_{j=1}^{m}\int\limits_{\dot\sigma_j}^{\dot\sigma_{j+1}}d\sigma\,A_\sigma(L,\sigma) \cdot(\sum\limits_{p\leq j}{k}_{n+p} - c)\ .\la{action2}
\end{align}
Integrating over ${y}$ and $l$ sets $A_\alpha = -\d_\alpha\tilde{ {x}}$, where $\tilde{ {x}}$ is pure gauge. The action then reduces to the original one from equation (\ref{wsaction}) with $ {x}$ replaced by $\tilde{ {x}}$. Hence, the two actions, $S_1$ in (\ref{wsaction}) and $S$ in (\ref{action2}), are equivalent. 

In order to construct the T-dual action ${A}_\alpha$ is integrated out in $S$. This is done by first integrating the term $ {y}\cdot {F}$ by parts, moving the derivatives from $A_\alpha$ to $y$. Then the holonomy term $\oint d\sigma\, {A}_\sigma\left(\tau,\sigma\right)$ is evaluated at the $\tau=L$ boundary.\footnote{This is made possible by the holonomy being independent of $\tau$.} After an overall rescaling of the fields, $\left( {A},\, {y},\,z\right) \rightarrow \left(\frac{\sqrt{\lambda}}{4\pi} {A},\,\frac{4\pi}{\sqrt{\lambda}} {y},\,\frac{\sqrt{\lambda}}{4\pi}z\right)$, the action takes the form,
\begin{align}
S =\frac{\sqrt{\lambda}}{4\pi}&\left[\,\int\limits_{\mathcal{D}} d\sigma d\tau\,\left(\frac{\left(\partial_\alpha z\right)^2 +  {A}_\alpha^2}{z^2} + i\, {A}_{[\sigma}\cdot\partial_{\tau]} {y} \right) \right.
\nn\\
&\qquad + i\int\limits_0^{L}d\tau\, {A}_\tau\left(\tau,\gamma\left(\tau\right)\right)\cdot[ {y}\left(\tau,\gamma\left(\tau\right) + 2\pi\right) - {y}\left(\tau,\gamma\left(\tau\right)\right) -  {q}]\nonumber\\
& \qquad- i\sum\limits_{i=1}^{n}\int\limits_{\sigma_i}^{\sigma_{i+1}}d\sigma\,  {A}_\sigma\left(0,\sigma\right)\cdot(\sum\limits_{p\leq i}k_p+c+y\left(0,\sigma\right))\label{FinalAction}\\
&\qquad\left. - i\sum\limits_{j=1}^{m}\int\limits_{\dot\sigma_j}^{\dot\sigma_{j+1}}d\sigma\, {A}_\sigma\left(L,\sigma\right)\cdot(\sum\limits_{p\leq j} {k}_{n+p} +l-  {c} -  {y}\left(L,\sigma\right))\right]\ .\nonumber
\end{align}
Note that the field $y$ can be discontinuous on a line along the cylinder without causing the action to diverge. As a result, a new boundary term arises from the integration by parts in $\sigma$. In (\ref{FinalAction}) this new boundary term was placed along the curve $\gamma\left(\tau\right)$ that was introduced in (\ref{DefGamma}).

Integrating out ${A}_\alpha$ in the bulk of the cylinder leads to the string action in the T-dual AdS background, $ds^2=(d\tilde z^2+dy^2)/\tilde z^2$, with $\tilde z=1/z$. The integration of $A_\tau$ along the curve $\sigma=\gamma(\tau)$ gives the periodicity constraint for the T-dual $y$ field,
\begin{align}\la{periodicconstraint}
\lim_{\epsilon\to0}\ {y}(\tau,\gamma\left(\tau\right) + 2\pi-\epsilon)= \lim_{\epsilon\to0}\  y(\tau,\gamma\left(\tau\right)+\epsilon) +q\ .
\end{align}
It implies that going around the cylinder changes the value of $y$ by ${q}$. Hence, the image of the cylinder in the T-dual AdS is not a cylinder, but its universal cover. In particular, the quantum fluctuations of the string at $(\tilde z,y)$ have an image at $(\tilde z,y+q)$ and are not independent.

Finally, integrating out the boundary values of $ {A}_\sigma$ gives the following Dirichlet boundary conditions for the T-dual $y$ coordinate,
\begin{align}
y(0,\sigma_i < \sigma <\sigma_{i+1}) &= - \sum\limits_{p\leq i}\,k_p -c\ ,\nn\\
y(L,\dot\sigma_{j} < \sigma <\dot\sigma_{j+1}) &= \ \ \ \sum\limits_{p\leq j}\,k_{n+p}-c+l\ .
\end{align}

\begin{figure}[t]
\centering
\def\svgwidth{0.8\textwidth}
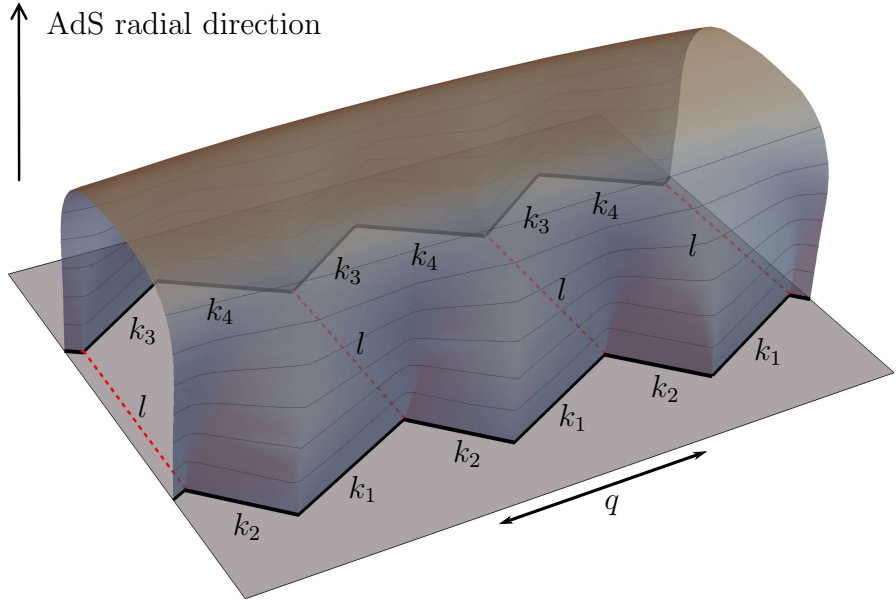
\caption{\small The T-dual of a string with a cylindrical topology that is holographically dual to a double trace amplitude. It ends at the boundary of $AdS_5$ along two periodic lines composed of the null external gluon momenta. The worldsheet is subject to a quantum periodicity constraint that restricts it to be periodic in the bulk as well. The period is equal to the total momentum flow through the cylinder, $q$ in (\ref{qmomenta}). The separation between the two lines, $l$, is T-dual to the momentum flow around the cylinder and it is being integrated over.}\label{WSD}
\end{figure}

These conditions imply that the T-dual string stretches between two periodic null polygonal Wilson lines. They are constructed from the ordered momenta $\{{k}_1,k_2,\dots,k_n\}$ and $\{k_{n+1},\dots,k_{n+m}\}$, respectively. The period of each of these lines is $q$. The vector $c$ corresponds to a simultaneous translation of the two lines and can be set to zero. The vector $l$ is the separation between the two lines and is being integrated over, see figure \ref{WSD}. This integration projects the total momentum flow between the two lines to zero. Under T-duality this momentum is mapped to the winding of the string state on the cylinder and this projection is the expected T-dual manifestation of fact that the string state on the amplitude side has zero winding. 

The periodic srting path integral obtained above is equivalent to the one of a string in a spacetime with the $q$ direction compactified on a circle of radius $q/(2\pi)$. Before T-duality the string has momentum number one and winding number zero around the circle. After T-duality, the string ends on two closed null polygons that wrap around the circle with edges $(k_1,\dots,k_n)$ and $(k_{n+1},\dots,k_{n+m})$. The string has winding number that is fixed to one and momentum number zero around the circle. The latter projection comes about due to the integration over the component of $l$ in the direction of $q$.

There is another interesting way of thinking about the integration over $l$. The vectors $q$ and $l$ span a two dimensional space. An orthogonal basis for this space is $\{q,l_\perp\}$, where $l_\perp=l-q(l\cdot q/q^2)$. The quantity $i|l_\perp|/|q|$ can be thought of as the spacetime modular parameter of the cylinder. At the semiclassical level the Virasoro constraint relates it to the worldsheet modular parameter $iL/(2\pi)$. Hence, the integration over $l$ can be converted into the integration over the worldsheet modular parameter.\footnote{{To make this relation precise it is convenient to use static gauge in the $\{q,l_\perp\}$ plane instead of the conformal one used here.}}

At strong coupling the periodic srting path integral is dominated by its minimal surface area saddle point. Because the boundary conditions are periodic, so is the minimal surface. Hence, the periodic constraint (\ref{periodicconstraint}) is automatically satisfied. This minimal surface area can be calculated using a simple generalization of the techniques of \cite{Alday:2010vh,Alday:2010ku,Maldacena:2010kp,unpublished} and leads to a periodic Y-system. It will be reported on in \cite{toappear}.

\section{One-loop duality test}\la{oneloopsec}

In this section, the duality will be tested explicitly for the four-point MHV amplitude at the leading order in perturbation theory. All tree-level amplitudes are single trace, so the double trace amplitude receives its first perturbative contribution at one-loop order. The cutting procedure strips away the loop integration and by that reduces the number of loops from one to zero. On the other side of the duality, one has the cylinder expectation value of a Wilson lines correlator with eight $\eta$ insertions. Similarly to the N$^2$MHV Wilson loop, this object starts at tree level. Specifically, the duality tested in this section is
\beq\la{duality4p}
{\mathbb M}_{2,2}^\text{tree}(l) ={\mathbb W}_{2,2}^{\,\text{tree}}(l)\ ,
\eeq
where
\beq\la{AtoR}
{\mathbb A}_{2,2}^\text{tree}(l)={g_{YM}^2\,\delta^4\(\sum k_i\)\delta^8\(\sum \lambda_i\tilde\eta_i\)\over\<1\,2\>^2\<3\,4\>^2}
\times{\mathbb M}_{2,2}^\text{tree}(l)\ .
\eeq

The left hand side of the duality equation (\ref{duality4p}) is evaluated in section \ref{cutoneloop} and the right hand side in section \ref{WL1l}.

\subsection{Cylindrically cut four point double trace amplitude at Born level}\la{cutoneloop}

The four point double trace amplitude was studied at one-loop order in \cite{Bern:1990ux,Feng:2011fja}. It can be expressed as a sum of three massless scalar box integrals,
\beq\la{oneloopdt}
\begin{split}
A_{2,2}^\text{1-loop} &= 2\,A_4^\text{1-loop}\left(1,2,3,4\right) + 2\,A_4^\text{1-loop}\left(1,3,2,4\right) + 2\,A_4^\text{1-loop}\left(1,3,4,2\)\\
&=2\,g_{YM}^2\,s_{12}s_{23}\,A_4^\text{tree}(1,2,3,4)\left(\hspace{5pt}\boxa + \boxb + \boxc\hspace{5pt}\right)\ ,
\end{split}
\eeq
where $A^\text{tree}$ is the Parke-Taylor tree-level partial amplitude,
\beq
A_4^\text{tree}(1,2,3,4)={g^2_\text{YM}(2\pi)^4\,\delta^4\(\sum k_i\)\delta^8\(\sum \lambda_i\tilde\eta_i\)
\over\<1\,2\>\<2\,3\>\<3\,4\>\<4\,1\>}\ ,
\eeq
and the massless scalar box integral is 
\beq
\planarbox = \int\frac{d^4l}{\left(2\pi\right)^4}\,\frac{1}{l^2\left(l + k_1\right)^2\left(l + k_1 + k_2\right)^2\left(l - k_4\right)^2}\ .
\eeq
In (\ref{oneloopdt}) the inner and the outer faces of the box represent the two traces. The box integrals only depend on the distribution of the external momenta on the cusps of the box. 

Next, this expression is rewritten in terms of the cylindrically cut amplitude,
\beq
A_{2,2}^\text{1-loop} ={\lambda\over N}\,\int d^4l\, \cA_{2,2}^\text{tree}\left(l\right)\ .
\eeq
For the special case considered here there are only two particles in each trace and, therefore, no relative ordering and no orientation on the cylinder. As opposed to the fishnet model, ${\cal N}=4$ SYM is CPT invariant and its 't Hooft string is orientable. Hence, when taking the cylindrical cut, one has to average over the two orientations. Consequently, each of the three box diagrams in (\ref{oneloopdt}) has two different cylindrical cuts. Appendix \ref{nm2app} expands on this point. 

The cylindrically cut amplitude $A_{2,2}^\text{tree}\left(l\right)$ is only defined up to a shift of $l$ by $q$, (\ref{cutsum}). One representative of this class of amplitudes is
\beq\la{cutamp}
\begin{split}
\cA_{2,2}^\text{tree}\left(l\right) =s_{12}s_{23}\,A_4^\text{tree}\times\Bigg(\hspace{5pt}&\cboxa + \cboxb + \cboxc\\
+\,\,&\cboxaa + \cboxbb + \cboxcc\hspace{5pt}\Bigg)\ ,
\end{split}
\eeq
where the blue dashed line represents the cut (\ref{deltainsert}). When drawn on the cylinder, the six cut boxes in equation (\ref{cutamp}) take the following form,
\beq
\includegraphics[width=.75\textwidth]{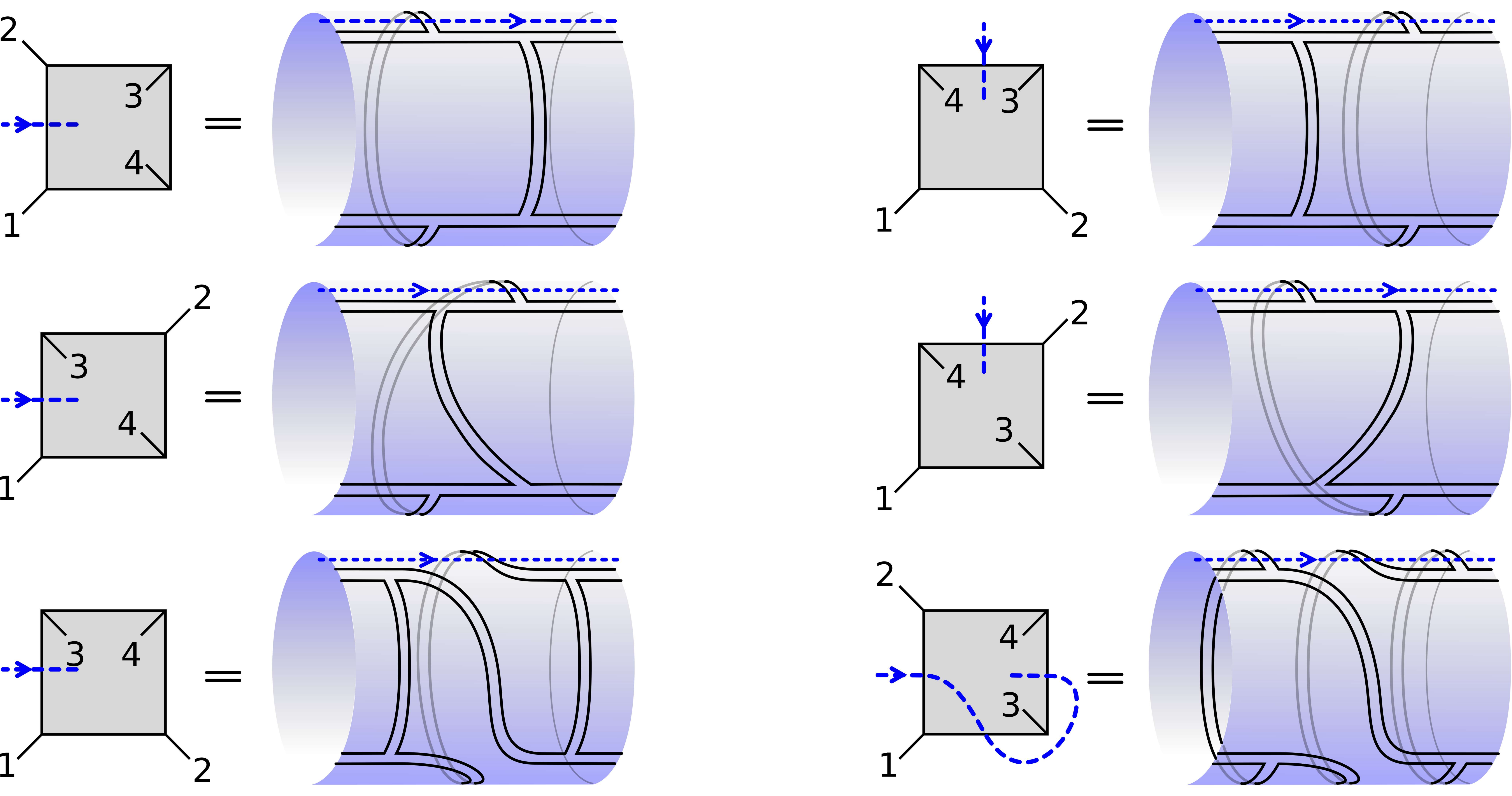}\quad.
\eeq
Explicitly,
\begin{align}
\cboxa &= \frac{1}{l^2\left(l - k_1\right)^2\left(l + k_2\right)^2\left(l + k_2 + k_3\right)^2},&\cboxaa &= \frac{1}{l^2\left(l + k_3\right)^2\left(l - k_4\right)^2\left(l + k_2 + k_3\right)^2},\nonumber\\
\cboxb &= \frac{1}{l^2\left(l - k_1\right)^2\left(l + k_3\right)^2\left(l + k_2 + k_3\right)^2},&\cboxbb &= \frac{1}{l^2\left(l + k_2\right)^2\left(l - k_4\right)^2\left(l + k_2 + k_3\right)^2},\nonumber\\
\cboxc &= \frac{1}{l^2\left(l - k_1\right)^2\left(l + k_3\right)^2\left(l - k_1 - k_2\right)^2},&\cboxcc &= \frac{\left(l - k_1 + k_3\right)^{-\,2}}{\left(l - k_1\right)^2\left(l + k_3\right)^2\left(l + k_2 + k_3\right)^2}.
\end{align}
The cylindrically cut double trace amplitude ${\mathbb M}_{2,2\ \text{tree}}^\text{MHV}(l)$ is obtained by summing (\ref{cutamp}) over all integer shifts, $l\to l+a\,q$, and stripping away the factor in (\ref{AtoR}). These infinite sums can be evaluated analytically in Mathematica. 

It is important to point out that the reduction procedure to scalar box diagrams that was used in \cite{Feng:2011fja} to derive the representation of the one-loop amplitude (\ref{oneloopdt}) does not involve integration by parts. Therefore, this procedure commutes with the cylindrical cutting prescription. Beyond one loop, however, this is no longer the case and one has to first take the cylinder cut of the Feynman diagram and only then apply reduction procedures to scalar integrals.\footnote{See, for example, \cite{Bern:1997nh} for a two-loop representation of the double trace amplitude in terms of scalar integrals. One cannot apply the cylindrical cutting procedure directly to these types of representations.}

\subsection{Wilson lines correlator at tree level}\la{WL1l}

Next, the result obtained above will be reproduced on the Wilson lines correlator side of the duality. First, the supercomponents of the Wilson lines will be discussed. These components determine the vertex and edge insertions along the periodic lines (\ref{superlines}). This part of the calculation is not limited to perturbation theory and therefore is valid to all orders.

We consider the four-point MHV amplitude (\ref{AtoR}). All the supercomponents of MHV amplitudes are related by supersymmetry and are accounted for by $\delta^8(\cQ)$ in (\ref{AtoR}). Therefore, all the $\tilde\eta$'s inside ${\mathbb W}_{2,2}$ can be set to zero. This implies that all the cusps of each line share the same supercoordiante,
\beq
\theta_1^{[a]}=\theta_2^{[b]}\ ,\qquad\dot\theta_3^{[a]}=\dot\theta_4^{[b]}\ ,\qquad\forall\,a,b \ . 
\eeq
The superseparation of the two lines, $\theta=\dot\theta_2-\theta_2$, still has to be integrated over,
\beq\la{4pMHVV}
{\mathbb W}_{2,2}(l) = \int\!d^8\theta\,\widehat{\cW}_{2,2}(l,\theta) \ .
\eeq
Due to the dual supersymmetry, the Wilson lines correlator is invariant under simultaneous shift of the $\theta_i$'s and the $\dot\theta_j$'s. This symmetry is trivialized by the map between the amplitude and the Wilson lines variables. It can be used to set either the $\theta_i$'s or the $\dot\theta_j$'s to zero. For example, for $\dot\theta_j=0$, $\theta_i=-\theta$ and one finds a simple expression for the $\eta$'s of the left line, $\eta_i=\<i\,\theta\>$. For this choice the $\theta$ integration (\ref{4pMHVV}) takes the form,
\beq\la{WL12}
{\mathbb W}_{2,2}(l)=\int\!d^8\theta\,\widehat\cW_{2,2}(-\theta,-\theta;0,0|l)=\int\!d^8\theta\,\widetilde\cW_{2,2}(\<1\,\theta\>,\<2\,\theta\>;0,0|l) \ ,
\eeq
where $\widetilde\cW_{2,2}(\eta_1,\eta_2;\eta_3,\eta_4|l)$ is a function of the $\eta$ variables and $\widehat\cW_{2,2}(\theta_1,\theta_2;\theta_3,\theta_4|\,l)$ is a function of the $\theta$'s. The integration over $\theta$ can now be converted into an integration over $\eta_1$ and $\eta_2$. For a generic kinematical configuration, $\lambda_1$ and $\lambda_2$ are independent and hence can be used as a basis for integrating over $\theta^{A=1,2,3,4}$. That is,
\beq\la{spinorbasis}
|\theta^A\>={|2\>\<1|-|1\>\<2|\over\<1\,2\>}|\theta^A\> \ .
\eeq
In the special case considered above ($\theta_1=\theta_2=-\theta$, $\dot\theta=0$) this relation becomes
\beq
|\theta^A\>={|2\>\eta_1^A-|1\>\eta_2^A\over\<1\,2\>} \ .
\eeq
Therefore,
\beq\la{duality12}
{\mathbb W}_{2,2}(l) = \<1\,2\>^4\int\!d^4\eta_1\,d^4\eta_2\,\widetilde\cW_{2,2}(\eta_1,\eta_2;0,0|l) \ .
\eeq

Alternatively, one could set $\theta_i=0$ and decompose $\dot\theta_j=\theta$ in the basis $\{\lambda_3,\lambda_4\}$. This results in
\beq\la{duality34}
{\mathbb W}_{2,2}(l) = \<3\,4\>^4\int\!d^4\eta_3\,d^4\eta_4\,\widetilde\cW_{2,2}(0,0;\eta_3,\eta_4|l) \ .
\eeq
Another possible choice is a hybrid parametrization for which $\dot\theta^{A=1,2}=0$ and $\theta^{A=3,4}=0$. This choice will turn out to be very convenient later on. It leads to the following parametrization of the Wilson loop,
\begin{align}\la{2222RF}
{\mathbb W}_{2,2}(l) = \langle1\,2\rangle^2\,\langle3\,4\rangle^2\int\!d\eta^1_1\,d\eta^2_1\,d\eta^1_2\,d\eta^2_2\,d\eta^3_3\,d\eta^4_3\,d\eta^3_4\,d\eta^4_4\,\widetilde\cW_{2,2}(\eta_1,\eta_2;\eta_3,\eta_4|l) \ .
\end{align}
For this parametrization the factors outside of the integral in (\ref{2222RF}) cancel the Parke-Taylor-like factor in (\ref{AtoR}) and the duality (\ref{duality4p}) takes the form,
\begin{align}\la{2222}
{\mathbb A}_{2,2}^\text{MHV}(l)&= g_{YM}^2\,\delta^4\(\sum k_i\)\delta^8\(\sum \lambda_i\tilde\eta_i\)\times\\
&\times \int\!d\eta^1_1\,d\eta^2_1\,d\eta^1_2\,d\eta^2_2\,d\eta^3_3\,d\eta^4_3\,d\eta^3_4\,d\eta^4_4\,\widetilde\cW_{2,2}(\eta_1,\eta_2;\eta_3,\eta_4|l)\nn \ .
\end{align}

\begin{figure}[t]
\centering
\def\svgwidth{.9\textwidth}
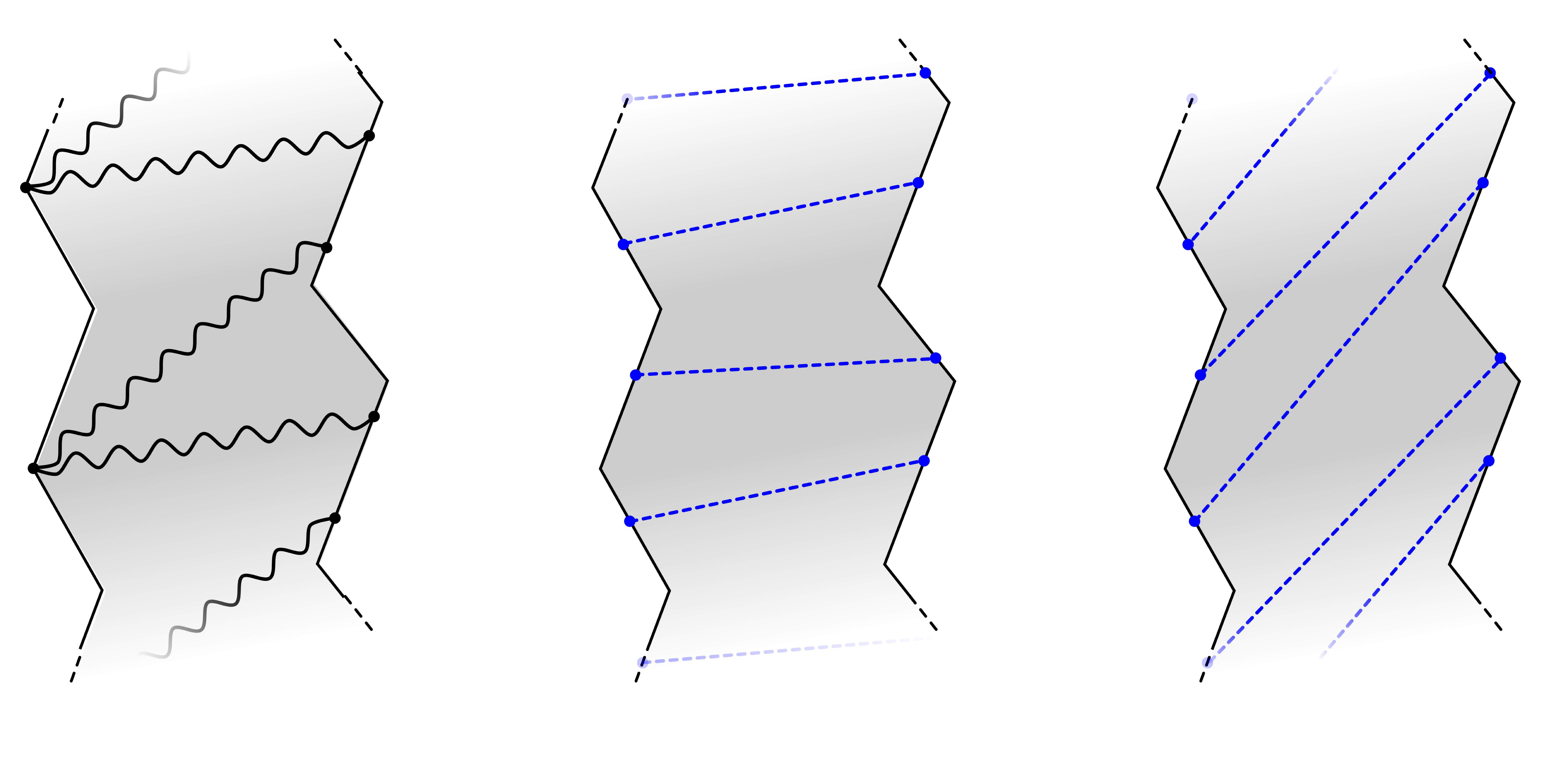
\caption{\small a) An example of a gluon contraction that contributes to $\mathbb{W}_{2,2}\left(l\right)$ for the parametrization (\ref{duality12}). On the left line one finds a square of the field strength component ($F^2$) inserted at a single cusp (it is not the same as the square of the $F$ component  that appears when it is inserted by itself). On the right line one finds two gauge fields that are integrated over an edge. These integrals are intertwined, making it harder to extract the result from known single trace amplitudes. b,c) Two specific edge-to-edge diagrams in the hybrid component parametrization (\ref{2222RF}). Diagram (b) is related to (c) by a shift of the right line by one period. Diagram (b) can be chosen to be the representative of this family in $\cB(\{Z_i\},\{\dot{Z}_j\})$. For that choice, diagram (c) will contribute to $\cB(\{Z_i\},\{\dot{Z}_j^{[+]}\})$ in the sum (\ref{2222c}).}\label{NLC}
\end{figure}

So far, the discussion has been valid at any loop order. We now focus on the calculation of the cylinder Wilson lines correlator at the leading order in perturbation theory. Any choice of the parametrization requires performing an independent calculation, all of which give the same result in the end. Here, making a convenient choice of the Wilson lines parametrization can simplify the calculation dramatically. For example, the choice (\ref{duality12}) includes the contribution that is shown in figure \ref{NLC}.a. For this contribution two intertwined integrations of a gauge field along the $k_4$ edge result in difficulty isolating the contribution of a single block. On the other hand, for the hybrid parametrization (\ref{2222}) all insertions are scalars and the blocks decouple. The calculation in this case is easier to perform because the result of each individual scalar contraction can be extracted from known single trace amplitudes.

\begin{figure}[t]
\centering
\def\svgwidth{0.95\textwidth}
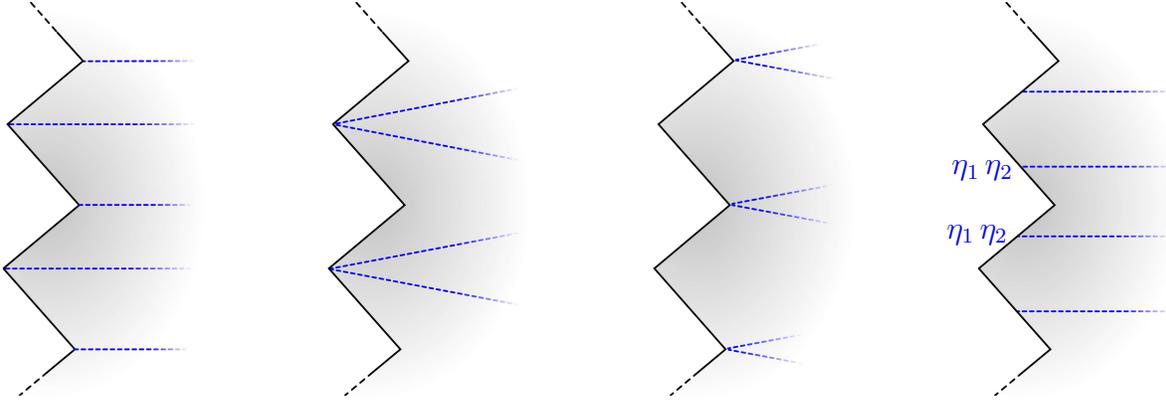
\caption{\small All possible ways how scalar propagators can attach themselves to a Wilson line for the hybrid parametrization (\ref{2222RF}). The second and third contributions belong to the same double cusp class.}\label{WLC}
\end{figure}

In the hybrid configuration (\ref{2222}) one finds two $\eta$'s on each edge. The left line has only R-charge components $\eta_i^A$ with $A=1$ and $A=2$, while the right line has only $A=3$ and $A=4$ components. As follows from the detailed analysis of the vertex and edge operators in \cite{CaronHuot:2010ek}, only scalar edge and vertex insertions contribute to this configuration. 
The two scalars on each side appear in three types of configurations that are shown in figure \ref{WLC}. We will refer to them as \textit{single cusp}, \textit{double cusp} and \textit{edge} configurations, respectively. To calculate the cylinder Wilson line correlator the scalars on the left line have to be contracted with the scalars on the right line in all possible planar periodic ways. 

Every such contraction is repeated with all possible integer period shifts of one line with respect to the other. For example, in figures \ref{NLC}.b and \ref{NLC}.c two different edge-to-edge contractions that are related to each other by a shift of the right line by a single period are shown. This property allows us to define a building block $\cB(\{Z_i\},\{\dot{Z}_j\})$, which includes a single representative of every type of contraction. This object is not uniquely defined -- any representative from the infinite family of contractions that are related by a shift $l\to l+a\,q$ can be used. In terms of $\cB$ the cylinder Wilson line correlator is given by
\beq\label{2222c}
{\mathbb W}_{2,2\ \text{tree}}^{\text{N}^2\text{MHV}}(l) = \langle1\,2\rangle^2\langle3\,4\rangle^2\sum\limits_{a\,=\,-\infty}^\infty \cB(\{Z_i\},\{\dot{Z}_j^{\left[a\right]}\}) \ .
\eeq

The building block $\cB$ is, in a sense, the dual space analog of the cylindrically cut amplitude $\cA$ in (\ref{cutsum}). However, as opposed to the fishnet theory discussed in section \ref{Kazsec}, where the two sums were matched on a diagram-by-diagram basis, here this seems to be just an analogy. We could not find a choice of representative $\cB$ and a cylindrically cut amplitude $\cA$, such that the two are equal prior to the summations in (\ref{2222c}) and (\ref{cutsum}). 

The advantage of working with the hybrid configuration is that any specific contraction of the scalars between the two lines, such as the edge one in figure \ref{NLC}.b, factors into a product of two planar single trace contractions. Therefore, the periodicity of the Wilson lines configuration becomes irrelevant as it is reduced to a product of expectation values of closed polygons. This allows the amplitude to be expressed through known tree-level single trace amplitudes. Take, for example, any edge-to-edge contraction, represented by the blue dashed lines in figure \ref{NLC}.b and  \ref{NLC}.c. It is equal to the product of two independent scalar propagators ending on two separate edges. Consider one such scalar contraction, say between the $i$'th edge of the left line, whose twistors are labelled as $\{\dots,Z_{i-1},Z_i,Z_{i+1},\dots\}$, and the $j$'th edge of the right line, whose twistors are labelled as $\{\dots,\dot Z_{j-1},\dot Z_j,\dot Z_{j+1},\dots\}$. This single edge-to-edge contraction can be evaluated as the $(i,i,j,j)$ component of the hexagon NMHV amplitude with the following six twistors, $\{Z_{i-1},Z_i,Z_{i+1},\dot Z_{j-1},\dot Z_j,\dot Z_{j+1}\}$,
\beq
EE(i,j) = R^{(i,i,j,j)}_{\text{tree}}=\,\,\,\ee \ .
\eeq

One convenient source of any tree-level and one-loop N$^k$MHV $n$-point amplitude is the Mathematica package \cite{Bourjaily:2013mma} by Bourjaily, Caron-Huot and Trnka. The tree-level Hexagon NMHV amplitude can be extracted using this package by defining the six twistors as above and running the command {\verb superComponent[{},{1,2},{},{},{3,4},{}]@treeAmp[6, 1]}. 

The other two types of single scalar contractions are cusp-to-cusp and cusp-to-edge. These two can likewise be extracted from the NMHV hexagons as
\begin{alignat}{3}
CC\left(i,\,j\right) & = R^{\left(i,\,i+1,\,j,\,j+1\right)}_{\text{tree}} &&=\,\,\,\cc \nonumber \ ,\\
CE\left(i,\,j\right) & = R^{\left(i,\,i+1,\,j,\,j\right)}_{\text{tree}} &&=\,\,\,\ce \ . \label{Hexx}
\end{alignat}

All the contributions to $\cB$ can be arranged into nine groups corresponding to all different pairings of the single cusp, double cusp or edge insertions between the lines. Any contraction with a single and double cusp configurations, shown in the left three diagrams in figure \ref{WLC}, comes with an extra factor of $-1$, which appears from rearranging the Grassmann $\eta$ parameters on the edges. Additionally, the single cusp configuration also generates a factor of 2, due to the fact that there are two different distributions of $\eta$ parameters that contribute to it. 

The first class of contractions is a single cusp contracted with a single cusp. One choice of a representative for this class is\footnote{Note that the six twistors that are used to evaluate $C(i,j)$ are $\{Z_{i-1},Z_i,Z_{i+1},\dot Z_{j-1},\dot Z_j,\dot Z_{j+1}\}$. For example, for $CC(1,4)$ these six twistors are $\{Z_{2}^{[-]},Z_1,Z_{2},\dot Z_{3},\dot Z_4,\dot Z_{3}^{[-]}\}$.}
\beq
CC1=4\[CC(1,4)CC(2,\,3) + CC(1,\,3)CC(2,4^{[+]})\] \ .
\eeq
The second one is the contraction of a single cusp on one line with a double cusp on the other. The following representative is chosen for it, 
\begin{align}
CC2=&\,\,2\[CC(1,4)CC(1,3^{[-]}) + CC(1,4)CC(1,3)\right.\nn\\
&+CC(1,4)CC(2^{[-]},4) + CC(1,4)CC(2,4)\nn\\
&+CC(2,3)CC(2,4) + CC(2,3)CC(2,4^{[+]})\nn\\
&+\left.CC(2,3)CC(1,3) + CC(2,3)CC(1^{[+]},3)\] \ .
\end{align}
The third one corresponds to contractions between double cusp configurations on both lines,
\begin{align}
CC3=&\,\,CC(1,4)^2 + CC(2,3)^2 + CC(1,3)^2 + CC(2,4)^2\nonumber\\
&+CC(1,4)CC(1,4^{[+]}) + CC(2,4)CC(2,4^{[+]}) \nonumber\\
&+CC(2^{[-]},4)CC(2,4) + CC(2^{[-]},3)CC(2,3) \ .
\end{align}
Combining the edge configuration with the single cusp one finds,
\begin{align}
CE1=&- 2\[CE(1,4)CE(2,3) + CE(1,3)CE(2,4^{[+]})\right.\nn\\
&\left.+\,\,CE(4,1)CE(3,2) + CE(4,2)CE(3,1^{[+]})\] \ .
\end{align}
The edge combined with the double cusp results in
\begin{align}
CE2=&- CE(2,4)CE(2,3) - CE(2,3)CE(2,4^{[+]})\nonumber\\
&- CE(1,4)CE(1,3) - CE(1,3)CE(1,4^{[+]})\\
&- CE(3,1)CE(3,2) - CE(3,2)CE(3,1^{[+]})\nonumber\\
&- CE(4,1)CE(4,2) - CE(4,2)CE(4,1^{[+]})\nonumber \ .
\end{align}
Finally, a representative for contractions of edge insertions on both sides is
\begin{align}
EE1=EE(1,4)EE(2,3) + EE(1,3)EE(2,4^{[+]}) \ .
\end{align}
Combining all the pieces together, one finds, 
\beq\la{Btotal}
\cB(\{Z_i\},\{\dot{Z}_j\}) = CC1+CC2+CC3+CE1+CE2+EE1 \ .
\eeq

The final step is to plug $\cB$ from (\ref{Btotal}) into the sum in (\ref{2222c}). We did not perform this sum analytically. Instead, we truncated it after a certain number of blocks and evaluated the result numerically with a generic randomly generated geometry. Summing over 10,000 blocks takes a few minutes on a standard laptop and gives a match with the cylindrically cut amplitude (\ref{duality4p}) up to the 18'th digit (that is an error of $10^{-16}\%$). It would be interesting to prove this match analytically.

\section{BCFW recursion relation at Born level}\la{BCFWsec}

To establish the duality of the cylindrically cut one-loop double trace amplitude and the cylinder Wilson lines correlator at Born level we use the BCFW on-shell recursion relation. Since the BCFW recursion relation was not established before for the cylindrically cut amplitude, it will have to be studied on both sides of the duality. 
The aim is to prove that
\beq\la{duality2}
{\mathbb M}_{\,n,m}^\text{tree}(l)={\mathbb W}^{\,\text{tree}}_{n,m}(l)\ ,
\eeq
by showing that both sides satisfy the same recursion relation with the same initial conditions. We begin by studying the poles in $l$ of both sides. These poles can be thought of as a specific type of Born level unitarity cuts.

First, we consider the pole at $l^2=0$ of ${\mathbb M}_{2,2}^\text{tree}(l)$ and $ {\mathbb W}^{\,\text{tree}}_{2,2}(l)$ and then generalize.

\subsection{The $l^2=0$ pole}\la{polesec}

At $l^2=0$ the cylindrically cut one-loop double trace amplitude has a multi-particle factorization pole. Kinematically, it corresponds to a single propagator carrying momentum $l$ around the cylinder going on-shell. In this limit (\ref{cutamp}) becomes
\beq\label{prescpole}
\lim_{l^2\to0}{\mathbb A}_{2,2}^\text{tree}(l) = {1\over l^2}\times\int d^4\tilde{\eta}_l\,A_6^{\text{NMHV}}\left(\tilde{\eta}_1,\tilde{\eta}_2,\tilde{\eta}_l,\tilde{\eta}_4,\tilde{\eta}_5,-\tilde{\eta}_l\right),
\eeq
where the convention $\tilde\eta_{-l}=-\tilde\eta_l$ and $|-l\>=|l\>$ is used.

Note that $l$ drops out of the $\delta^8(\cQ)$ on the r.h.s. Hence, it can also be written as
\beq\la{poleM}
A_6^{\text{NMHV}}\left(\tilde{\eta}_1,\tilde{\eta}_2,\tilde{\eta}_l,\tilde{\eta}_4,\tilde{\eta}_5,-\tilde{\eta}_l\right)={{g_{YM}^2}\,\delta^4\(\sum_{i=1}^4 k_i\)\delta^8\(\sum_{i=1}^4 \lambda_i\tilde\eta_i\)\over\<1\,2\>\<2\,l\>\<l\,3\>\<3\,4\>\<4\,l\>\<l\,1\>}\times\widetilde\cM_6\(0,0,\tilde{\eta}_l,0,0,-\tilde{\eta}_l\)\ ,
\eeq
where $\widetilde\cM$ is the amplitude in the $\tilde\eta$ variables. One can now switch to the $\eta$ variables using
\beq
\tilde\eta_i={\<i-1\,i\>\eta_{i+1}+\<i+1\,i-1\>\eta_i+\<i\,i+1\>\eta_{i-1}\over\<i-1\,i\>\<i\,i+1\>}\ .
\eeq
Since only $\tilde\eta_l=-\tilde\eta_{-l}$ are non zero in $\cM^\text{NMHV}$, it is possible to set $\eta_l=\eta_{-l}=\eta_3=\eta_4=0$. For this choice,
\beq\label{E74}
\eta_1=\<1\,\theta\>\ ,\qquad\eta_2=\<2\,\theta\>\ ,\qquad\theta\equiv|l\>\,\tilde\eta_l\ .
\eeq
Combining (\ref{prescpole}), (\ref{poleM}) and (\ref{E74}) gives 
\beq\label{pole}
\lim_{l^2\to0}{\mathbb A}_{2,2}^\text{1-loop}(l)
= {1\over l^2}\times A_6^{\text{tree}}(1,2,l,3,4,-l)\times\int d^4\tilde\eta_l\,\cM\(\<1\,\theta\>,\<2\,\theta\>,0,0,0,0\)\ ,
\eeq
where $\cM$ is the amplitude in the $\eta$ variables.

Similarly to the cylindrically cut amplitude, the tree-level Wilson lines correlator $\mathbb{W}_{2,2}$ in (\ref{duality12}) has a pole at $l^2=0$. At this pole the cusps at $x_2$ and at $\dot x _2$ become null separated. The factorization of the Wilson lines correlator happens inside one building block and its dynamics are therefore the same as in the single trace Wilson loop case that has been worked out in detail in \cite{CaronHuot:2010ek}. In this kinematical limit the correlator becomes
\beq\la{factorizationWL}
\widetilde\cW_{2,2}(1,2;3,4|l)\quad\rightarrow\quad[4^{[-]}\,3\,2\,1^{[+]}]\times W_6(1,2,l,3,4,\check l)\ ,
\eeq
where $W_6$ is the single trace hexagon super Wilson loop, $\widetilde\cW_{2,2}$ has been defined in (\ref{WL12}) and
\beq
\cZ_l=\<k\,2\,1^{[+]}\,[4^{[-]}\>\cZ_{3]}\ ,\qquad \check\cZ_{l}=\<k\,4\,3^+\,[2^{[-]}\>\cZ_{1]}\ ,
\eeq	
are super momentum twistors (\ref{supertwistors}) that correspond to factorization edges. The supersymmetric version of a cut propagator is
\beq\la{superfactorization}
[4\,3\,2\,1]=\delta(\<1\,2\,3\,4\>){\delta^{0|4}(\<k\,1\,2\,3\>\eta_4+\<4\,k\,1\,2\>\eta_3+\<3\,4\,k\,1\>\eta_2+\<1\,3\,4\,k\>\eta_1)\over\<k\,1\,2\,3\>\,\<4\,k\,1\,2\>\,\<3\,4\,k\,1\>\,\<1\,3\,4\,k\>}\ .
\eeq
Plugging (\ref{factorizationWL}) into (\ref{duality12}), then into the duality equation (\ref{AtoR}), and using the relation 
\beq
{1\over l^2}\ \rightarrow\ \delta(l^2)=\<1\,2\>\,\<3\,4\>\,\delta(\<1^{[+]}\,2\,3\,4^{[-]}\>)\ ,
\eeq
gives
\beq
\lim_{l^2\to0}{\mathbb A}_{2,2}^\text{1-loop}(l)= {\<1\,2\>\,\<3\,4\>\<l\,1\>^4\over\<1^{[+]}\,2\,3\,4^{[-]}\>}\times A_6^{\text{tree}}(1,2,l,3,4,-l)\times\int d^4\eta\,\cM\(\eta,{\<2\,l\>\over\<1\,l\>}\eta,0,0,0,0\)\ ,
\eeq
which is equal to (\ref{pole}). Similarly, plugging (\ref{factorizationWL}) into (\ref{2222}) and (\ref{AtoR}) gives the hybrid form of the same equation,
\beqa\la{pole2}
\lim_{l^2\to0}{\mathbb A}_{2,2}^\text{tree}(l)&=& {\<1\,2\>\,\<3\,4\>\<l\,1\>^2\<l\,3\>^2\over\<1^{[+]}\,2\,3\,4^{[-]}\>}\times A_6^{\text{tree}}(1,2,l,3,4,-l)\\
&\times&\int d\eta_1^1\,d\eta_1^2\int d\eta_3^3\,d\eta_3^4\,\cM\(\eta_1,{\<2\,l\>\over\<1\,l\>}\eta_1,0,\eta_3,{\<4\,l\>\over\<3\,l\>}\eta_3,0\)\ .\nn
\eeqa
We conclude that both ${\mathbb M}_{\,2,2}^\text{tree}(l)$ and ${\mathbb W}^{\,\text{tree}}_{2,2}(l)$ have a pole at $l^2=0$ with the same residue,
\beq\la{polematch}
\lim_{l^2\to0}{\mathbb M}_{\,2,2}^\text{tree}(l)=\lim_{l^2\to0}{\mathbb W}^{\,\text{tree}}_{2,2}(l)\ .
\eeq

The same manipulations directly apply to any kinematical point where a cusp on the left line becomes null separated from a cusp on the right line. These are located at $(l+a\,q)^2=0$, $(l+k_2+a\,q)^2=0$, $(l-k_4+a\,q)^2=0$ and $(l+k_2-k_4+a\,q)^2=0$. In order to generalize (\ref{polematch}) for these poles one simply has to redefine $l$ accordingly. It is also straightforward to make this generalization for any number of particles in the traces. For example, for $n$ particles in the first trace, instead of $\{\<\theta\,1\>,\<\theta\,2\>\}$ in (\ref{pole}) and $\{\eta_1,{\<2\, l\>\over\<1\,l\>}\eta_1\}$ in (\ref{pole2}), one has $\{\<\theta\,1\>,\dots,\<\theta\,n\>\}$ and $\{\eta_1,{\<2\, l\>\over\<1\,l\>}\eta_1,\dots,{\<n\, l\>\over\<1\,l\>}\eta_1\}$, respectively.

\subsection{BCFW recursion relation at Born level}\la{treeBCFW}

We choose the BCFW deformation,
\beq\la{BCFWshift}
\cZ_1\rightarrow \cZ_1+z \cZ_{n^{[-]}}\qquad\Rightarrow\qquad \cZ_{1^{[a]}}\rightarrow \cZ_{1^{[a]}}+z \cZ_{n^{[a-1]}}\ ,
\eeq
where $n$ is the number of particles in the first trace and the superscript represents the block shift as in (\ref{blockshift}). Clearly, the momentum flow through the cylinder $q$ is independent of this deformation and, therefore, it is not a function of $z$. The momentum $l$, as defined in (\ref{distance}), is also independent of $z$. Of all the Wilson lines' cusps, only $x_1\to\hat x_1(z)$ is a function of $z$. As a result, distances such as $l-k_1=\dot x_m-x_1$ do depend on $z$.

Any representative of the unsummed cylindrically cut one-loop amplitude $\cA(l)$ is a manifestly rational function. Shifting $l$ by $q$ shifts all the poles of $\cA(l)$. Hence, the summation over $n$ in (\ref{cutsum}) produces a meromorphic function of $z$ 	with an infinite set of well separated poles. A dispersion integral can be written for such a function,
\beq\la{dtdispersion}
{\mathbb M}_{n,m}^\text{tree}(l)
=\oint{dz\over2\pi i\, z}\widehat{\mathbb M}_{n,m}^\text{tree}(l;z)\ ,
\eeq
where the contour encircles the pole at $z=0$, but no other pole. 

Similarly, the tree-level Wilson lines correlator can be written as an infinite sum of rational functions, with $l$ shifted by an integer multiple of $q$. Such a shifted block decomposition of ${\mathbb W}_{2,2}$ has been worked out explicitly in section \ref{oneloopsec}. It is a meromorphic function of $z$ with an infinite set of well separated poles and hence,
\beq\la{BCFWint}
{\mathbb W}_{n,m}^{\,\text{tree}}(l)=\oint{dz\over2\pi i\, z}\widehat{\mathbb W}_{n,m}^{\,\text{tree}}(l;z)\ .
\eeq

\begin{figure}
\begin{center}
\includegraphics[width=0.84\textwidth]{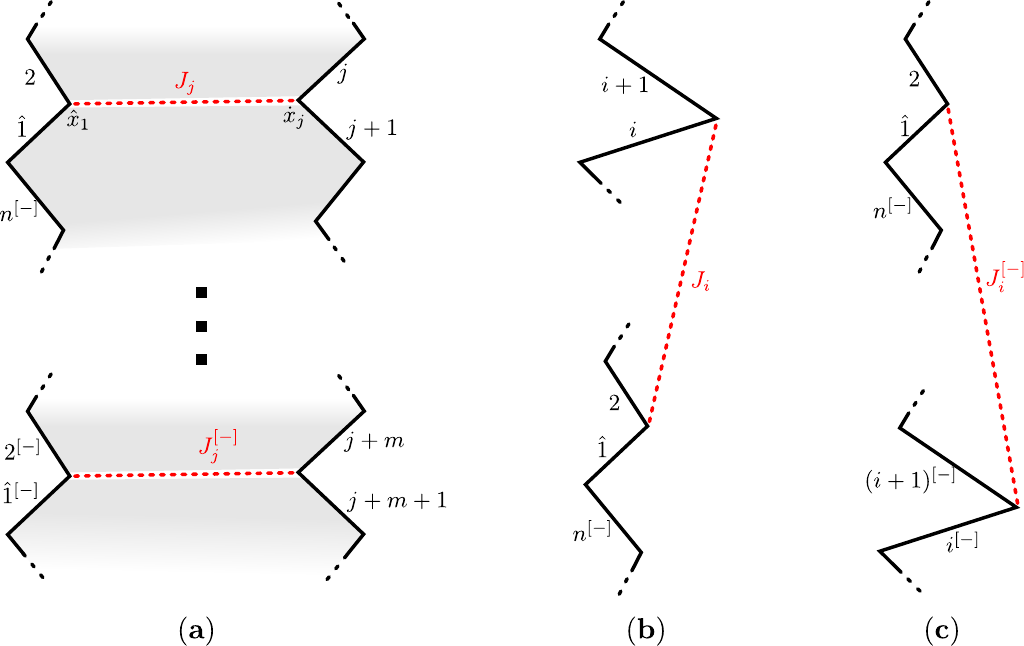}
\caption{\small Factorization poles of the Wilson lines correlator. a) Poles for which the cusp $\hat x_1(z)$ becomes null separated from a cusp on the right Wilson line. b,c) Poles for which the cusp $\hat x_1(z)$ becomes null separated from a cusp on the same Wilson line. That cusp can either be located above $\hat{x}_1(z)$, (b), or below $\hat{x}_1(z)$, (c), and the two must lie within one period.}\label{BCFWtree}
\end{center}
\end{figure}
This integral can be evaluated by summing over the residues at the poles of $\widehat{\mathbb W}_{n,m}^{\,\text{tree}}(l;z)$. There are two types of poles. First, there are the factorization poles for which the cusp $\hat x_1$ on the left Wilson line becomes null separated from a cusp on the right Wilson line, see figure \ref{BCFWtree}.a. These poles are located at
\beq\la{factorizationpoles}
z_j=-{\<1\,2\,j\,j+1\>\over\<n^{[-]}\,2\,j\,j+1\>}\ ,\qquad (\hat x_1-\dot x_j)^2\to0\ .
\eeq
The match of the $l^2=0$ pole (\ref{polematch}), shown in section \ref{polesec}, generalizes to all the poles in (\ref{factorizationpoles}), as explained in section \ref{polesec}. On the amplitude side of the duality, these correspond to the kinematical points were $l+k_2 +a\,q$ or $l+k_2-k_4+a\,q$ go on-shell.

The second type are the poles that correspond to two cusps on the same Wilson line becoming null separated, see figures \ref{BCFWtree}.b and \ref{BCFWtree}.c. These poles can be divided into three groups as follows,
\begin{itemize}
\item Two cusps on the same Wilson line that are more than one block away from each other become null separated. That is, $(\hat x_1-x_{1+r})^2\to0$ where $|r|>n$ is not a multiple of $n$. Such factorization poles are absent on the amplitude side. On the Wilson lines side they are inconsistent with planarity and the periodicity constraint. That is, the color contractions of such poles cross one of their periodic images, making their contribution non-planar.
\item The sum of all momenta in one trace goes on shell, that is, $q^2\to0$. Since $q=\hat x_1-\hat x_{1+n}$ is independent of $z$, this pole does not contribute to the integral (\ref{BCFWint}). For the four-point amplitude one may naively expect the pole at $z=\infty$ to be of this type. However, since $q$ is independent of $z$, such a pole at infinity cannot be present. We have confirmed this by an explicit calculation, see appendix \ref{zinfapp}. Note that individual building blocks of the Wilson lines correlator {with $n=m=2$} do have poles at $z=\infty$, but these cancel out after summation over all possible shifts of $l$ by $q$.
\item Finally, there are the factorization poles for which two cusps that fit inside a single building block become null separated. They are located at (see figures \ref{BCFWtree}.b and \ref{BCFWtree}.c)
\beq
z_i=-{\<1\,2\,i\,i+1\>\over\<n^{[-]}\,2\,i\,i+1\>}\ ,\qquad (\hat x_1- x_i)^2\to0\ ,\qquad |i-1|<n\ .
\eeq
Such factorizations are localized inside each block separately and therefore are insensitive to the periodic arrangement. Hence, the factorization on these poles works in the same way as in the single trace case and the matching between the poles is automatic. For $n>2$, this set of poles includes the $i=-1=(n-1)^{[-]}$ case for which the deformed cusp $\hat x_1$ becomes null separated from $x_{n-1}^{[-]}$. At this collinear limit $z=\infty$ and the number of edges in the left Wilson line is reduced by one due to $\cZ_1$ being omitted. This results in a lower-point Wilson lines correlator ${\mathbb W}_{n-1,m}$.\\
Note that supersymmetry can be used to avoid all factorization channels of this type by choosing a parametrization of the Wilson lines correlator for which any subset of the $\eta$'s in the left trace cannot form a singlet. Since the supersymmetric $\delta$-function on the factorization pole (\ref{superfactorization}) is a combination of singlets made of four $\eta$'s, it cannot appear if the two null separated cusps are on the same Wilson line. One such choice for $n=m=2$ is (\ref{2222}). 
\end{itemize} 

Summing up these two types of residues results in the following representation of the cylindrically cut amplitude,
\beqa\la{BCFWdt}
&&\!\!\!\!\!\!\!\!\!\!\!\!\!\!\!\!\!\!\!\!{\mathbb W}_{n,m}(1,\dots,n;n+1,\dots,n+m;l)\nn\\
&=&\!\!\sum_{j=-\infty}^{\infty}\int d^8\theta\,[n^{[-]}\,1\,2\,j\,j+1]\,W_{n+m+2}(J_j,j+1,\dots,j+m,J_j^{[-]},2^{[-]},\dots,\hat 1_j)\nn\\
&+&{\mathbb W}_{n-1,m}(2,3,4,\dots,n;n+1,\dots,n+m;l)\\
&+&\!\!\sum_{i=2}^{n-2}[n^{[-]}\,1\,2\,i^{[-]}\,(i+1)^{[-]}]\,W_{n+2-i}(J_{i^{[-]}},(i+1)^{[-]},\dots,n^{[-]},\hat 1_{i^{[-]}})\,\times\nn\\
&&\qquad\qquad\qquad\qquad\quad\qquad\ \times\,{\mathbb W}_{i,m}(J_{i^{[-]}},2,\dots,i;n+1,\dots,n+m;l)\nn\\
&+&\!\!\sum_{i=4}^{n}[n^{[-]}\,1\,2\,i\,i+1]\,W_i(J_i,2,3,\dots,i)\,{\mathbb W}_{n-i+2,m}(\hat 1_i,J_i,i+1,\dots,n;n+1,\dots,n+m;l)\ ,\nn
\eeqa
where 
\beq\la{hat1}
\hat 1_j=\<2\,j\,j+1\,[1\>\,n^{[-]}]\ ,\qquad J_j=\<n^{[-]}\,1\,2\,[j\>\,j+1]\ ,
\eeq
and $W$ is the super Wilson loop from \cite{CaronHuot:2010ek} or its dual single trace amplitude. The first line on the right hand side of (\ref{BCFWdt}) corresponds to the factorization pictured in figure \ref{BCFWtree}.a, the second and third lines to figure \ref{BCFWtree}.c and the last line to figure \ref{BCFWtree}.b. We checked this relation explicitly in Mathematica for the case of the four-point correlation function ${\mathbb W}_{2,2}$, for which equation (\ref{BCFWdt}) reads
\begin{align}\la{BCFWdtExample}
{\mathbb W}_{2,2}(1,2;3,4;l) = \sum_{j\,=\,-\infty}^{\infty}\int d^8\theta\,[2^{[-]}\,1\,2\,j\,j+1]\,W_{6}(J_j,j+1,j+2,J_j^{[-]},2^{[-]},\hat 1_j)\ .
\end{align}
The $l^2 = 0$ pole of ${\mathbb W}_{2,2}$ has been specifically examined in section \ref{polesec}.\par
The same recursion relation holds for ${\mathbb M}$, with the single trace amplitude $M_n$ instead of $W_n
$ (the two are, of course, equal). One can use it to reduce the number of particles (edges) in each of the traces (Wilson lines) up to $n = m = 2$, for which the duality has been checked explicitly. This process cannot be continued up to the point where there is only one particle (edge) in one of the two traces (lines) because $q$ is independent of the BCFW deformation (\ref{BCFWshift}).\footnote{One may imagine a more complex deformation for which $q$ is a function of $z$ and use it to go all the way up to the point where there is only one particle (edge) in one of the two traces (lines). On the amplitude side such an amplitude is proportional to the trace of a single $SU(N)$ generator and, therefore, vanishes (one can only exchange an on shell color singlet that is dual to a closed string state). On the Wilson lines side one would find an infinite null line. We expect that any propagator ending on it will result in an integral over a total derivative. The boundaries of such integrals are infinitely far away and therefore the integral vanishes.}

\section{The cylinder loop integrand}\la{loopintegrandsec}

The loop integrand has been defined for planar amplitudes \cite{ArkaniHamed:2010kv} and their dual Wilson loops \cite{CaronHuot:2010ek}. It is a rational function which upon integration and regularization gives the loop amplitude. It is important to point out that it is fully determined by its poles and asymptotic behaviour, which allows one to find a recursion relation that enables a systematic construction of the integrand at any loop order \cite{ArkaniHamed:2010kv,CaronHuot:2010ek}.  

We introduce a new class of objects that, we claim, are the natural generalization of the planar integrands, henceforth referred to as {\it cylinder integrands}. Upon integration and regularization, the cylinder integrands give the cylindrically cut double trace amplitude ${\mathbb M}_{n,m}(l)$ and its dual cylinder Wilson lines correlator ${\mathbb W}_{n,m}(l)$. As opposed to the planar case, it is not a rational function. Instead, it is an infinite sum of rational functions that arises from the universal cover of the cylinder. Similarly to the planar loop integrand, it is fully determined by its poles and asymptotic behaviour. In the next section both cylinder integrands, the amplitude one and the Wilson lines one, will be shown to satisfy the same recursion relation and hence to be equal.

The existence of a loop integrand is made possible by the fact that both the cylindrically cut double trace amplitude ${\mathbb M}_{n,m}(l)$ and the cylinder Wilson lines correlator ${\mathbb W}_{n,m}(l)$ are effectively planar objects. Therefore, there is a meaningful way of identifying integrand loop momenta of different Feynman diagrams. In fact, the cylindrical loop integrands have already 	made an appearance in the fishnet model, see (\ref{WLsum}). The same construction that was applied to a single Feynman diagram in the fishnet model generalizes to the corresponding sum of diagrams in ${\cal N}=4$ SYM theory.

\subsection{The cylinder integrand of the Wilson lines correlator}\la{WLintegrand}

The cylindrical Wilson lines integrand can be constructed by following the procedure in \cite{CaronHuot:2010ek} in a straightforward way. The Wilson lines correlator is computed using chiral Lagrangian insertions with the periodicity constraint imposed on them. Every Lagrangian insertion of the form,
\beq
\cL_\text{on-shell}(y)={1\over4}F_{\alpha\,\beta}F^{\alpha\,\beta}-{1\over4}\phi_{A\,B}[\psi^A_\alpha,\psi^{\alpha\,A}]-{1\over64}[\phi_{A\,B},\phi_{C\,D}][\phi^{A\,B},\phi^{C\,D}]\ ,
\eeq
comes with infinitely many images shifted by an integer multiple of $q$,
\beq
\cL_\text{on-shell}^{[a]}(y+a\,q)=\cL_\text{on-shell}(y)\ .
\eeq
These images come about as a diagram in the theory compactified on a circle is lifted to a periodic flat space diagram as described in section \ref{dualitysec}. At $L$-loop order one finds $L$ such Lagrangian insertions with their periodic images,
\beqa
&&{(i\lambda)^L\over L!}\int d^4y_1\dots d^4y_L\times\\
&&\dots\times\[{\cL_\text{on-shell}(y_1)\over\lambda^2}\dots{\cL_\text{on-shell}(y_L)\over\lambda^2}\]\times\[{\cL^{[+]}_\text{on-shell}(y_1+q)\over\lambda^2}\dots{\cL^{[+]}_\text{on-shell}(y_L+q)\over\lambda^2}\]\times\dots\ .\nn
\eeqa
\begin{figure}[t]
\centering
\def\svgwidth{6cm}
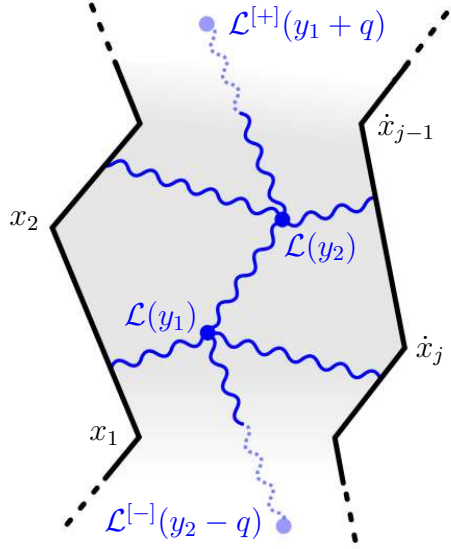
\caption{\small The cylinder integrand of the cylinder Wilson lines correlator. At $L$ loops there are $L$ chiral Lagrangian insertions. Each lagrangian insertion has infinitely many images separated by $q$. These insertions are then contracted with the Wilson lines in all planar periodic ways.}\label{WLintegrand}
\end{figure}
Similarly to $l$ in equation (\ref{modq}), the dual coordinates are only defined modulo a shift,
\beq
y_i\simeq y_i+q\ .
\eeq
Finally, these Lagrangian insertions are contracted at tree level, keeping only planar diagrams that respect the quantum periodicity constraint and taking the contribution of one period. 

The cylinder integrand is then defined as the sum of all such periodic diagrams for a fixed set of points, $\{y_1,\dots,y_L\}$. Shifting any of these points by one period results in the same cylinder integrand. In other words, the set $\{y_1,y_2,\dots,y_L\}$ and the set $\{y_1+q,y_2,\dots,y_L\}$ correspond to the same cylinder integrand.

Each specific periodic contraction, such as the one in figure \ref{WLintegrand}, contains a finite number of propagators and is, therefore, a rational function. The cylinder integrand, on the other hand, is not a rational function. It has poles corresponding to Lagrangian insertion points becoming null separated from cusps of the periodic Wilson lines, $1/(y-x_i-a\,q)^2$. Since a periodic null polygonal Wilson line has infinitely many cusps that are separated by the periodic shift and each insertion point has infinitely many images, the integrand has infinitely many poles at $1/(y-x_i)^2$, $1/(y-x_i-q)^2$, ... . These poles are related by a constant periodic shift and are therefore well separated.

We denote the cylinder integrand of the cylinder Wilson lines correlator at $L$-loop order by
\beq\la{WLci}
\cW_{n,m;L}(k_1,\dots,k_n;k_{n+1},\dots,k_{n+m};l;\{y_i\}_L)\ .
\eeq

When discussing the loop integrand and, in particular, its dual conformal transformations, it is useful to change variables from the chiral Lagrangian insertion points $\{y_i\}$ to their corresponding lines in twistor space $(A_i\,B_i)$,
\beq
d^4y={d^{4}Z_A\,d^{4}Z_B\over\text{vol[GL(2)]}\times\<\lambda_A\,\lambda_B\>^4}\ .
\eeq
It is important to point out that when translating from $d^4y_i$ to ${d^4Z_A\,d^4Z_B\over\text{vol[GL(2)]}}$ one has to absorb the Jacobian, $\<A\,B\>^4=\<\lambda_A\,\lambda_B\>^4$, into the definition of the integrand, see \cite{ArkaniHamed:2010kv,CaronHuot:2010ek} for details. The periodic shift by $q$ acts in twistor space as an $SL(4)$ matrix $\mathbb P(q)$,
\beq\la{Pdefinition}
{\mathbb P}=\(\!\begin{array}{cc}\One&0\\ q&\One\end{array}\!\)\ .
\eeq
The corresponding cylinder integrand is denoted by
\beqa\la{WLciTS}
&&\cW_{n,m;L}(1,\dots,n;n+1,\dots,n+m;{\mathbb P};\{AB\}_L)\\
&&\qquad\qquad\qquad\qquad=\(\prod_{a=1}^L\<A_a\,B_a\>^{-4}\)\times\cW_{n,m;L}(k_1,\dots,k_n;k_{n+1},\dots,k_{n+m};l;\{y_i\}_L)\ .\nn
\eeqa
Here, $1,\dots,n$ and $n+1,\dots,n+m$ represent the twistors $Z_1,\dots,Z_n$ and $\dot Z_1,\dots,\dot Z_m$, respectively. Under the periodic shift, they transform as 
\beq
Z_i^{[a]}={\mathbb P}^{\,a}\cdot Z_i\ ,\qquad \dot Z_j^{[a]}={\mathbb P}^{\,a}\cdot\dot Z_j\ .
\eeq
Shifting all the $Z_i's$ or all the $\dot Z_j$'s by one period results in the same integrand.

Let us consider, for example, the cylindrically cut two-loop amplitude in the fishnet model (\ref{Kaz2loopAmp}). The corresponding integrand is given by the sum, see figure \ref{KZ2LD}, 
\begin{align}\la{KKzintegrand}
&\cW_{2,2;2}^\text{fishnet}(k_1,k_2;k_3,k_4;l;\{y_1,y_2\})\\
 &={1\over2!\,(2\pi)^8} \sum_{a,b,c}\frac{1}{(x_1 - y_1^{[a]})^2(x_2 - y_2^{[b]})^2\,(y_1^{[a]}-y_2^{[b]})^2(y_1^{[a]} - y_2^{[b-1]})^2}\nn\\
&\qquad\quad\qquad\quad\qquad\times\({1\over (y_1^{[a]}-\dot{x}_2^{[c]} )^2(y_2^{[b]}-\dot{x}_1^{[c]})^2}+{1\over (y_1^{[a]}-\dot{x}_1^{[c]} )^2(y_2^{[b]}-\dot{x}_2^{[c+1]})^2}\)\nn\\
&\qquad\quad\qquad\quad+(y_1\leftrightarrow y_2)\ ,\nn
\end{align}
where the factor of $1/2!$ comes from the symmetrization in $y_1$ and $y_2$. Here, the sum over $a$ and $b$ accounts for all shifts of $y_1$ and $y_2$ by an integer multiple of $q$. The sum over $c$ corresponds to all shifts of $l$ by an integer multiple of $q$. The cylindrically cut two-loop amplitude is then given by
\beq
{\mathbb W}_{2,2\,\text{2-loop}}^\text{fishnet}(l)=\lambda^2\int\limits_{y_1\simeq y_1+q}\!\!\!\!d^4y_1\int\limits_{y_2\simeq y_2+q}\!\!\!\!d^4y_2\ \ \cW_{2,2;2}^\text{fishnet}(k_1,k_2;k_3,k_4;l;\{y_1,y_2\})\ .
\eeq
Note that even though the cylinder integrand (\ref{KKzintegrand}) is not a rational function, it is determined by summing over images of a rational function, which can be chosen to be any term in the sum (\ref{KKzintegrand}).

In terms of the twistor variables,
\begin{align}\la{Kzintegrand}
&\cW_{2,2;2}^\text{fishnet}(1,2;3,4;{\mathbb P};\{A\,B\}_2)/(\<1\,2\>^2\,\<3\,4\>^2)\\
 &=\frac{1}{2!} \sum_{a,b,c}\frac{1}{\<1\,2\,A_1^{[a]}\,B_1^{[a]}\>\,\<2\,1^+\,A_2^{[b]}\,B_2^{[b]}\>\,\<A_1^{[a]}\,B_1^{[a]}\,A_2^{[b]}\,B_2^{[b]}\>\,\<A_1^{[a]}\,B_1^{[a]}\, A_2^{[b-1]}\,B_2^{[b-1]}\>}\times\nonumber\\
 &\qquad\qquad\times\({1\over\<A_1^{[a]}\,B_1^{[a]}\,4^{[c]}\,3^{[c-1]}\>\<A_2^{[b]}\,B_2^{[b]}\,3^{[c]}\,4^{[c]}\>}+{1\over\<A_1^{[a]}\,B_1^{[a]}\,3^{[c]}\,4^{[c]}\>\<A_2^{[b]}\,B_2^{[b]}\,4^{[c+1]}\,3^{[c]}\>}\)\nn\\
&+\((A_1,B_1)\leftrightarrow(A_2,B_2)\)\ .\nn
\end{align}
Note that on the left hand side the factor $\<1\,2\>^2\,\<3\,4\>^2$ has been stripped away. This factor will be discussed in section \ref{symmetriessec}.

In order to construct the recursion relation in section \ref{BCFWloopsec} a new object, $\cW_n$, has to be introduced. It is constructed by summing over all periodic shifts of Lagrangian insertion points of the $L$-loop single trace integrand, $W_n$, 
\beq\la{STsum}
\cW_{n}(k_1,\dots,k_n;q;\{y_i\}_L)=\sum_{a_1,\dots,a_L}W_{n}(k_1,\dots,k_n;\{y_i+a_iq\}_L)\ ,
\eeq
or, in terms of twistors,
\beq
\cW_{n}(1,\dots,n;{\mathbb P};\{A,B\}_L)=\sum_{a_1,\dots,a_L}W_{n}(1,\dots,n;\{A^{[a]},B^{[a]}\}_L)\ .
\eeq

\subsection{The cylinder integrand of the cylindrically cut double trace amplitude}\la{cutampinegrand}
\begin{figure}[h]
\centering
\def\svgwidth{12cm}
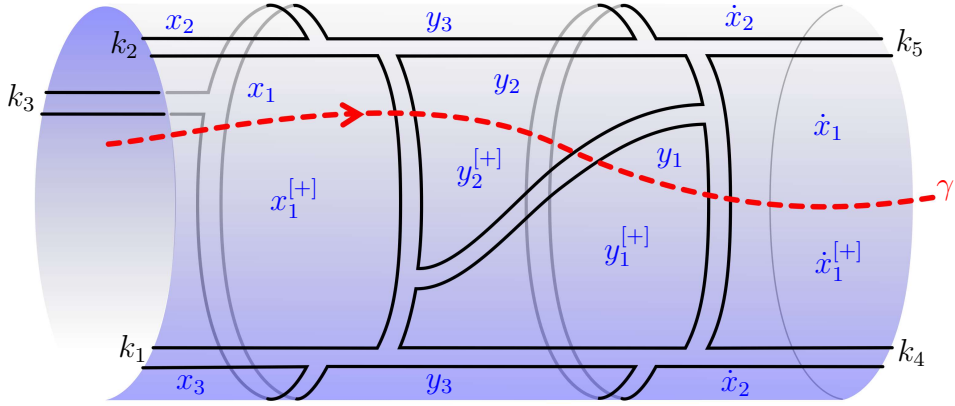
\caption{\small To construct the cylinder integrand of the cylindrically cut double trace amplitude, one has to associate a set of dual coordinates with the faces of any Feynman diagram. Internal faces of the diagram are associated with loop integration variables $y_i$ and external ones with the dual coordinates $x_i$. Circling around the cylinder in the direction of the ordering of the momenta on the first trace results in a shift of the dual coordinates by $q$, $y_i\to y^{[+]}_i$ and $x_i\to x^{[+]}_i$. The points $\{y_i\}_{i=1}^L$ are then symmetriezed.}\label{ampintegrand}
\end{figure}

Let us consider the cylinder cut of a leading color Feynman diagram that contributes to the double trace amplitude, as defined in (\ref{deltainsert}), (\ref{cutamplitude}). If there are more than two particles in each
trace, there are two relative orderings of the traces. The cylinder integrand is defined independently for each of these by following the same procedure as in the planar case. After a change of variables, the cylindrically cut diagrams are written in terms of the dual coordinates $\{x_i\}_{i=1}^n$ instead of the external momenta, see (\ref{xdef}), (\ref{distance}). Similarly, a set of loop integration points $\{y_i\}_{i=1}^L$ is used instead of the loop integration variables $\{l_i\}_{i=1}^L$. The dual $y_i$ variables correspond to the faces of the cut leading color diagrams. They are defined such that the propagator between the faces $y_j$ and $y_k$ has momentum $P_{j\,k}=y_j-y_k$. Similarly, a propagator on the boundary of the leading color diagram carries momentum $P^\text{boundary}_{j\,k}=x_j-y_k$. 

After going around the cylinder and returning to the same face, the corresponding dual $y$-coordinate is shifted by $q$. Hence, the dual coordinates live on the universal cover of the cylinder, see figure \ref{ampintegrand}. For any diagram, the set of points $\{y_i\}_{i=1}^L$ depends on the choice of the cut. To obtain the cylinder integrand that is cut independent and hence physical, the integrand must be symmetrized in $\{y_i\}_{i=1}^L$ and summed over all possible shifts of each of the points $\{y_i\}_{i=1}^L$ and $\{\dot x_j\}_{j=1}^m$ by an integer multiple of $q$. 

The summation over integer shifts of the integration points can be viewed as performing part of the loop integration. That is, shifting a point $y_i\to y_i+q$ amounts to shifting the momentum that is running around the corresponding face by $q$. Because the amplitude is UV finite, the resulting cylinder integrand is finite. 

Similarly to the Wilson lines cylinder integrand introduced in (\ref{WLci}), we denote the cylinder integrand of the cylindrically cut double trace amplitude at $L$ loops by
\beq\la{cutampci}
\cM_{n,m;L}(1,\dots,n;n+1,\dots,n+m;l;\{AB\}_L)\ .
\eeq
As in (\ref{STsum}), the periodic sum of the single trace amplitude integrands is defined as 
\beq\la{STsum2}
\cM_{n}(k_1,\dots,k_n;q;\{y_i\}_L)=\sum_{a_1,\dots,a_L}M_{n}(k_1,\dots,k_n;\{y_i+a_iq\}_L) \ .
\eeq
It is equal to (\ref{STsum}) because the single trace amplitude and Wilson loop integrands are identical.

\section{BCFW recursion relation at loop level}\la{BCFWloopsec}

It has been shown in \cite{CaronHuot:2010ek} that the Wilson loop integrand satisfies the same recursion relation as the loop integrand of the planar amplitude \cite{ArkaniHamed:2010kv} and that the two are equal. To promote this into a proof of the planar amplitude -- Wilson loop duality, one has to integrate the integrand, construct a regularization independent ratio and match the two. Here, we will generalize the integrand construction of \cite{CaronHuot:2010ek} to the double trace duality. The arguments of \cite{CaronHuot:2010ek} apply almost unchanged to the cylinder integrands. Namely, the cylinder integrands of the cylindrically cut amplitude and of the cylinder Wilson lines correlator satisfy the same recursion relations that determine them uniquely. 

At tree level ${\mathbb M}_{n,m}(l)$ and ${\mathbb W}_{n,m}(l)$ satisfy the same BCFW recursion relation as was found in section \ref{BCFWsec}. This recursion relation is generalized to loop level, starting with the periodic Wilson lines picture. 

\subsection{The Wilson lines recursion relation}

\begin{figure}[t]
\centering
\def\svgwidth{9cm}
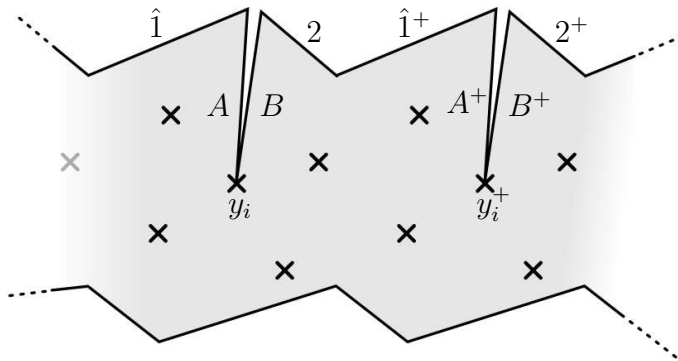
\caption{\small Single-cut factorization poles for the cylinder loop integrand $\cW_{2,2;L}(l)$. The cusp $\hat x_1$ becomes null separated from a chiral Lagrangian insertion point $y_i$. The periodic images of $y_i$ are likewise null separated from images of $\hat x_1$, $(y^{[a]}_i -\hat x^{[a]}_1)^2 = 0$, and are well separated from each other.}\label{singlecut}
\end{figure}

Borrowing the arguments of \cite{CaronHuot:2010ek} is straightforward. As explained in section \ref{WLintegrand}, the chiral Lagrangian insertions are used to construct the Wilson lines cylinder integrand. It is an infinite sum of rational functions of the insertion points and external data with well separated poles. Therefore, the BCFW deformation prescription used at tree-level is still applicable,
\beq\la{WLoint}
{\cal W}_{n,m;L}(l)=\oint{dz\over2\pi i\, z}\widehat {\cal W}_{n,m;L}(l;z)\ ,
\eeq
where the BCFW deformed integrand $\widehat {\cal W}_{n,m;L}(l;z)$ is evaluated on the deformed external supertwistors (\ref{BCFWshift}). 

The integral is evaluated by summing the residues of all the poles of $\widehat {\cal W}_{n,m;L}(l;z)$. The two types of contributions discussed in section \ref{treeBCFW} remain unscathed. They produce either single trace integrands or products of lower-point double trace integrands and single trace ones, see figure \ref{BCFWtree}. At loop level the Lagrangian insertions have to be distributed between the two integrands in each product in all possible ways. Every single trace integrand with Lagrangian insertions has to be summed over all its periodic images. Hence, these single trace loop integrand factors are all of the form (\ref{STsum}). 

The new feature of the recursion relation at loop level is the possibility of the deformed cusp $\hat x_1$ becoming null separated from a Lagrangian insertion point $y_i$. All the periodic images of this cusp, $\hat x_1^{[a]}$, become null separated from the corresponding images of the Lagrangian insertion point, $y_i^{[a]}$. These contributions are referred to as single-cut terms and coincide with the forward limit of lower-loop higher-point cylinder integrands. All the images of this forward limit are well separated in dual coordinate space, see figure \ref{singlecut}.\footnote{Here we take $q$ to be a generic non-null momentum. Hence, the forward limit cannot align with $q$.} Hence, the analysis of \cite{CaronHuot:2010ek} is still applicable with no modifications. It leads to the following recursion relation,
\beqa\la{BCFWdt2}
&&\!\!\!\!\!\!\!\!\!\!\!\!\!\!\!\!\!\!\!\!\cW_{n,m;L}(1,\dots,n;n+1,\dots,n+m;l;\{AB\}_L)\nn\\
&=&\!\!\sum_{j=-\infty}^{\infty}\int d^8\theta\,[n^{[-]}\,1\,2\,j\,j+1]\,\cW_{n+m+2}(J_j,j+1,\dots,j+m,J_j^{[-]},2^{[-]},\dots,\hat 1_j;q;\{AB\}_L)\nn\\
&+&\cW_{n-1,m;L}(2,3,4,\dots,n;n+1,\dots,n+m;l;\{AB\}_L)\\
&+&\!\!{1\over L!}\sum_{\sigma_L}\sum_{i=2}^{n-2}[n^{[-]}\,1\,2\,i^{[-]}\,(i+1)^{[-]}]\,\cW_{n+2-i}(J_{i^{[-]}},(i+1)^{[-]},\dots,n^{[-]},\hat 1_{i^{[-]}};q;\{AB\}_{L/R})\nn\\
&&\qquad\qquad\qquad\qquad\quad\ \times\,\cW_{i,m;R}(J_{i^{[-]}},2,\dots,i;n+1,\dots,n+m;l;\{AB\}_R)\nn\\
&+&\!\!{1\over L!}\sum_{\sigma_L}\sum_{i=4}^{n}[n^{[-]}\,1\,2\,i\,i+1]\,\cW_i(J_i,2,3,\dots,i;q;\{AB\}_{L/R})\nn\\
&&\qquad\qquad\qquad\qquad\quad\ \times\,\cW_{n-i+2,m;R}(\hat 1_i,J_i,i+1,\dots,n;n+1,\dots,n+m;l;\{AB\}_R)\nn\\
&+&\!\!{1\over L}\sum_{j=1}^L\oint\limits_{GL(2)}[A_j\,B_j\,2\,1\,n^{[-]}]\,\cW_{n+2,m;L-1}(\hat 1_{AB_j},\hat A_j,B_j,2,\dots,n;n+1,\dots,n+m;l;\{AB\}_{L/j})\ , \nn
\eeqa
where the shifted momentum supertwistors are given by (\ref{hat1}) and
\beq
\hat 1_{AB_j}=\<A_j\,B_j\,n^{[-]}\,[1\>\,2]\ ,\qquad\hat A_j=\<n^{[-]}\,1\,2\,[A_j\>\,B_j]\ .
\eeq
The sum over $\sigma_L$ is a sum over all partitions of the loop momenta into two sets $\{AB\}_L=\{AB\}_R\cup \{AB\}_{L/R}$. The contour integral over GL(2) matrices $G$, that send $(A,B)$ to $(A',B') = (A,B)G$, is done by taking the residues of the poles located around $A'\propto B'\propto\hat A$.

By repeatedly using the recursion relation (\ref{BCFWdt2}) the double trace integrand can be reduced to a linear combination of products of single trace integrands and $\cW_{2,2}$. 
In a sense, the perturbative double trace amplitude can be constructed by gluing together dual conformal invariant single trace objects. It turns out that a similar structure is also present at finite coupling. There, one can compute the cylindrically cut double trace amplitude using the dual conformal invariant single trace pentagon transitions, see discussion section and \cite{toappear}.   

As opposed to the planar integrand recursion relation, the right hand side of (\ref{BCFWdt2}) contains an infinite sum. The terms in this sum are all related by the periodicity constraint and therefore, they can be generated by a single term in the sum. In other words, the cylinder integrand is generated by summing over all the periodic images of a single rational function, see (\ref{Kzintegrand}) for example. However, there is no unique way of choosing that rational function for the same reason that a cylindrically cut Feynman diagram is not well defined prior to summation over images, see (\ref{cutsum}).

\subsection{The cylindircally cut amplitude recursion relation}

As explained in section \ref{cutampinegrand}, the integrand of the cylindrically cut amplitude is an infinite convergent sum of rational functions with well separated poles. Hence, similarly to the Wilson loop integrand (\ref{WLoint}), it can be written as 
\beq\la{dtdispersion}
{\cal M}_{n,m;L}(l)=\oint{dz\over2\pi i\, z}\widehat{\cal M}_{n,m;L}(l;z)\ ,
\eeq
and evaluated by summing over all the residues of the poles of $\widehat{\cal M}_{n,m;L}(l;z)$. 
The two types of poles that appeared at tree level carry over to loop level, see section \ref{cutampinegrand} and figure \ref{BCFWtree}. The only new type of poles at loop level is single-cut, for which the loop momentum running between the external face of the Feynman diagrams marked by $\hat x_1$ and the internal face marked by $y_i$ goes on-shell. Similarly to the single trace integrand, the residues of these poles are equal to forward limits of lower loop higher-point cylinder integrands. Similarly to all other poles, the forward limit ones enter the cylinder integrands with all their integer periodic shifts.

The summation over all poles of $\widehat{\cal M}_{n,m;L}(l;z)$ leads to the same recursion relation as in (\ref{BCFWdt2}), with $\cM$ instead of $\cW$. Using these recursion relations repeatedly can reduce the integrands $\cM_{n,m;L}$ and $\cW_{n,m;L}$ to tree-level four point integrands $\cM_{2,2;0}$ and $\cW_{2,2;0}$ and single trace integrands (\ref{STsum}) and (\ref{STsum2}). The integrands $\cM_{2,2;0}\propto {\mathbb M}_{2,2}^\text{tree}$ and $\cW_{2,2;0}\propto {\mathbb W}_{2,2}^\text{tree}$ have been shown to be equal in section \ref{BCFWsec}. Hence, we conclude that the loop level cylinder integrands of the cylindrically cut double trace amplitude and the cylinder Wilson lines correlator are the same.

\section{The role of broken dual superconformal symmetry}\la{symmetriessec}

The planar S-matrix of ${\cal N}=4$ SYM and its dual description in terms of polygonal Wilson loops possess a large amount of symmetries. Some of these are anomalous, but the anomaly is well understood and is under control \cite{Drummond:2007au,CaronHuot:2011kk}. These symmetries are so powerful that they fix the result uniquely for any value of the cusp anomalous dimension \cite{CaronHuot:2011kk}. It is therefore useful to understand which of these symmetries are preserved when considering the double trace amplitude. 

Let us consider, for example, the dual conformal generators. Their action on the single trace amplitude is sensitive to the ordering of the external particles in the trace (it is a level-one generator of the Yangian algebra). Hence, the generalization of this type of symmetries to the double trace amplitude is quite interesting. Specifically, in this section we will focus on extending the dual superconformal symmetry. Other symmetries of the planar S-matrix can be similarly extended to symmetries of the cylinder Wilson lines correlator and will not be discussed here. 

The dual conformal generators act locally in dual coordinate space. That is, they are represented by a sum over generators that act on a single vertex of the polygon or a single dual momentum supertwistor. Therefore, the map between the cylindrically cut double trace amplitude and the Wilson lines correlator gives a generalization of dual conformal transformations. Two periodic Wilson lines can be viewed as if they were a single infinite Wilson loop acted on with the standard dual conformal generators. The single trace Wilson loop is, of course, invariant under such transformations (up to the well understood dual conformal anomaly localized at the cusps). One has to keep in mind, however, that in the double trace case the Wilson lines correlator is subject to the quantum periodicity constraint ${\mathbb P}$. 
This constraint is not invariant under dual conformal transformations. Instead, under a dual conformal transformation $\cK$ it transforms into a new ``twisted'' periodicity constraint,
\beq\la{Ptrans}
{\mathbb P}\ \rightarrow\ \widetilde{\mathbb P}={\cal K}\cdot{\mathbb P}\cdot{\cal K}^{-1}\ .
\eeq
As a result, dual conformal transformations map one periodic Wilson lines correlator to a new Wilson lines correlator that is subject to the twisted periodicity constraint $\widetilde{\mathbb P}$. We will argue that this transformation is a symmetry of the cylinder Wilson lines correlators. The original cylinder Wilson lines correlator and the twisted one correspond to the same double trace amplitude. In other words, this symmetry associates different periodic Wilson lines correlators with the same double trace amplitude and, in this sense, can be thought of as a sort of gauge symmetry of the amplitude instead of a global symmetry. Similarly to the planar case, we expect this set of symmetries to uniquely determine the cylinder correlator for any value of the cusp anomalous dimension.

\subsection{Wilson lines correlators with twisted periodicity}

The definition of the cylinder Wilson lines correlator (\ref{superdistance}) contains integration over $\theta$, the superseparation between the lines (\ref{superdistance}). Integrating over it extracts a $\theta$ component of the Wilson lines correlator (\ref{superlines}). Like in the case of the single trace Wilson loop the dual conformal invariant objects are the $\eta$ components of the Wilson lines correlators. The $\eta$ and the $\theta$ components are related by a simple Jacobian, see section \ref{WL1l}.\footnote{This Jacobian factor can be combined with the measure $d^4l$ to produce a manifestly dual conformal invariant integration measure.} For example, changing variables from the eight $\theta$'s to $\eta_i^{A}$ and $\eta_k^{A}$ results in the following expression for the $\theta$ integration in (\ref{cutduality}),
\beq\la{thetatoeta}
\int d^8\theta\,\widehat\cW_{n,m}=\<i\,k\>^4 \int d\eta_i^4d\,\eta_k^4\,\widetilde\cW_{n,m}\ ,
\eeq
where $\widetilde\cW_{n,m}$ is a function of the $\eta$'s and the integration over $\eta_i$ and $\eta_k$ amounts to extracting a specific component of it, see section \ref{WL1l}. It is a function of the supertwistors $\{\cZ_i\}_{i=1}^n$, $\{\dot\cZ_j\}_{j=1}^m$ and the superperiodicity constraint $\cP$, 
\beq\la{CDIWL}
\widetilde\cW_{n,m}=\widetilde\cW_{n,m}(\{\cZ_i\}_{i=1}^n,\{\dot\cZ_j\}_{j=1}^m;\cP)\ ,
\eeq
where $\cP$ is the $6\times 6$ matrix whose upper $4\times4$ block is $\mathbb P$,
\beq\la{superperiod}
\cP=\(\begin{array}{ccc}\One&0&0\\ q&\One&0\\ Q&0&\One\end{array}\!\)\ ,\qquad\cZ^{[a]}=\cP^a\cdot\cZ\ .
\eeq
Here $Q$ is the total supercharge going through the cylinder (\ref{Qperiod}). Similarly to the cylinder Wilson lines correlator in (\ref{CDIWL}), the corresponding integrand will also be denoted by $\widetilde\cW_{n,m}$ with added dependence on the loop integration points, 
\beq\la{CDIWLint}
\widetilde\cW_{n,m;L}=\widetilde\cW_{n,m;L}(\{\cZ_i\}_{i=1}^n,\{\dot\cZ_j\}_{j=1}^m;\{A,B\}_L;\cP)\ .
\eeq
It is related to the loop integrand discussed in section \ref{loopintegrandsec} by the same relation as in (\ref{thetatoeta}).

The cylinder Wilson lines correlator and its integrand can now be generalized to the ``twisted'' ones. Under a dual superconformal transformation ${\cal K}$ the twistors and the periodicity constrint $\cP$ transform to their twisted counterparts, 
\beq\la{DCtransformation}
\cP\ \rightarrow\ \widetilde\cP={\cal K}\cdot\cP\cdot{\cal K}^{-1}\ ,\qquad\cZ^{[a]}\ \rightarrow\ \widetilde\cZ^{[a]}={\cal K}\cdot\cZ^{[a]}=\widetilde\cP^{a}\cdot\widetilde\cZ\ .
\eeq
The twisted cylinder Wilson lines correlator and its integrand are defined as their untwisted versions, with the periodicity constraint $\cP$ replaced by $\widetilde\cP$. This implies that the periodic images of any external twistor and any twistor that parametrizes a chiral Lagrangian insertion point are given by $\widetilde \cZ^{[a]}=\widetilde\cP^{\,a}\cdot\cZ$.

We conjecture that the cylinder integrand (\ref{CDIWLint}) is invariant under the transformations (\ref{DCtransformation}), 
\beq\la{integrandinvariance}
\widetilde\cW_{n,m;L}(\{\cZ_i\}_{i=1}^n,\{\dot\cZ_j\}_{j=1}^m;\{A,B\}_L;\cP)=\widetilde\cW_{n,m;L}(\{\widetilde\cZ_i\}_{i=1}^n,\{\dot{\widetilde\cZ_j}\}_{j=1}^m;\{\widetilde A,\widetilde B\}_L;\widetilde\cP)\ .
\eeq
For example, the right hand side of (\ref{Kzintegrand}) only depends on four-brackets and is, therefore, manifestly invariant under dual conformal transformations (\ref{DCtransformation}). It coincides with the two-loop cylinder integrand $\widetilde\cW_{2,2;2}^{\,\text{fishnet}}(Z_1,Z_2;\dot Z_3,\dot Z_4;\{A,B\}_2;\cP)$ in the fishnet model. 

In general, for (\ref{integrandinvariance}) to hold true all factors of the form $\<\lambda_A\,\lambda_B\>$ must cancel. These spinor helicity two-brackets can be written as $\<\lambda_A\,\lambda_B\>=\<Z_A\,Z_B\, I_\infty\>$, where $(I_\infty)^{KL}$ is the so-called infinity twistor, which is a block diagonal matrix with the only non-zero diagonal element equal to $\epsilon_{ab}$, see for example \cite{ArkaniHamed:2010gh}. It is important to point out that the left and right hand sides of (\ref{integrandinvariance}) are to be computed with the {\it same} infinity twistor.\footnote{If instead one also transforms the infinity twistor, dual conformal invariance becomes a meaningless tautology.} A violation of dual conformal symmetry can be characterised by an explicit dependence on the infinity twistor.

Any specific realization of the matrix $\cP$, such as the one in (\ref{superperiod}), can be thought of as fixing a gauge. Different gauges are related by dual superconformal transformaions (\ref{DCtransformation}). The non-trivial statement in equation (\ref{integrandinvariance}) is that it relates the integrands of two different cylinder Wilson lines correlators that differ both in their geometries and in the periodicity constraint imposed on them. The similarities to gauge theories can be further illustrated by the following argument. Given a specific form of $\cP$, such as the one in (\ref{superperiod}), one can promote any Lorentz invariant function into a function that is invariant under the dual conformal transformation (\ref{DCtransformation}). A similar situation arises for any gauge symmetry. Any observable in a gauged fixed form of the theory can be promoted to a gauge invariant observable in the unfixed theory. For example, the ratio of two-brackets $\<i\,i+1\>^2/\<k\,k+1\>^2$ depends on the infinity twistor and is therefore not dual conformal invariant. However, provided that the gauge has already been fixed by choosing $\cP$ to be the one in (\ref{superperiod}), this ratio can be promoted to the dual conformal invariant ratio $\<i\,i+1\>^2/\<k\,k+1\>^2\to\<i\,i+1\,i^{[+]}\,(i+1)^{[+]}\>/\<k\,k+1\,k^{[+]}\,(k+1)^{[+]}\>$.

Similarly to the cylinder integrand, we conjecture that the cylinder Wilson lines correlator $\widetilde{\cW}_{n,m}$ in (\ref{CDIWL}) is invariant under dual conformal transformations, up to the dual conformal anomaly \cite{Drummond:2007au}. Considering an infinitesimal dual conformal transformation,
\beq
\cK=\epsilon\, \beta_\mu K^\mu\ ,
\eeq
results in the following anomaly equation,
\beqa\la{DCanomaly}
&&\!\!\!{1\over\epsilon}\({\widetilde\cW_{n,m}^\text{finite}(\{\widetilde\cZ_i\}_{i=1}^n,\{\dot{\widetilde\cZ}_j\}_{j=1}^m;\widetilde\cP)_\alpha\over \widetilde\cW_{n,m}^\text{finite}(\{\cZ_i\}_{i=1}^n,\{\dot\cZ_j\}_{j=1}^m;\cP)_\alpha}-1\)\\
&&\qquad\qquad\qquad={1\over2}\Gamma_\text{cusp}(g)\,\beta_\mu\(\sum_{i=1}^n x^\mu_{i,i+1}\log{x_{i,i+2}^2\over x_{i-1,i+1}^2}+\sum_{j=1}^m \dot x^\mu_{j,j+1}\log{\dot x_{j,j+2}^2\over\dot x_{j-1,j+1}^2}\)+\cO(\epsilon)\ ,\nn
\eeqa
where $\widetilde\cW_{n,m}^\text{finite}$ is the regulator-independent part of the Wilson lines correlator $\widetilde\cW_{n,m}$. The subscript $\alpha$ labels the supercomponents. This equation implies that the dual conformal invariance is only broken by local anomalies at the cusps. These anomalies are insensitive to the periodicity constraint that only affects well-separated points. Equation (\ref{DCanomaly}) can be thought of as a gauge anomaly. It is therefore useful to consider anomaly free, and hence dual conformal invariant, ratios instead of $\widetilde\cW_{n,m}$, see \cite{Alday:2010ku,toappear}. Such ratios only depends on $3M-7$ independent conformal cross ratios, where $M$ is the total number of particles.

To summarize, we conjecture that the cylindrically cut amplitude is dual to a family of cylinder Wilson lines correlators, with the generalized periodicity constraint (\ref{Ptrans}). Different Wilson lines correletors with different periodicity constraints that are related by a dual conformal transformation are all different representations of the same cylindrically cut double trace amplitude. In this sense dual conformal symmetry is gauged.

\section{Discussion}\la{discussionsec}

In this paper we have extended the duality between planar scattering amplitudes in ${\cal N}=4$ SYM theory and polygonal Wilson loops to the first $1/N$ correction to the amplitude (\ref{duality1}). This correction corresponds to the leading color double trace contribution to the amplitude. On the other side of the duality one finds the correlation function of two periodic null polygonal Wilson lines subject to a quantum periodicity constraint. Two ideas were necessary to establish this new duality. The first was the recognition of the momentum flow around the cylinder $l$ as a well defined physical quantity that arises in the 't Hooft limit, see (\ref{deltainsert}). The second was cutting open the cylinder by considering its universal cover for any given value of $l$. In particular, the second step allowed us to map the single trace planar duality into a new double trace one. Under this duality, the cylinder momentum $l$ is mapped to the separation between the two periodic Wilson lines. Some of the applications and extensions of this duality will now be discussed.

One application is the extension of the loop integrand to the double trace amplitude discussed in section \ref{loopintegrandsec}. The cylinder loop integrand is given by a sum over all periodic images of a rational function. Similarly to the planar loop integrand, the cylinder integrand satisfies a recursion relation. This relation, along with the planar loop integrand, uniquely determines the cylinder integrand. As opposed to the planar loop integrand, the cylinder loop integrand is not dual conformal invariant. However, the dual conformal invariance of the planar loop integrand has implications on the cylinder integrand. As explained in section \ref{symmetriessec}, these implications can be thought of as gauging of the dual superconformal symmetry in the non-planar case. Similarly to the planar case, the cylinder Wilson lines correlator satisfies a dual conformal anomaly equation (\ref{DCanomaly}).

\begin{figure}[t]
\centering
\def\svgwidth{12cm}
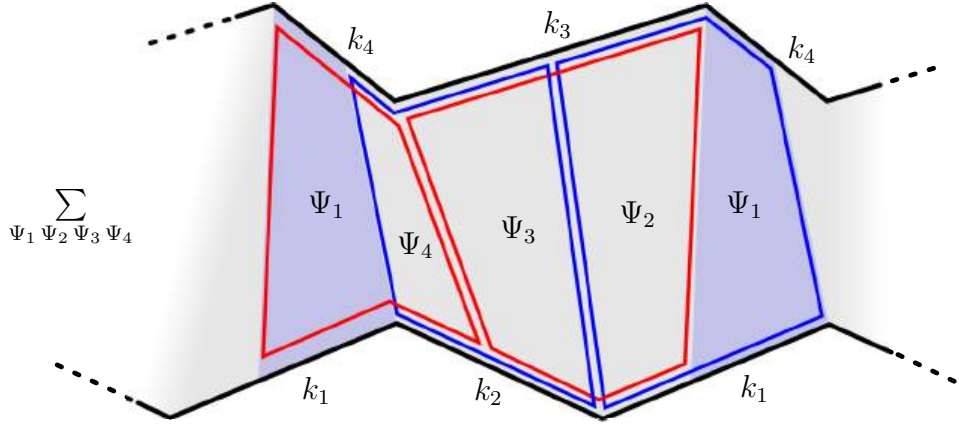
\caption{\small The cylinder Wilson lines correlator can be computed at finite coupling using the pentagon operator product expansion. In this approach the correlator is decomposed into a sequence of GKP flux-tubes that correspond to the null squares in the figure. The quantum periodicity constraint is imposed by identifying the flux-tube state $\Psi_1$ in two channels that are related by the periodicity. This identification corresponds to cutting and gluing the holographic dual string in $AdS_5\times S^5$ spacetime.}\label{POPE}
\end{figure}

Another application of the double trace duality is the extension of the pentagon OPE finite coupling approach to the double trace amplitude \cite{Alday:2010ku,Basso:2013vsa,POPEprogram}. This extension, which was the original motivation for this project, will be reported on in a future publication \cite{toappear} and briefly summarized here. The most unusual feature of the calculation of the Wilson lines correlator in perturbation theory is the fact that the periodicity constraint has to be imposed not only on the geometry of the Wilson lines, but also at the quantum level on each Feynman diagram, see section \ref{dualitysec}. In the POPE approach imposing the same periodicity constraint becomes a natural and simple process. The POPE approach requires one to sum over all possible flux-tube excitations, which can be interpreted as inserting a complete basis of states of the planar flux-tube. The correlation function between two null polygonal Wilson lines can be decomposed into a sequence of flux-tubes, see figure \ref{POPE}. In this case imposing the quantum periodicity constraint amounts to identifying the flux-tube state in a given channel with its periodic image, resulting in a periodic sequence of OPE channels. This is in contrast to the POPE of a single trace null polygonal Wilson loop, for which the sequence of OPE channels starts with the vacuum at the bottom of the polygon and ends with the vacuum at the top. 

Let us consider, for example, the case of the four point double trace amplitude, $n=m=2$. The periodic Wilson lines correlator is decomposed into four POPE channels in a way that is consistent with the periodicity constraint, see figure \ref{WL^2}.\footnote{The general case, in which the number of particles in the two traces is different, will be considered in \cite{toappear}.} The POPE decomposition is then wrapped on a cylinder by identifying the flux-tube state in the next, fifth, channel with the one in the first channel, $|\Psi_5\>=|\Psi_1\>$ in figure \ref{POPE}. Each POPE channel comes with three independent conformal cross ratios, $\{\tau_i,\sigma_i,\phi_i\}_{i=1}^4$, that are repeated periodically \cite{Basso:2013vsa}. However, due to the periodicity constraint, only five of these are independent.\footnote{In general, there are $3M-7$ independent conformal cross ratios, where $M=n+m$ is the total number of particles.} 

\begin{figure}[t]
\centering
\def\svgwidth{16cm}
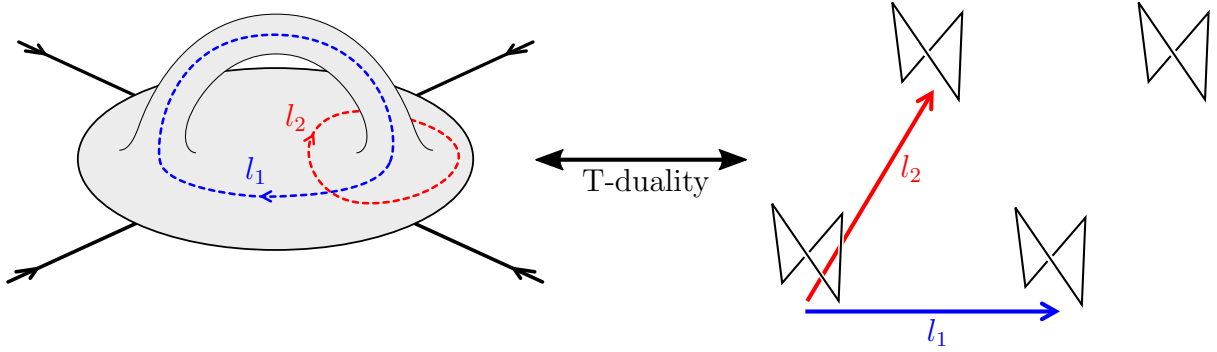
\caption{\small An extension of the double trace duality to a duality between the $1/N^2$ correction to the single trace partial amplitude (on the left) and a periodic correlation function of a Wilson loop-like object (on the right). In this case there are two cuts $\gamma_1$ and $\gamma_2$ with momenta $l_1$ and $l_2$, that correspond to the two cycles of the torus. They are non-self intersecting cuts of the Feynman diagrams that have intersection number one with each other. The generalization of the cylinder cut (\ref{cutsum}) now take the form ${\mathbb A}(l_1,l_2)=\sum_{g\in SL(2,{\mathbb Z})}\cA^{\gamma_1\,\gamma_2}\(g.(l_1,l_2)^{\scriptscriptstyle T}\)$. On the Wilson loop side one finds a lattice of closed polygonal Wilson loops, which is the universal cover of the torus and has two periods, $l_1$ and $l_2$.}\label{torus}
\end{figure}

In addition to these applications, this work can be further extended in several ways, two of which will be discussed below. First, while only the first $1/N$ correction to the planar amplitude in the 't Hooft large $N$ expansion was considered in this paper, the same idea can be applied to any $1/N^L$ order. Consider, for example, the next correction at order $1/N^2$, where there are two types of contributions -- the triple trace pants amplitude and the single trace torus amplitude. The duality for the triple trace is similar to the double trace one, so we will focus on the single trace torus contribution. It is given by the sum of all large 
$N$ 't Hooft Feynman diagrams with torus topology. Hence, it requires two cuts with momenta $l_1^\mu$ and $l_2^\mu$ that correspond to two independent cycles of the torus, see figure \ref{torus}. Similarly to the double trace case, $l_1$ is only defined modulo a shift by $l_2$ and $l_2$ is only defined modulo a shift by $l_1$. In other words, the vectors $(l_1,l_2)$ parametrize a torus in the dual coordinate space. They span a plane that can be parametrized by a complex coordinate. In this two dimensional coordinate system, the modular parameter of the T-dual torus is
\beq
\tau=l_2/l_1 \ .
\eeq
Performing a T-duality transformation gives a two-dimensional lattice of identical closed polygonal Wilson loops, see figure \ref{torus}. There are two quantum periodicity constraints that correspond to the modular transformations of the T-dual torus. The spacetime integration over $l_1$ and $l_2$ can be rearranged into an integration over the modular parameter of the T-dual torus $\tau$, an overall rescaling of the torus, and the spacetime rotations of the torus in four dimensions. In general, one may think of the cut amplitude as the integrand of the string loop expansion in $1/N$. It would be interesting to find out if it satisfies any recursion relations.

The ideas discussed in this paper can also be extended to the computation of form factors in the planar limit. This will be reported on in detail in a future publication \cite{inprogress} and briefly outlined below. Similarly to double trace amplitudes, planar form factors live in momentum space and are evaluated by summing over diagrams with cylinderical topology. Therefore, the same ideas discussed in this paper can be applied to their computation. For example, we claim the form factor of the Lagrangian to be T-dual to the expectation value of a single periodic null polygonal Wilson line subject to a quantum periodicity constraint.\footnote{Similar ideas were discussed in \cite{Maldacena:2010kp,Brandhuber:2010ad}.} This duality would allow us to compute form factors at finite coupling using an extension of the integrability based POPE approach \cite{inprogress}.

\section*{Acknowledgements}

We thank Tom Yahav for collaboration at initial stages of this work. We are grateful to Freddy Cachazo, Claude Duhr and Shota Komatsu for useful discussions. We thank Shota Komatsu for comments on the manuscript and Josua Groeger for useful correspondence. AS has been supported by the I-CORE Program of the Planning and Budgeting Committee, The Israel Science Foundation (grant No. 1937/12) and by the Israel Science Foundation (grant number 968/15) and the EU-FP7 Marie Curie, CIG fellowship. AT is supported by Israel Science Foundation (grant number 968/15). AT and RBI and would like to thank CERN Theoretical Physics Department for their hospitality.

\appendix

\section{Special self symmetry for $n=m=2$}\la{nm2app}
The four-point double trace amplitude contains a special self-symmetry, resulting in a subtlety that warrants clarification. For any $n$ and $m$ the cylinder can be inverted, turning it inside-out. This has the effect of reversing the ordering of the particles in the two traces and the sign of $l$. Therefore, only the relative ordering of the particles in the trace is physical. On the Wilson lines side of the duality reversing the orderings of the edges on the two periodic lines and the sign of $l$ maps the original configuration to a different one that is related to it by the CPT symmetry of the theory.\footnote{In the fishnet model CPT is not a symmetry and, therefore, the 't Hooft string is not orientable.}

For the special case of $n=m=2$ there is no distinct ordering of either of the two traces. As a result, the flipping of the two traces' orderings and the CPT symmetry of the Wilson lines correlator become self-symmetries. This may lead to confusion regarding the definition of the cutting procedure that will now be clarified.  

Consider a leading color Feynman diagram in the case of $n=m=2$. The cut $\gamma$ is defined such that it starts between particles 2 and 1 on one trace and ends between particles 4 and 3 on the other, where the orderings $\dots\to1\to2\to1\to\dots$ and $\dots\to3\to4\to3\to\dots$ are correlated through the cylinder. Alternatively, one may consider a cut $\gamma'$ that starts between particles 1 and 2 on one trace and ends between particles 3 and 4 on the other. By turning the cylinder inside-out, the cut $\gamma$ at a given value of $l$ is mapped to the cut $\gamma'$ at $-l$. Therefore, they are equivalent. The left Wilson line has the sequence of cusps $x^{[a]}_1,x^{[a]}_2$ and the other line has the sequence $\dot x^{[a]}_1,\dot x^{[a]}_2$. The cut $\gamma$ corresponds to the configuration with $l=\dot x_ 2-x_2$ while the cut $\gamma'$ at $-l$ corresponds to the configuration with $-l=\dot x_ 1-x_1$. Despite the fact that these are two different geometrical configurations, they are related by CPT and therefore lead to the same result.

\section{Absence of the pole at $z=\infty$ for $n=m=2$}\la{zinfapp}

In this appendix we will show how the pole at $z=\infty$ vanishes for the deformed tree-level Wilson lines correlator $\widehat{\mathbb W}_{2,2}^{\,\text{tree}}(l;z)$ in (\ref{BCFWint}). In section \ref{oneloopsec} a representation of the Wilson lines correlator in terms of the building block (\ref{2222c}) has been derived. Instead of considering the BCFW deformation of the correlator as a whole, the contribution of each shifted block $\cB(\{Z_i\},\{\dot{Z}_j^{\left[a\right]}\})$ will be evaluated. It will then become apparent that these add up to zero after summation.

Applying the BCFW-deformation (\ref{BCFWshift}) to the building block results in a new object henceforth denoted by $\widehat{\cB}(l;a;z)\equiv \cB(\{\hat{Z}_1(z),Z_i\},\{\dot{Z}_j^{\left[a\right]}\})$. An explicit calculation shows that the $z=\infty$ residue of this object has a remarkably simple form,
\begin{align}
\underset{z\,=\,\infty}{\rm res} \frac{\widehat{\cB}(l;a;z)}{z} = \frac{{\rm b}(l;a) - {\rm b}(l;a-1)}{2}\ ,
\end{align}
where
\begin{align}
{\rm b}(l;a) = \frac{1}{\<1\,2\,3^{[a]}\,4^{[a]}\>\,\<1\,2\,3^{[a+1]}\,4^{[a+1]}\>}\left(1 - \frac{\<1\,2\,3^{[a]}\,3^{[a+1]}\>\,\<1\,2\,4^{[a]}\,4^{[a+1]}\>}{\<1\,2\,3^{[a]}\,4^{[a+1]}\>^2}\right)\ .
\end{align}
One can clearly see that after summation the contributions of all blocks add up to zero, 
\begin{align}
\underset{z\,=\,\infty}{\rm res} \frac{\widehat{\mathbb{W}}_{2,2}^{\text{tree}}(l;z)}{z} = \<1\,2\>^2\,\<3\,4\>^2\sum\limits_{a\,=\,-\,\infty}^{\infty}\underset{z\,=\,\infty}{\rm res} \frac{\widehat{\cB}(l;a;z)}{z} = 0\ .
\end{align}

\end{document}

%% file: Hex_cc.eps_tex
\begingroup%
  \makeatletter%
  \providecommand\color[2][]{%
    \errmessage{(Inkscape) Color is used for the text in Inkscape, but the package 'color.sty' is not loaded}%
    \renewcommand\color[2][]{}%
  }%
  \providecommand\transparent[1]{%
    \errmessage{(Inkscape) Transparency is used (non-zero) for the text in Inkscape, but the package 'transparent.sty' is not loaded}%
    \renewcommand\transparent[1]{}%
  }%
  \providecommand\rotatebox[2]{#2}%
  \ifx\svgwidth\undefined%
    \setlength{\unitlength}{141.88875213bp}%
    \ifx\svgscale\undefined%
      \relax%
    \else%
      \setlength{\unitlength}{\unitlength * \real{\svgscale}}%
    \fi%
  \else%
    \setlength{\unitlength}{\svgwidth}%
  \fi%
  \global\let\svgwidth\undefined%
  \global\let\svgscale\undefined%
  \makeatother%
  \begin{picture}(1,0.63124915)%
    \put(0,0){\includegraphics[width=\unitlength,page=1]{Hex_cc.pdf}}%
    \put(0.08764964,0.20873033){\color[rgb]{0,0,0}\makebox(0,0)[lt]{\begin{minipage}{0.12156851\unitlength}\raggedright $i$\end{minipage}}}%
    \put(-0.0364833,0.5269722){\color[rgb]{0,0,0}\makebox(0,0)[lt]{\begin{minipage}{0.20983064\unitlength}\raggedright $i+1$\end{minipage}}}%
    \put(0.360175,0.67788726){\color[rgb]{0,0,0}\makebox(0,0)[lt]{\begin{minipage}{0.32973373\unitlength}\raggedright $j-1$\end{minipage}}}%
    \put(0.8060024,0.25999699){\color[rgb]{0,0,0}\makebox(0,0)[lt]{\begin{minipage}{0.2231531\unitlength}\raggedright $j+1$\end{minipage}}}%
    \put(0.8080719,0.53880639){\color[rgb]{0,0,0}\makebox(0,0)[lt]{\begin{minipage}{0.17319345\unitlength}\raggedright $j$\end{minipage}}}%
    \put(0.36920697,0.07542009){\color[rgb]{0,0,0}\makebox(0,0)[lt]{\begin{minipage}{0.31974185\unitlength}\raggedright $i-1$\end{minipage}}}%
    \put(0,0){\includegraphics[width=\unitlength,page=2]{Hex_cc.pdf}}%
  \end{picture}%
\endgroup%

%% file: Hex_ce.eps_tex
\begingroup%
  \makeatletter%
  \providecommand\color[2][]{%
    \errmessage{(Inkscape) Color is used for the text in Inkscape, but the package 'color.sty' is not loaded}%
    \renewcommand\color[2][]{}%
  }%
  \providecommand\transparent[1]{%
    \errmessage{(Inkscape) Transparency is used (non-zero) for the text in Inkscape, but the package 'transparent.sty' is not loaded}%
    \renewcommand\transparent[1]{}%
  }%
  \providecommand\rotatebox[2]{#2}%
  \ifx\svgwidth\undefined%
    \setlength{\unitlength}{141.88875213bp}%
    \ifx\svgscale\undefined%
      \relax%
    \else%
      \setlength{\unitlength}{\unitlength * \real{\svgscale}}%
    \fi%
  \else%
    \setlength{\unitlength}{\svgwidth}%
  \fi%
  \global\let\svgwidth\undefined%
  \global\let\svgscale\undefined%
  \makeatother%
  \begin{picture}(1,0.63124915)%
    \put(0,0){\includegraphics[width=\unitlength,page=1]{Hex_ce.pdf}}%
    \put(0.08764964,0.20873033){\color[rgb]{0,0,0}\makebox(0,0)[lt]{\begin{minipage}{0.12156851\unitlength}\raggedright $i$\end{minipage}}}%
    \put(-0.0364833,0.5269722){\color[rgb]{0,0,0}\makebox(0,0)[lt]{\begin{minipage}{0.20983064\unitlength}\raggedright $i+1$\end{minipage}}}%
    \put(0.360175,0.67788726){\color[rgb]{0,0,0}\makebox(0,0)[lt]{\begin{minipage}{0.32973373\unitlength}\raggedright $j-1$\end{minipage}}}%
    \put(0.8060024,0.25999699){\color[rgb]{0,0,0}\makebox(0,0)[lt]{\begin{minipage}{0.2231531\unitlength}\raggedright $j+1$\end{minipage}}}%
    \put(0.8080719,0.53880639){\color[rgb]{0,0,0}\makebox(0,0)[lt]{\begin{minipage}{0.17319345\unitlength}\raggedright $j$\end{minipage}}}%
    \put(0.36920697,0.07542009){\color[rgb]{0,0,0}\makebox(0,0)[lt]{\begin{minipage}{0.31974185\unitlength}\raggedright $i-1$\end{minipage}}}%
    \put(0,0){\includegraphics[width=\unitlength,page=2]{Hex_ce.pdf}}%
  \end{picture}%
\endgroup%

%% file: Hex_ee.eps_tex
\begingroup%
  \makeatletter%
  \providecommand\color[2][]{%
    \errmessage{(Inkscape) Color is used for the text in Inkscape, but the package 'color.sty' is not loaded}%
    \renewcommand\color[2][]{}%
  }%
  \providecommand\transparent[1]{%
    \errmessage{(Inkscape) Transparency is used (non-zero) for the text in Inkscape, but the package 'transparent.sty' is not loaded}%
    \renewcommand\transparent[1]{}%
  }%
  \providecommand\rotatebox[2]{#2}%
  \ifx\svgwidth\undefined%
    \setlength{\unitlength}{141.88875213bp}%
    \ifx\svgscale\undefined%
      \relax%
    \else%
      \setlength{\unitlength}{\unitlength * \real{\svgscale}}%
    \fi%
  \else%
    \setlength{\unitlength}{\svgwidth}%
  \fi%
  \global\let\svgwidth\undefined%
  \global\let\svgscale\undefined%
  \makeatother%
  \begin{picture}(1,0.63124915)%
    \put(0,0){\includegraphics[width=\unitlength,page=1]{Hex_ee.pdf}}%
    \put(0.08764964,0.20873033){\color[rgb]{0,0,0}\makebox(0,0)[lt]{\begin{minipage}{0.12156851\unitlength}\raggedright $i$\end{minipage}}}%
    \put(-0.0364833,0.5269722){\color[rgb]{0,0,0}\makebox(0,0)[lt]{\begin{minipage}{0.20983064\unitlength}\raggedright $i+1$\end{minipage}}}%
    \put(0.360175,0.67788726){\color[rgb]{0,0,0}\makebox(0,0)[lt]{\begin{minipage}{0.32973373\unitlength}\raggedright $j-1$\end{minipage}}}%
    \put(0.8060024,0.25999699){\color[rgb]{0,0,0}\makebox(0,0)[lt]{\begin{minipage}{0.2231531\unitlength}\raggedright $j+1$\end{minipage}}}%
    \put(0.8080719,0.53880639){\color[rgb]{0,0,0}\makebox(0,0)[lt]{\begin{minipage}{0.17319345\unitlength}\raggedright $j$\end{minipage}}}%
    \put(0.36920697,0.07542009){\color[rgb]{0,0,0}\makebox(0,0)[lt]{\begin{minipage}{0.31974185\unitlength}\raggedright $i-1$\end{minipage}}}%
    \put(0,0){\includegraphics[width=\unitlength,page=2]{Hex_ee.pdf}}%
  \end{picture}%
\endgroup%

%% file: cutamp.eps_tex
\begingroup%
  \makeatletter%
  \providecommand\color[2][]{%
    \errmessage{(Inkscape) Color is used for the text in Inkscape, but the package 'color.sty' is not loaded}%
    \renewcommand\color[2][]{}%
  }%
  \providecommand\transparent[1]{%
    \errmessage{(Inkscape) Transparency is used (non-zero) for the text in Inkscape, but the package 'transparent.sty' is not loaded}%
    \renewcommand\transparent[1]{}%
  }%
  \providecommand\rotatebox[2]{#2}%
  \ifx\svgwidth\undefined%
    \setlength{\unitlength}{1408.30435685bp}%
    \ifx\svgscale\undefined%
      \relax%
    \else%
      \setlength{\unitlength}{\unitlength * \real{\svgscale}}%
    \fi%
  \else%
    \setlength{\unitlength}{\svgwidth}%
  \fi%
  \global\let\svgwidth\undefined%
  \global\let\svgscale\undefined%
  \makeatother%
  \begin{picture}(1,0.50241026)%
    \put(0,0){\includegraphics[width=\unitlength,page=1]{cutamp.pdf}}%
    \put(0.3375298,0.34646485){\color[rgb]{0,0,0}\makebox(0,0)[lb]{\smash{$\color{blue}\gamma$}}}%
    \put(0.33637553,0.27357132){\color[rgb]{0,0,0}\makebox(0,0)[lb]{\smash{$P^{-2}_{\color{blue}\gamma[1]}$}}}%
    \put(0.60948869,0.25171994){\color[rgb]{0,0,0}\makebox(0,0)[lb]{\smash{$P^{-2}_{\color{blue}\gamma[2]}$}}}%
    \put(0.77429981,0.14771136){\color[rgb]{0,0,0}\makebox(0,0)[lb]{\smash{$P^{-2}_{\color{blue}\gamma[3]}$}}}%
    \put(0.38297451,0.48279904){\color[rgb]{0,0,0}\makebox(0,0)[lb]{\smash{$\color{red}\gamma'$}}}%
  \end{picture}%
\endgroup%

%% file: WL2.eps_tex
\begingroup%
  \makeatletter%
  \providecommand\color[2][]{%
    \errmessage{(Inkscape) Color is used for the text in Inkscape, but the package 'color.sty' is not loaded}%
    \renewcommand\color[2][]{}%
  }%
  \providecommand\transparent[1]{%
    \errmessage{(Inkscape) Transparency is used (non-zero) for the text in Inkscape, but the package 'transparent.sty' is not loaded}%
    \renewcommand\transparent[1]{}%
  }%
  \providecommand\rotatebox[2]{#2}%
  \ifx\svgwidth\undefined%
    \setlength{\unitlength}{207.95462541bp}%
    \ifx\svgscale\undefined%
      \relax%
    \else%
      \setlength{\unitlength}{\unitlength * \real{\svgscale}}%
    \fi%
  \else%
    \setlength{\unitlength}{\svgwidth}%
  \fi%
  \global\let\svgwidth\undefined%
  \global\let\svgscale\undefined%
  \makeatother%
  \begin{picture}(1,1.25111313)%
    \put(0,0){\includegraphics[width=\unitlength]{WL2.pdf}}%
    \put(0.03423177,0.47004402){\color[rgb]{0,0,0}\makebox(0,0)[lb]{\smash{$i$}}}%
    \put(-0.05040208,0.95476509){\color[rgb]{0,0,0}\makebox(0,0)[lb]{\smash{$i+1$}}}%
    \put(0.8805703,0.75472147){\color[rgb]{0,0,0}\makebox(0,0)[lb]{\smash{$j$}}}%
    \put(0.84979435,0.33924627){\color[rgb]{0,0,0}\makebox(0,0)[lb]{\smash{$j+1$}}}%
    \put(0.1804175,0.52390192){\color[rgb]{0,0,0}\makebox(0,0)[lb]{\smash{$k_1$}}}%
    \put(0.17272352,0.87013125){\color[rgb]{0,0,0}\makebox(0,0)[lb]{\smash{$k_2$}}}%
    \put(0.7113026,0.72394553){\color[rgb]{0,0,0}\makebox(0,0)[lb]{\smash{$k_3$}}}%
    \put(0.66513869,0.40079815){\color[rgb]{0,0,0}\makebox(0,0)[lb]{\smash{$k_4$}}}%
    \put(0.04961974,0.2700004){\color[rgb]{0,0,0}\makebox(0,0)[lb]{\smash{$x_{i-1}$}}}%
    \put(0.44514558,1.07287098){\color[rgb]{0,0,0}\makebox(0,0)[lb]{\smash{$\color{blue}l\color{black}$}}}%
    \put(-0.08887201,0.77780343){\color[rgb]{0,0,0}\makebox(0,0)[lb]{\smash{$x_i$}}}%
    \put(0.91134624,0.47004402){\color[rgb]{0,0,0}\makebox(0,0)[lb]{\smash{$\dot{x}_j$}}}%
    \put(0.81132442,1.000929){\color[rgb]{0,0,0}\makebox(0,0)[lb]{\smash{$\dot{x}_{j-1}$}}}%
  \end{picture}%
\endgroup%

%% file: KazPlanar21.eps_tex
\begingroup%
  \makeatletter%
  \providecommand\color[2][]{%
    \errmessage{(Inkscape) Color is used for the text in Inkscape, but the package 'color.sty' is not loaded}%
    \renewcommand\color[2][]{}%
  }%
  \providecommand\transparent[1]{%
    \errmessage{(Inkscape) Transparency is used (non-zero) for the text in Inkscape, but the package 'transparent.sty' is not loaded}%
    \renewcommand\transparent[1]{}%
  }%
  \providecommand\rotatebox[2]{#2}%
  \ifx\svgwidth\undefined%
    \setlength{\unitlength}{566.64595476bp}%
    \ifx\svgscale\undefined%
      \relax%
    \else%
      \setlength{\unitlength}{\unitlength * \real{\svgscale}}%
    \fi%
  \else%
    \setlength{\unitlength}{\svgwidth}%
  \fi%
  \global\let\svgwidth\undefined%
  \global\let\svgscale\undefined%
  \makeatother%
  \begin{picture}(1,1.00050028)%
    \put(0,0){\includegraphics[width=\unitlength]{KazPlanar21.pdf}}%
    \put(-0.46431021,2.04046126){\color[rgb]{0,0,0}\makebox(0,0)[lt]{\begin{minipage}{0.23564944\unitlength}\raggedright \end{minipage}}}%
    \put(-0.06343033,0.29835853){\color[rgb]{0,0,0}\makebox(0,0)[lt]{\begin{minipage}{0.19766441\unitlength}\raggedright $k_1$\end{minipage}}}%
    \put(-0.06284554,0.74331876){\color[rgb]{0,0,0}\makebox(0,0)[lt]{\begin{minipage}{0.19766441\unitlength}\raggedright $k_2$\end{minipage}}}%
    \put(0.2534013,1.06238924){\color[rgb]{0,0,0}\makebox(0,0)[lt]{\begin{minipage}{0.19766438\unitlength}\raggedright $k_3$\end{minipage}}}%
    \put(0.70235889,1.06309515){\color[rgb]{0,0,0}\makebox(0,0)[lt]{\begin{minipage}{0.19766438\unitlength}\raggedright $k_4$\end{minipage}}}%
    \put(0.50666435,0.54228017){\color[rgb]{0,0,0}\makebox(0,0)[lt]{\begin{minipage}{0.21164059\unitlength}\raggedright $y$\end{minipage}}}%
    \put(0.41398437,0.0721449){\color[rgb]{0,0,0}\makebox(0,0)[lt]{\begin{minipage}{0.66534379\unitlength}\raggedright ${\Red{\phi_2^\dagger(}x_7\Red{)}\over\<7\,8\>c_0^2}$\end{minipage}}}%
    \put(0.15725887,0.83505236){\color[rgb]{0,0,0}\makebox(0,0)[lt]{\begin{minipage}{0.19766438\unitlength}\raggedright $x_2$\end{minipage}}}%
    \put(0.80528255,0.81952238){\color[rgb]{0,0,0}\makebox(0,0)[lt]{\begin{minipage}{0.19766438\unitlength}\raggedright $x_4$\end{minipage}}}%
    \put(0.80810618,0.19691134){\color[rgb]{0,0,0}\makebox(0,0)[lt]{\begin{minipage}{0.19766438\unitlength}\raggedright $x_6$\end{minipage}}}%
    \put(0.16149431,0.18702861){\color[rgb]{0,0,0}\makebox(0,0)[lt]{\begin{minipage}{0.19766438\unitlength}\raggedright $x_8$\end{minipage}}}%
    \put(0.70518252,-0.01412062){\color[rgb]{0,0,0}\makebox(0,0)[lt]{\begin{minipage}{0.19766438\unitlength}\raggedright $k_7$\end{minipage}}}%
    \put(0.25904855,-0.01412057){\color[rgb]{0,0,0}\makebox(0,0)[lt]{\begin{minipage}{0.19766438\unitlength}\raggedright $k_8$\end{minipage}}}%
    \put(1.0057369,0.29411039){\color[rgb]{0,0,0}\makebox(0,0)[lt]{\begin{minipage}{0.19766438\unitlength}\raggedright $k_6$\end{minipage}}}%
    \put(1.0057369,0.73870472){\color[rgb]{0,0,0}\makebox(0,0)[lt]{\begin{minipage}{0.19766438\unitlength}\raggedright $k_5$\end{minipage}}}%
    \put(0.89751873,0.54874688){\color[rgb]{0,0,0}\makebox(0,0)[lt]{\begin{minipage}{0.66812688\unitlength}\raggedright ${\Red{\phi_1^\dagger(}x_5\Red{)}\over\<5\,6\>c_0^2}$\end{minipage}}}%
    \put(0.41867495,1.00026426){\color[rgb]{0,0,0}\makebox(0,0)[lt]{\begin{minipage}{0.59576973\unitlength}\raggedright ${\Red{\phi_2(}x_3\Red{)}\over\<3\,4\>c_0^2}$\end{minipage}}}%
    \put(-0.06670628,0.5354771){\color[rgb]{0,0,0}\makebox(0,0)[lt]{\begin{minipage}{0.59576969\unitlength}\raggedright ${\Red{\phi_1(}x_1\Red{)}\over\<1\,2\>c_0^2}$\end{minipage}}}%
  \end{picture}%
\endgroup%

%% file: Kaz1loop2.eps_tex
\begingroup%
  \makeatletter%
  \providecommand\color[2][]{%
    \errmessage{(Inkscape) Color is used for the text in Inkscape, but the package 'color.sty' is not loaded}%
    \renewcommand\color[2][]{}%
  }%
  \providecommand\transparent[1]{%
    \errmessage{(Inkscape) Transparency is used (non-zero) for the text in Inkscape, but the package 'transparent.sty' is not loaded}%
    \renewcommand\transparent[1]{}%
  }%
  \providecommand\rotatebox[2]{#2}%
  \ifx\svgwidth\undefined%
    \setlength{\unitlength}{2373.91643855bp}%
    \ifx\svgscale\undefined%
      \relax%
    \else%
      \setlength{\unitlength}{\unitlength * \real{\svgscale}}%
    \fi%
  \else%
    \setlength{\unitlength}{\svgwidth}%
  \fi%
  \global\let\svgwidth\undefined%
  \global\let\svgscale\undefined%
  \makeatother%
  \begin{picture}(1,0.29584264)%
    \put(0,0){\includegraphics[width=\unitlength]{Kaz1loop2.pdf}}%
    \put(0.28882364,0.15356446){\color[rgb]{0,0,0}\makebox(0,0)[lt]{\begin{minipage}{0.17201706\unitlength}\raggedright $l+k_2+k_3$\end{minipage}}}%
    \put(0.15197697,0.15164609){\color[rgb]{0,0,0}\makebox(0,0)[lt]{\begin{minipage}{0.17201706\unitlength}\raggedright $l$\end{minipage}}}%
    \put(0.96006644,0.05207848){\color[rgb]{0,0,0}\makebox(0,0)[lt]{\begin{minipage}{0.17201709\unitlength}\raggedright $k_3$\end{minipage}}}%
    \put(0.96084494,0.25544026){\color[rgb]{0,0,0}\makebox(0,0)[lt]{\begin{minipage}{0.17201706\unitlength}\raggedright $k_4$\end{minipage}}}%
    \put(0.08634219,0.28709612){\color[rgb]{0,0,0}\makebox(0,0)[lt]{\begin{minipage}{0.13217551\unitlength}\raggedright $\Blue{\gamma}$\end{minipage}}}%
    \put(0.62553534,0.28709612){\color[rgb]{0,0,0}\makebox(0,0)[lt]{\begin{minipage}{0.13217551\unitlength}\raggedright $\Blue{\gamma}$\end{minipage}}}%
    \put(0.84001246,0.15269306){\color[rgb]{0,0,0}\makebox(0,0)[lt]{\begin{minipage}{0.17201706\unitlength}\raggedright $l+k_2$\end{minipage}}}%
    \put(0.6473406,0.15268677){\color[rgb]{0,0,0}\makebox(0,0)[lt]{\begin{minipage}{0.17201706\unitlength}\raggedright $l-k_4$\end{minipage}}}%
    \put(0.58321878,0.05373401){\color[rgb]{0,0,0}\makebox(0,0)[lt]{\begin{minipage}{0.17201707\unitlength}\raggedright $k_1$\end{minipage}}}%
    \put(0.59106534,0.25340814){\color[rgb]{0,0,0}\makebox(0,0)[lt]{\begin{minipage}{0.17201709\unitlength}\raggedright $k_2$\end{minipage}}}%
    \put(0.41345923,0.05207848){\color[rgb]{0,0,0}\makebox(0,0)[lt]{\begin{minipage}{0.17201709\unitlength}\raggedright $k_4$\end{minipage}}}%
    \put(0.41423778,0.25544026){\color[rgb]{0,0,0}\makebox(0,0)[lt]{\begin{minipage}{0.17201706\unitlength}\raggedright $k_3$\end{minipage}}}%
    \put(0.03661156,0.05373401){\color[rgb]{0,0,0}\makebox(0,0)[lt]{\begin{minipage}{0.17201707\unitlength}\raggedright $k_1$\end{minipage}}}%
    \put(0.04445812,0.25340814){\color[rgb]{0,0,0}\makebox(0,0)[lt]{\begin{minipage}{0.17201709\unitlength}\raggedright $k_2$\end{minipage}}}%
  \end{picture}%
\endgroup%

%% file: KazDRCC2.eps_tex
\begingroup%
  \makeatletter%
  \providecommand\color[2][]{%
    \errmessage{(Inkscape) Color is used for the text in Inkscape, but the package 'color.sty' is not loaded}%
    \renewcommand\color[2][]{}%
  }%
  \providecommand\transparent[1]{%
    \errmessage{(Inkscape) Transparency is used (non-zero) for the text in Inkscape, but the package 'transparent.sty' is not loaded}%
    \renewcommand\transparent[1]{}%
  }%
  \providecommand\rotatebox[2]{#2}%
  \ifx\svgwidth\undefined%
    \setlength{\unitlength}{2615.4538901bp}%
    \ifx\svgscale\undefined%
      \relax%
    \else%
      \setlength{\unitlength}{\unitlength * \real{\svgscale}}%
    \fi%
  \else%
    \setlength{\unitlength}{\svgwidth}%
  \fi%
  \global\let\svgwidth\undefined%
  \global\let\svgscale\undefined%
  \makeatother%
  \begin{picture}(1,0.54618822)%
    \put(0,0){\includegraphics[width=\unitlength]{KazDRCC2.pdf}}%
    \put(0.04962938,0.30900827){\color[rgb]{0,0,0}\makebox(0,0)[lt]{\begin{minipage}{0.05105416\unitlength}\raggedright \end{minipage}}}%
    \put(0.17706752,0.29850258){\color[rgb]{0,0,0}\makebox(0,0)[lt]{\begin{minipage}{0.15613128\unitlength}\raggedright $\Blue{l}$\end{minipage}}}%
    \put(0.1202966,0.15896598){\color[rgb]{0,0,0}\rotatebox{15.47856514}{\makebox(0,0)[lt]{\begin{minipage}{0.20742091\unitlength}\raggedright $\Blue{l+k_2+k_3}$\end{minipage}}}}%
    \put(0.05578088,0.32505271){\color[rgb]{0,0,0}\makebox(0,0)[lt]{\begin{minipage}{0.15613126\unitlength}\raggedright $k_1$\end{minipage}}}%
    \put(0.05179263,0.17790906){\color[rgb]{0,0,0}\makebox(0,0)[lt]{\begin{minipage}{0.15613128\unitlength}\raggedright $k_2$\end{minipage}}}%
    \put(0.29341492,0.23235637){\color[rgb]{0,0,0}\makebox(0,0)[lt]{\begin{minipage}{0.15613126\unitlength}\raggedright $k_3$\end{minipage}}}%
    \put(0.28712536,0.37973942){\color[rgb]{0,0,0}\makebox(0,0)[lt]{\begin{minipage}{0.15613128\unitlength}\raggedright $k_4$\end{minipage}}}%
    \put(-0.08236168,0.1100495){\color[rgb]{0,0,0}\makebox(0,0)[lt]{\begin{minipage}{0.28946266\unitlength}\raggedright $\Red{\phi_1(}x_1\Red{)}$\end{minipage}}}%
    \put(-0.0205751,0.26359826){\color[rgb]{0,0,0}\makebox(0,0)[lt]{\begin{minipage}{0.28946266\unitlength}\raggedright $\Red{\phi_1(}x_2\Red{)}$\end{minipage}}}%
    \put(-0.11050209,0.40613539){\color[rgb]{0,0,0}\makebox(0,0)[lt]{\begin{minipage}{0.32296315\unitlength}\raggedright $\Red{\phi_1(}x_1^{[+]}\Red{)}$\end{minipage}}}%
    \put(0.31221581,0.30152662){\color[rgb]{0,0,0}\makebox(0,0)[lt]{\begin{minipage}{0.28946266\unitlength}\raggedright $\Red{\phi_1^\dagger(}\dot x_2\Red{)}$\end{minipage}}}%
    \put(0.38073157,0.48994496){\color[rgb]{0,0,0}\makebox(0,0)[lt]{\begin{minipage}{0.28946266\unitlength}\raggedright $\Red{\phi_1^\dagger(}\dot x_1\Red{)}$\end{minipage}}}%
    \put(0.39174296,0.21037609){\color[rgb]{0,0,0}\makebox(0,0)[lt]{\begin{minipage}{0.30865875\unitlength}\raggedright $\Red{\phi_1^\dagger(}\dot x_1^{[-]}\Red{)}$\end{minipage}}}%
    \put(0.77748659,0.36597893){\color[rgb]{0,0,0}\rotatebox{36.14636846}{\makebox(0,0)[lt]{\begin{minipage}{0.09082535\unitlength}\raggedright $\Blue{l-k_4}$\end{minipage}}}}%
    \put(0.7448693,0.21222134){\color[rgb]{0,0,0}\rotatebox{32.12808159}{\makebox(0,0)[lt]{\begin{minipage}{0.09905437\unitlength}\raggedright $\Blue{l+k_2}$\end{minipage}}}}%
    \put(0.67181152,0.32872317){\color[rgb]{0,0,0}\makebox(0,0)[lt]{\begin{minipage}{0.15613126\unitlength}\raggedright $k_1$\end{minipage}}}%
    \put(0.67271722,0.19136752){\color[rgb]{0,0,0}\makebox(0,0)[lt]{\begin{minipage}{0.15613128\unitlength}\raggedright $k_2$\end{minipage}}}%
    \put(0.92351575,0.22623889){\color[rgb]{0,0,0}\makebox(0,0)[lt]{\begin{minipage}{0.15613126\unitlength}\raggedright $k_3$\end{minipage}}}%
    \put(0.91477922,0.37851593){\color[rgb]{0,0,0}\makebox(0,0)[lt]{\begin{minipage}{0.15613128\unitlength}\raggedright $k_4$\end{minipage}}}%
    \put(0.53611588,0.10943775){\color[rgb]{0,0,0}\makebox(0,0)[lt]{\begin{minipage}{0.28946266\unitlength}\raggedright $\Red{\phi_1(}x_1\Red{)}$\end{minipage}}}%
    \put(0.59423198,0.26359826){\color[rgb]{0,0,0}\makebox(0,0)[lt]{\begin{minipage}{0.28946266\unitlength}\raggedright $\Red{\phi_1(}x_2\Red{)}$\end{minipage}}}%
    \put(0.50797551,0.39695924){\color[rgb]{0,0,0}\makebox(0,0)[lt]{\begin{minipage}{0.32296315\unitlength}\raggedright $\Red{\phi_1(}x_1^{[+]}\Red{)}$\end{minipage}}}%
    \put(0.92885824,0.30030313){\color[rgb]{0,0,0}\makebox(0,0)[lt]{\begin{minipage}{0.28946266\unitlength}\raggedright $\Red{\phi_1^\dagger(}\dot x_2\Red{)}$\end{minipage}}}%
    \put(1.00104451,0.48933322){\color[rgb]{0,0,0}\makebox(0,0)[lt]{\begin{minipage}{0.28946266\unitlength}\raggedright $\Red{\phi_1^\dagger(}\dot x_1\Red{)}$\end{minipage}}}%
    \put(1.00349144,0.21098784){\color[rgb]{0,0,0}\makebox(0,0)[lt]{\begin{minipage}{0.30865875\unitlength}\raggedright $\Red{\phi_1^\dagger(}\dot x_1^{[-]}\Red{)}$\end{minipage}}}%
  \end{picture}%
\endgroup%

%% file: Kaz2loop2.eps_tex
\begingroup%
  \makeatletter%
  \providecommand\color[2][]{%
    \errmessage{(Inkscape) Color is used for the text in Inkscape, but the package 'color.sty' is not loaded}%
    \renewcommand\color[2][]{}%
  }%
  \providecommand\transparent[1]{%
    \errmessage{(Inkscape) Transparency is used (non-zero) for the text in Inkscape, but the package 'transparent.sty' is not loaded}%
    \renewcommand\transparent[1]{}%
  }%
  \providecommand\rotatebox[2]{#2}%
  \ifx\svgwidth\undefined%
    \setlength{\unitlength}{2322.06512211bp}%
    \ifx\svgscale\undefined%
      \relax%
    \else%
      \setlength{\unitlength}{\unitlength * \real{\svgscale}}%
    \fi%
  \else%
    \setlength{\unitlength}{\svgwidth}%
  \fi%
  \global\let\svgwidth\undefined%
  \global\let\svgscale\undefined%
  \makeatother%
  \begin{picture}(1,0.27818514)%
    \put(0,0){\includegraphics[width=\unitlength]{Kaz2loop2.pdf}}%
    \put(0.13409434,0.66619451){\color[rgb]{0,0,0}\makebox(0,0)[lt]{\begin{minipage}{0.05750476\unitlength}\raggedright \end{minipage}}}%
    \put(0.03990696,0.0424837){\color[rgb]{0,0,0}\makebox(0,0)[lt]{\begin{minipage}{0.17585818\unitlength}\raggedright $k_1$\end{minipage}}}%
    \put(0.04371909,0.2474649){\color[rgb]{0,0,0}\makebox(0,0)[lt]{\begin{minipage}{0.1758582\unitlength}\raggedright $k_2$\end{minipage}}}%
    \put(0.42182792,0.04131386){\color[rgb]{0,0,0}\makebox(0,0)[lt]{\begin{minipage}{0.1758582\unitlength}\raggedright $k_4$\end{minipage}}}%
    \put(0.42251793,0.24849814){\color[rgb]{0,0,0}\makebox(0,0)[lt]{\begin{minipage}{0.17585818\unitlength}\raggedright $k_3$\end{minipage}}}%
    \put(0.07987389,0.28834117){\color[rgb]{0,0,0}\makebox(0,0)[lt]{\begin{minipage}{0.13512697\unitlength}\raggedright $\Blue{\gamma}$\end{minipage}}}%
    \put(0.57735947,0.0424837){\color[rgb]{0,0,0}\makebox(0,0)[lt]{\begin{minipage}{0.17585818\unitlength}\raggedright $k_1$\end{minipage}}}%
    \put(0.58117154,0.2474649){\color[rgb]{0,0,0}\makebox(0,0)[lt]{\begin{minipage}{0.1758582\unitlength}\raggedright $k_2$\end{minipage}}}%
    \put(0.9592803,0.04131388){\color[rgb]{0,0,0}\makebox(0,0)[lt]{\begin{minipage}{0.04744586\unitlength}\raggedright $k_3$\end{minipage}}}%
    \put(0.95997025,0.24849814){\color[rgb]{0,0,0}\makebox(0,0)[lt]{\begin{minipage}{0.17585818\unitlength}\raggedright $k_4$\end{minipage}}}%
    \put(0.61732635,0.28834117){\color[rgb]{0,0,0}\makebox(0,0)[lt]{\begin{minipage}{0.13512697\unitlength}\raggedright $\Blue{\gamma}$\end{minipage}}}%
  \end{picture}%
\endgroup%

%% file: Kaz2loopDual2.eps_tex
\begingroup%
  \makeatletter%
  \providecommand\color[2][]{%
    \errmessage{(Inkscape) Color is used for the text in Inkscape, but the package 'color.sty' is not loaded}%
    \renewcommand\color[2][]{}%
  }%
  \providecommand\transparent[1]{%
    \errmessage{(Inkscape) Transparency is used (non-zero) for the text in Inkscape, but the package 'transparent.sty' is not loaded}%
    \renewcommand\transparent[1]{}%
  }%
  \providecommand\rotatebox[2]{#2}%
  \ifx\svgwidth\undefined%
    \setlength{\unitlength}{2514.52967366bp}%
    \ifx\svgscale\undefined%
      \relax%
    \else%
      \setlength{\unitlength}{\unitlength * \real{\svgscale}}%
    \fi%
  \else%
    \setlength{\unitlength}{\svgwidth}%
  \fi%
  \global\let\svgwidth\undefined%
  \global\let\svgscale\undefined%
  \makeatother%
  \begin{picture}(1,0.54877809)%
    \put(0,0){\includegraphics[width=\unitlength]{Kaz2loopDual2.pdf}}%
    \put(0.05407335,0.31640047){\color[rgb]{0,0,0}\makebox(0,0)[lt]{\begin{minipage}{0.0531033\unitlength}\raggedright \end{minipage}}}%
    \put(-0.02811667,0.1056286){\color[rgb]{0,0,0}\makebox(0,0)[lt]{\begin{minipage}{0.07200258\unitlength}\raggedright $x_1$\end{minipage}}}%
    \put(0.19253815,0.19343612){\color[rgb]{0,0,0}\makebox(0,0)[lt]{\begin{minipage}{0.16239782\unitlength}\raggedright $y_1$\end{minipage}}}%
    \put(0.20526418,0.32006006){\color[rgb]{0,0,0}\makebox(0,0)[lt]{\begin{minipage}{0.05553172\unitlength}\raggedright $y_2$\end{minipage}}}%
    \put(0.18845928,0.49439033){\color[rgb]{0,0,0}\makebox(0,0)[lt]{\begin{minipage}{0.10284648\unitlength}\raggedright $y_1^{[+]}$\end{minipage}}}%
    \put(0.75679105,0.18422883){\color[rgb]{0,0,0}\makebox(0,0)[lt]{\begin{minipage}{0.16239782\unitlength}\raggedright $y_1$\end{minipage}}}%
    \put(0.78058718,0.34259576){\color[rgb]{0,0,0}\makebox(0,0)[lt]{\begin{minipage}{0.05553172\unitlength}\raggedright $y_2$\end{minipage}}}%
    \put(0.73598632,0.48920987){\color[rgb]{0,0,0}\makebox(0,0)[lt]{\begin{minipage}{0.10284648\unitlength}\raggedright $y_1^{[+]}$\end{minipage}}}%
    \put(0.02935638,0.26639464){\color[rgb]{0,0,0}\makebox(0,0)[lt]{\begin{minipage}{0.07200258\unitlength}\raggedright $x_2$\end{minipage}}}%
    \put(-0.04696721,0.41299348){\color[rgb]{0,0,0}\makebox(0,0)[lt]{\begin{minipage}{0.12048276\unitlength}\raggedright $x_1^{[+]}$\end{minipage}}}%
    \put(0.37831241,0.47980824){\color[rgb]{0,0,0}\makebox(0,0)[lt]{\begin{minipage}{0.07200257\unitlength}\raggedright $\dot x_1$\end{minipage}}}%
    \put(0.31316722,0.29596753){\color[rgb]{0,0,0}\makebox(0,0)[lt]{\begin{minipage}{0.07200257\unitlength}\raggedright $\dot x_2$\end{minipage}}}%
    \put(0.38891745,0.20914761){\color[rgb]{0,0,0}\makebox(0,0)[lt]{\begin{minipage}{0.12502775\unitlength}\raggedright $\dot x_1^{[-]}$\end{minipage}}}%
    \put(0.58909578,0.1056286){\color[rgb]{0,0,0}\makebox(0,0)[lt]{\begin{minipage}{0.07200258\unitlength}\raggedright $x_1$\end{minipage}}}%
    \put(0.64656883,0.26639464){\color[rgb]{0,0,0}\makebox(0,0)[lt]{\begin{minipage}{0.07200257\unitlength}\raggedright $x_2$\end{minipage}}}%
    \put(0.57024524,0.41299348){\color[rgb]{0,0,0}\makebox(0,0)[lt]{\begin{minipage}{0.12048276\unitlength}\raggedright $x_1^{[+]}$\end{minipage}}}%
    \put(0.99552486,0.47980824){\color[rgb]{0,0,0}\makebox(0,0)[lt]{\begin{minipage}{0.07200257\unitlength}\raggedright $\dot x_1$\end{minipage}}}%
    \put(0.93037967,0.29596753){\color[rgb]{0,0,0}\makebox(0,0)[lt]{\begin{minipage}{0.07200257\unitlength}\raggedright $\dot x_2$\end{minipage}}}%
    \put(1.0061299,0.20914761){\color[rgb]{0,0,0}\makebox(0,0)[lt]{\begin{minipage}{0.12502775\unitlength}\raggedright $\dot x_1^{[-]}$\end{minipage}}}%
  \end{picture}%
\endgroup%

%% file: cylinder.eps_tex
\begingroup%
  \makeatletter%
  \providecommand\color[2][]{%
    \errmessage{(Inkscape) Color is used for the text in Inkscape, but the package 'color.sty' is not loaded}%
    \renewcommand\color[2][]{}%
  }%
  \providecommand\transparent[1]{%
    \errmessage{(Inkscape) Transparency is used (non-zero) for the text in Inkscape, but the package 'transparent.sty' is not loaded}%
    \renewcommand\transparent[1]{}%
  }%
  \providecommand\rotatebox[2]{#2}%
  \ifx\svgwidth\undefined%
    \setlength{\unitlength}{418.65483398bp}%
    \ifx\svgscale\undefined%
      \relax%
    \else%
      \setlength{\unitlength}{\unitlength * \real{\svgscale}}%
    \fi%
  \else%
    \setlength{\unitlength}{\svgwidth}%
  \fi%
  \global\let\svgwidth\undefined%
  \global\let\svgscale\undefined%
  \makeatother%
  \begin{picture}(1,0.54682542)%
    \put(0,0){\includegraphics[width=\unitlength]{cylinder.pdf}}%
    \put(0.26077051,0.07992042){\color[rgb]{0,0,0}\makebox(0,0)[lb]{\smash{$k_1$}}}%
    \put(0.27987933,0.19075158){\color[rgb]{0,0,0}\makebox(0,0)[lb]{\smash{$k_2$}}}%
    \put(0.26154301,0.38566155){\color[rgb]{0,0,0}\makebox(0,0)[lb]{\smash{$k_n$}}}%
    \put(0.83403512,0.07992042){\color[rgb]{0,0,0}\makebox(0,0)[lb]{\smash{$k_{n+m}$}}}%
    \put(0.85396759,0.28323167){\color[rgb]{0,0,0}\makebox(0,0)[lb]{\smash{$k_{n+2}$}}}%
    \put(0.83480762,0.38566155){\color[rgb]{0,0,0}\makebox(0,0)[lb]{\smash{$k_{n+1}$}}}%
    \put(0.49930131,0.25809373){\color[rgb]{0,0,0}\makebox(0,0)[lb]{\smash{$\color{blue}\gamma\color{black}$}}}%
    \put(0.01312835,0.37801802){\color[rgb]{0,0,0}\makebox(0,0)[lb]{\smash{$\sigma$}}}%
    \put(0.2730083,0.52706682){\color[rgb]{0,0,0}\makebox(0,0)[lb]{\smash{$\tau$}}}%
  \end{picture}%
\endgroup%

%% file: Dual_string_worldsheet.eps_tex
\begingroup%
  \makeatletter%
  \providecommand\color[2][]{%
    \errmessage{(Inkscape) Color is used for the text in Inkscape, but the package 'color.sty' is not loaded}%
    \renewcommand\color[2][]{}%
  }%
  \providecommand\transparent[1]{%
    \errmessage{(Inkscape) Transparency is used (non-zero) for the text in Inkscape, but the package 'transparent.sty' is not loaded}%
    \renewcommand\transparent[1]{}%
  }%
  \providecommand\rotatebox[2]{#2}%
  \ifx\svgwidth\undefined%
    \setlength{\unitlength}{1270.99999616bp}%
    \ifx\svgscale\undefined%
      \relax%
    \else%
      \setlength{\unitlength}{\unitlength * \real{\svgscale}}%
    \fi%
  \else%
    \setlength{\unitlength}{\svgwidth}%
  \fi%
  \global\let\svgwidth\undefined%
  \global\let\svgscale\undefined%
  \makeatother%
  \begin{picture}(1,0.63771477)%
    \put(0,0){\includegraphics[width=\unitlength,page=1]{Dual_string_worldsheet.pdf}}%
    \put(0.51621301,0.18253278){\color[rgb]{0,0,0}\makebox(0,0)[lt]{\begin{minipage}{0.05895445\unitlength}\raggedright $k_2$\end{minipage}}}%
    \put(0.62242289,0.21812307){\color[rgb]{0,0,0}\makebox(0,0)[lt]{\begin{minipage}{0.05895445\unitlength}\raggedright $k_1$\end{minipage}}}%
    \put(0.39589861,0.366799){\color[rgb]{0,0,0}\makebox(0,0)[lt]{\begin{minipage}{0.05895445\unitlength}\raggedright $k_3$\end{minipage}}}%
    \put(0.47318009,0.38082194){\color[rgb]{0,0,0}\makebox(0,0)[lt]{\begin{minipage}{0.05895445\unitlength}\raggedright $k_4$\end{minipage}}}%
    \put(0.29443882,0.10877196){\color[rgb]{0,0,0}\makebox(0,0)[lt]{\begin{minipage}{0.05895445\unitlength}\raggedright $k_2$\end{minipage}}}%
    \put(0.41097521,0.1463292){\color[rgb]{0,0,0}\makebox(0,0)[lt]{\begin{minipage}{0.05895445\unitlength}\raggedright $k_1$\end{minipage}}}%
    \put(0.18948887,0.3063478){\color[rgb]{0,0,0}\makebox(0,0)[lt]{\begin{minipage}{0.05895445\unitlength}\raggedright $k_3$\end{minipage}}}%
    \put(0.26726209,0.32577987){\color[rgb]{0,0,0}\makebox(0,0)[lt]{\begin{minipage}{0.05547736\unitlength}\raggedright $k_4$\end{minipage}}}%
    \put(0.71536722,0.24793404){\color[rgb]{0,0,0}\makebox(0,0)[lt]{\begin{minipage}{0.05895445\unitlength}\raggedright $k_2$\end{minipage}}}%
    \put(0.8191184,0.28499954){\color[rgb]{0,0,0}\makebox(0,0)[lt]{\begin{minipage}{0.05895445\unitlength}\raggedright $k_1$\end{minipage}}}%
    \put(0.58598811,0.41771443){\color[rgb]{0,0,0}\makebox(0,0)[lt]{\begin{minipage}{0.05895445\unitlength}\raggedright $k_3$\end{minipage}}}%
    \put(0.65540178,0.43222911){\color[rgb]{0,0,0}\makebox(0,0)[lt]{\begin{minipage}{0.05895445\unitlength}\raggedright $k_4$\end{minipage}}}%
    \put(0.75273931,0.39004653){\color[rgb]{0,0,0}\makebox(0,0)[lt]{\begin{minipage}{0.05895449\unitlength}\raggedright $l$\end{minipage}}}%
    \put(0.62095333,0.34579004){\color[rgb]{0,0,0}\makebox(0,0)[lt]{\begin{minipage}{0.05895449\unitlength}\raggedright $l$\end{minipage}}}%
    \put(0.41835697,0.29071531){\color[rgb]{0,0,0}\makebox(0,0)[lt]{\begin{minipage}{0.05895449\unitlength}\raggedright $l$\end{minipage}}}%
    \put(0.1990415,0.22777275){\color[rgb]{0,0,0}\makebox(0,0)[lt]{\begin{minipage}{0.05895449\unitlength}\raggedright $l$\end{minipage}}}%
    \put(0,0){\includegraphics[width=\unitlength,page=2]{Dual_string_worldsheet.pdf}}%
    \put(0.66857671,0.12615755){\color[rgb]{0,0,0}\makebox(0,0)[lt]{\begin{minipage}{0.05895445\unitlength}\raggedright $q$\end{minipage}}}%
    \put(0,0){\includegraphics[width=\unitlength,page=3]{Dual_string_worldsheet.pdf}}%
    \put(0.10709525,0.61357628){\color[rgb]{0,0,0}\makebox(0,0)[lt]{\begin{minipage}{0.31015902\unitlength}\raggedright $\text{AdS radial  direction}$\end{minipage}}}%
    \put(0,0){\includegraphics[width=\unitlength,page=4]{Dual_string_worldsheet.pdf}}%
  \end{picture}%
\endgroup%

%% file: WL_NLGluon_Sum2.eps_tex
\begingroup%
  \makeatletter%
  \providecommand\color[2][]{%
    \errmessage{(Inkscape) Color is used for the text in Inkscape, but the package 'color.sty' is not loaded}%
    \renewcommand\color[2][]{}%
  }%
  \providecommand\transparent[1]{%
    \errmessage{(Inkscape) Transparency is used (non-zero) for the text in Inkscape, but the package 'transparent.sty' is not loaded}%
    \renewcommand\transparent[1]{}%
  }%
  \providecommand\rotatebox[2]{#2}%
  \ifx\svgwidth\undefined%
    \setlength{\unitlength}{3909.50441375bp}%
    \ifx\svgscale\undefined%
      \relax%
    \else%
      \setlength{\unitlength}{\unitlength * \real{\svgscale}}%
    \fi%
  \else%
    \setlength{\unitlength}{\svgwidth}%
  \fi%
  \global\let\svgwidth\undefined%
  \global\let\svgscale\undefined%
  \makeatother%
  \begin{picture}(1,0.49944447)%
    \put(0,0){\includegraphics[width=\unitlength]{WL_NLGluon_Sum2.pdf}}%
    \put(0.04070279,0.34028973){\color[rgb]{0,0,0}\makebox(0,0)[lt]{\begin{minipage}{0.03415518\unitlength}\raggedright \end{minipage}}}%
    \put(0.45984431,0.23323115){\color[rgb]{0,0,0}\makebox(0,0)[lt]{\begin{minipage}{0.17299292\unitlength}\raggedright $\color{blue}{a=0}$\end{minipage}}}%
    \put(0.04130361,0.34063957){\color[rgb]{0,0,0}\makebox(0,0)[lt]{\begin{minipage}{0.03415518\unitlength}\raggedright \end{minipage}}}%
    \put(-0.01391349,0.21990408){\color[rgb]{0,0,0}\makebox(0,0)[lt]{\begin{minipage}{0.20760596\unitlength}\raggedright $F^2$\end{minipage}}}%
    \put(-0.01986635,0.40667502){\color[rgb]{0,0,0}\makebox(0,0)[lt]{\begin{minipage}{0.20760596\unitlength}\raggedright $F^2$\end{minipage}}}%
    \put(0.11518904,0.03015692){\color[rgb]{0,0,0}\makebox(0,0)[lt]{\begin{minipage}{0.06548145\unitlength}\raggedright $({\bf a})$\end{minipage}}}%
    \put(-0.05893198,0.16260781){\color[rgb]{0,0,0}\makebox(0,0)[lt]{\begin{minipage}{0.20760595\unitlength}\raggedright $\eta_1\eta_2\eta_3\eta_4$\end{minipage}}}%
    \put(-0.06302457,0.26901516){\color[rgb]{0,0,0}\makebox(0,0)[lt]{\begin{minipage}{0.20760595\unitlength}\raggedright $\eta_1\eta_2\eta_3\eta_4$\end{minipage}}}%
    \put(0.35032679,0.16260781){\color[rgb]{0,0,0}\makebox(0,0)[lt]{\begin{minipage}{0.20760595\unitlength}\raggedright $\eta_1\eta_2$\end{minipage}}}%
    \put(0.3462342,0.26901516){\color[rgb]{0,0,0}\makebox(0,0)[lt]{\begin{minipage}{0.20760595\unitlength}\raggedright $\eta_1\eta_2$\end{minipage}}}%
    \put(0.59178944,0.19534853){\color[rgb]{0,0,0}\makebox(0,0)[lt]{\begin{minipage}{0.20760595\unitlength}\raggedright $\eta_3\eta_4$\end{minipage}}}%
    \put(0.59178944,0.30175587){\color[rgb]{0,0,0}\makebox(0,0)[lt]{\begin{minipage}{0.20760595\unitlength}\raggedright $\eta_3\eta_4$\end{minipage}}}%
    \put(0.8227428,0.29222806){\color[rgb]{0,0,0}\rotatebox{44.80732042}{\makebox(0,0)[lt]{\begin{minipage}{0.09244606\unitlength}\raggedright $\color{blue}{a=1}$\end{minipage}}}}%
    \put(0.71047519,0.16260781){\color[rgb]{0,0,0}\makebox(0,0)[lt]{\begin{minipage}{0.20760595\unitlength}\raggedright $\eta_1\eta_2$\end{minipage}}}%
    \put(0.7063826,0.26901516){\color[rgb]{0,0,0}\makebox(0,0)[lt]{\begin{minipage}{0.20760595\unitlength}\raggedright $\eta_1\eta_2$\end{minipage}}}%
    \put(0.95193751,0.19534853){\color[rgb]{0,0,0}\makebox(0,0)[lt]{\begin{minipage}{0.20760595\unitlength}\raggedright $\eta_3\eta_4$\end{minipage}}}%
    \put(0.95193751,0.30175587){\color[rgb]{0,0,0}\makebox(0,0)[lt]{\begin{minipage}{0.20760595\unitlength}\raggedright $\eta_3\eta_4$\end{minipage}}}%
    \put(0.47942931,0.03015692){\color[rgb]{0,0,0}\makebox(0,0)[lt]{\begin{minipage}{0.06548145\unitlength}\raggedright $({\bf b})$\end{minipage}}}%
    \put(0.84366712,0.03015692){\color[rgb]{0,0,0}\makebox(0,0)[lt]{\begin{minipage}{0.06548145\unitlength}\raggedright $({\bf c})$\end{minipage}}}%
  \end{picture}%
\endgroup%

%% file: Config_3.eps_tex
\begingroup%
  \makeatletter%
  \providecommand\color[2][]{%
    \errmessage{(Inkscape) Color is used for the text in Inkscape, but the package 'color.sty' is not loaded}%
    \renewcommand\color[2][]{}%
  }%
  \providecommand\transparent[1]{%
    \errmessage{(Inkscape) Transparency is used (non-zero) for the text in Inkscape, but the package 'transparent.sty' is not loaded}%
    \renewcommand\transparent[1]{}%
  }%
  \providecommand\rotatebox[2]{#2}%
  \ifx\svgwidth\undefined%
    \setlength{\unitlength}{2673.96603562bp}%
    \ifx\svgscale\undefined%
      \relax%
    \else%
      \setlength{\unitlength}{\unitlength * \real{\svgscale}}%
    \fi%
  \else%
    \setlength{\unitlength}{\svgwidth}%
  \fi%
  \global\let\svgwidth\undefined%
  \global\let\svgscale\undefined%
  \makeatother%
  \begin{picture}(1,0.33053076)%
    \put(0,0){\includegraphics[width=\unitlength]{Config_3.pdf}}%
    \put(0.80062854,0.1458392){\color[rgb]{0,0,0}\makebox(0,0)[lt]{\begin{minipage}{0.68294857\unitlength}\raggedright $\Blue{\eta_1\,\eta_2}$\end{minipage}}}%
    \put(0.80501588,0.20034872){\color[rgb]{0,0,0}\makebox(0,0)[lt]{\begin{minipage}{0.71510461\unitlength}\raggedright $\Blue{\eta_1\,\eta_2}$\end{minipage}}}%
  \end{picture}%
\endgroup%

%% file: WLintegrand.eps_tex
\begingroup%
  \makeatletter%
  \providecommand\color[2][]{%
    \errmessage{(Inkscape) Color is used for the text in Inkscape, but the package 'color.sty' is not loaded}%
    \renewcommand\color[2][]{}%
  }%
  \providecommand\transparent[1]{%
    \errmessage{(Inkscape) Transparency is used (non-zero) for the text in Inkscape, but the package 'transparent.sty' is not loaded}%
    \renewcommand\transparent[1]{}%
  }%
  \providecommand\rotatebox[2]{#2}%
  \ifx\svgwidth\undefined%
    \setlength{\unitlength}{207.95462541bp}%
    \ifx\svgscale\undefined%
      \relax%
    \else%
      \setlength{\unitlength}{\unitlength * \real{\svgscale}}%
    \fi%
  \else%
    \setlength{\unitlength}{\svgwidth}%
  \fi%
  \global\let\svgwidth\undefined%
  \global\let\svgscale\undefined%
  \makeatother%
  \begin{picture}(1,1.25111313)%
    \put(0,0){\includegraphics[width=\unitlength]{WLintegrand.pdf}}%
    \put(0.15733555,0.27769439){\color[rgb]{0,0,0}\makebox(0,0)[lb]{\smash{$x_{1}$}}}%
    \put(-0.01962613,0.75472147){\color[rgb]{0,0,0}\makebox(0,0)[lb]{\smash{$x_2$}}}%
    \put(0.8805703,0.46235003){\color[rgb]{0,0,0}\makebox(0,0)[lb]{\smash{$\dot{x}_j$}}}%
    \put(0.79593645,0.95476509){\color[rgb]{0,0,0}\makebox(0,0)[lb]{\smash{$\dot{x}_{j-1}$}}}%
    \put(0.2344076,0.53219746){\color[rgb]{0,0,0}\makebox(0,0)[lb]{\smash{$\color{blue}\cL(y_1)$}}}%
    \put(0.58833102,0.68607717){\color[rgb]{0,0,0}\makebox(0,0)[lb]{\smash{$\color{blue}\cL(y_2)$}}}%
    \put(0.46522724,1.17849222){\color[rgb]{0,0,0}\makebox(0,0)[lb]{\smash{$\color{blue}\cL^{[+]}(y_1+q)$}}}%
    \put(0.19061113,0.07588495){\color[rgb]{0,0,0}\makebox(0,0)[lb]{\smash{$\color{blue}\cL^{[-]}(y_2-q)$}}}%
  \end{picture}%
\endgroup%

%% file: ampintegrand.eps_tex
\begingroup%
  \makeatletter%
  \providecommand\color[2][]{%
    \errmessage{(Inkscape) Color is used for the text in Inkscape, but the package 'color.sty' is not loaded}%
    \renewcommand\color[2][]{}%
  }%
  \providecommand\transparent[1]{%
    \errmessage{(Inkscape) Transparency is used (non-zero) for the text in Inkscape, but the package 'transparent.sty' is not loaded}%
    \renewcommand\transparent[1]{}%
  }%
  \providecommand\rotatebox[2]{#2}%
  \ifx\svgwidth\undefined%
    \setlength{\unitlength}{1408.30435685bp}%
    \ifx\svgscale\undefined%
      \relax%
    \else%
      \setlength{\unitlength}{\unitlength * \real{\svgscale}}%
    \fi%
  \else%
    \setlength{\unitlength}{\svgwidth}%
  \fi%
  \global\let\svgwidth\undefined%
  \global\let\svgscale\undefined%
  \makeatother%
  \begin{picture}(1,0.50241026)%
    \put(0,0){\includegraphics[width=\unitlength]{ampintegrand.pdf}}%
    \put(0.23527915,0.35782603){\color[rgb]{0,0,0}\makebox(0,0)[lb]{\smash{$\color{blue}x_1$}}}%
    \put(0.09297308,0.07015521){\color[rgb]{0,0,0}\makebox(0,0)[lb]{\smash{$k_1$}}}%
    \put(0.08350543,0.40699951){\color[rgb]{0,0,0}\makebox(0,0)[lb]{\smash{$k_2$}}}%
    \put(-0.03077642,0.34800873){\color[rgb]{0,0,0}\makebox(0,0)[lb]{\smash{$k_3$}}}%
    \put(0.14438971,0.43394595){\color[rgb]{0,0,0}\makebox(0,0)[lb]{\smash{$\color{blue}x_2$}}}%
    \put(0.15688701,0.03403245){\color[rgb]{0,0,0}\makebox(0,0)[lb]{\smash{$\color{blue}x_3$}}}%
    \put(0.26076898,0.23299864){\color[rgb]{0,0,0}\makebox(0,0)[lb]{\smash{$\color{blue}x_1^{[+]}$}}}%
    \put(0.95583972,0.06529033){\color[rgb]{0,0,0}\makebox(0,0)[lb]{\smash{$k_4$}}}%
    \put(0.95388784,0.40944652){\color[rgb]{0,0,0}\makebox(0,0)[lb]{\smash{$k_5$}}}%
    \put(0.86276549,0.31573136){\color[rgb]{0,0,0}\makebox(0,0)[lb]{\smash{$\color{blue}\dot x_1$}}}%
    \put(0.76470979,0.43735428){\color[rgb]{0,0,0}\makebox(0,0)[lb]{\smash{$\color{blue}\dot x_2$}}}%
    \put(0.76016525,0.03176045){\color[rgb]{0,0,0}\makebox(0,0)[lb]{\smash{$\color{blue}\dot x_2$}}}%
    \put(0.86430931,0.16276346){\color[rgb]{0,0,0}\makebox(0,0)[lb]{\smash{$\color{blue}\dot x^{[+]}_1$}}}%
    \put(0.6890273,0.28636713){\color[rgb]{0,0,0}\makebox(0,0)[lb]{\smash{$\color{blue}y_1$}}}%
    \put(0.50838451,0.36816767){\color[rgb]{0,0,0}\makebox(0,0)[lb]{\smash{$\color{blue}y_2$}}}%
    \put(0.43340086,0.43633477){\color[rgb]{0,0,0}\makebox(0,0)[lb]{\smash{$\color{blue}y_3$}}}%
    \put(0.43340086,0.03528523){\color[rgb]{0,0,0}\makebox(0,0)[lb]{\smash{$\color{blue}y_3$}}}%
    \put(0.63088128,0.17910595){\color[rgb]{0,0,0}\makebox(0,0)[lb]{\smash{$\color{blue}y^{[+]}_1$}}}%
    \put(0.46730952,0.265917){\color[rgb]{0,0,0}\makebox(0,0)[lb]{\smash{$\color{blue}y^{[+]}_2$}}}%
    \put(0.9991754,0.25019845){\color[rgb]{0,0,0}\makebox(0,0)[lb]{\smash{$\color{red}\gamma$}}}%
  \end{picture}%
\endgroup%

%% file: singlecut.eps_tex
\begingroup%
  \makeatletter%
  \providecommand\color[2][]{%
    \errmessage{(Inkscape) Color is used for the text in Inkscape, but the package 'color.sty' is not loaded}%
    \renewcommand\color[2][]{}%
  }%
  \providecommand\transparent[1]{%
    \errmessage{(Inkscape) Transparency is used (non-zero) for the text in Inkscape, but the package 'transparent.sty' is not loaded}%
    \renewcommand\transparent[1]{}%
  }%
  \providecommand\rotatebox[2]{#2}%
  \ifx\svgwidth\undefined%
    \setlength{\unitlength}{392.59319155bp}%
    \ifx\svgscale\undefined%
      \relax%
    \else%
      \setlength{\unitlength}{\unitlength * \real{\svgscale}}%
    \fi%
  \else%
    \setlength{\unitlength}{\svgwidth}%
  \fi%
  \global\let\svgwidth\undefined%
  \global\let\svgscale\undefined%
  \makeatother%
  \begin{picture}(1,0.51869634)%
    \put(0,0){\includegraphics[width=\unitlength]{singlecut.pdf}}%
    \put(0.2938264,0.36108421){\color[rgb]{0,0,0}\makebox(0,0)[lb]{\smash{$A$}}}%
    \put(0.37223427,0.3603409){\color[rgb]{0,0,0}\makebox(0,0)[lb]{\smash{$B$}}}%
    \put(0.32577754,0.21728688){\color[rgb]{0,0,0}\makebox(0,0)[lb]{\smash{$y_i$}}}%
    \put(0.20844852,0.47786076){\color[rgb]{0,0,0}\makebox(0,0)[lb]{\smash{$\hat 1$}}}%
    \put(0.44185038,0.4701734){\color[rgb]{0,0,0}\makebox(0,0)[lb]{\smash{$2$}}}%
    \put(0.6483919,0.36515968){\color[rgb]{0,0,0}\makebox(0,0)[lb]{\smash{$A^+$}}}%
    \put(0.73902623,0.3603409){\color[rgb]{0,0,0}\makebox(0,0)[lb]{\smash{$B^+$}}}%
    \put(0.69256943,0.21728688){\color[rgb]{0,0,0}\makebox(0,0)[lb]{\smash{$y_i^+$}}}%
    \put(0.57524041,0.47786076){\color[rgb]{0,0,0}\makebox(0,0)[lb]{\smash{$\hat 1^{+}$}}}%
    \put(0.80864221,0.4701734){\color[rgb]{0,0,0}\makebox(0,0)[lb]{\smash{$2^+$}}}%
  \end{picture}%
\endgroup%

%% file: POPE4.eps_tex
\begingroup%
  \makeatletter%
  \providecommand\color[2][]{%
    \errmessage{(Inkscape) Color is used for the text in Inkscape, but the package 'color.sty' is not loaded}%
    \renewcommand\color[2][]{}%
  }%
  \providecommand\transparent[1]{%
    \errmessage{(Inkscape) Transparency is used (non-zero) for the text in Inkscape, but the package 'transparent.sty' is not loaded}%
    \renewcommand\transparent[1]{}%
  }%
  \providecommand\rotatebox[2]{#2}%
  \ifx\svgwidth\undefined%
    \setlength{\unitlength}{324.78077325bp}%
    \ifx\svgscale\undefined%
      \relax%
    \else%
      \setlength{\unitlength}{\unitlength * \real{\svgscale}}%
    \fi%
  \else%
    \setlength{\unitlength}{\svgwidth}%
  \fi%
  \global\let\svgwidth\undefined%
  \global\let\svgscale\undefined%
  \makeatother%
  \begin{picture}(1,0.43028318)%
    \put(0,0){\includegraphics[width=\unitlength]{POPE4.pdf}}%
    \put(0.75946213,0.02973288){\color[rgb]{0,0,0}\makebox(0,0)[lb]{\smash{$k_1$}}}%
    \put(0.46469146,0.02649279){\color[rgb]{0,0,0}\makebox(0,0)[lb]{\smash{$k_2$}}}%
    \put(0.54125293,0.43342456){\color[rgb]{0,0,0}\makebox(0,0)[lb]{\smash{$k_3$}}}%
    \put(0.32361345,0.41585961){\color[rgb]{0,0,0}\makebox(0,0)[lb]{\smash{$k_4$}}}%
    \put(0.27385987,0.02762163){\color[rgb]{0,0,0}\makebox(0,0)[lb]{\smash{$k_1$}}}%
    \put(0.81203079,0.40530302){\color[rgb]{0,0,0}\makebox(0,0)[lb]{\smash{$k_4$}}}%
    \put(0.62574561,0.22327012){\color[rgb]{0,0,0}\makebox(0,0)[lb]{\smash{$\Psi_2$}}}%
    \put(0.49343658,0.20637952){\color[rgb]{0,0,0}\makebox(0,0)[lb]{\smash{$\Psi_3$}}}%
    \put(0.37942565,0.18878525){\color[rgb]{0,0,0}\makebox(0,0)[lb]{\smash{$\Psi_4$}}}%
    \put(0.28089763,0.23382667){\color[rgb]{0,0,0}\makebox(0,0)[lb]{\smash{$\Psi_1$}}}%
    \put(-0.05128248,0.23453042){\color[rgb]{0,0,0}\makebox(0,0)[lb]{\smash{$\sum\limits_{\Psi_1\,\Psi_2\,\Psi_3\,\Psi_4}$}}}%
    \put(0.74397919,0.23382667){\color[rgb]{0,0,0}\makebox(0,0)[lb]{\smash{$\Psi_1$}}}%
  \end{picture}%
\endgroup%

%% file: torus5.pdf_tex
\begingroup%
  \makeatletter%
  \providecommand\color[2][]{%
    \errmessage{(Inkscape) Color is used for the text in Inkscape, but the package 'color.sty' is not loaded}%
    \renewcommand\color[2][]{}%
  }%
  \providecommand\transparent[1]{%
    \errmessage{(Inkscape) Transparency is used (non-zero) for the text in Inkscape, but the package 'transparent.sty' is not loaded}%
    \renewcommand\transparent[1]{}%
  }%
  \providecommand\rotatebox[2]{#2}%
  \ifx\svgwidth\undefined%
    \setlength{\unitlength}{938.00395815bp}%
    \ifx\svgscale\undefined%
      \relax%
    \else%
      \setlength{\unitlength}{\unitlength * \real{\svgscale}}%
    \fi%
  \else%
    \setlength{\unitlength}{\svgwidth}%
  \fi%
  \global\let\svgwidth\undefined%
  \global\let\svgscale\undefined%
  \makeatother%
  \begin{picture}(1,0.28366637)%
    \put(0,0){\includegraphics[width=\unitlength,page=1]{torus5.pdf}}%
    \put(0.76491811,0.00135876){\color[rgb]{0,0,0}\makebox(0,0)[lb]{\smash{$\Blue{l_1}$}}}%
    \put(0.74316481,0.13808136){\color[rgb]{0,0,0}\makebox(0,0)[lb]{\smash{$\Red{l_2}$}}}%
    \put(0.47860214,0.1262167){\color[rgb]{0,0,0}\makebox(0,0)[lb]{\smash{$\text{T-duality}$}}}%
    \put(0,0){\includegraphics[width=\unitlength,page=2]{torus5.pdf}}%
    \put(0.19850637,0.13658607){\color[rgb]{0,0,0}\makebox(0,0)[lb]{\smash{$\Blue{l_1}$}}}%
    \put(0.23187226,0.18049346){\color[rgb]{0,0,0}\makebox(0,0)[lb]{\smash{$\Red{l_2}$}}}%
  \end{picture}%
\endgroup%